\documentclass[12pt]{article}

\usepackage{epsfig}
\usepackage{amsfonts}
\usepackage{amssymb,amsmath}

\begin{document}

\def\identity{{\rlap{1} \hskip 1.6pt \hbox{1}}}
\def\half{{\textstyle{1\over2}}} 

\newcommand{\NP}{{\em Nucl.\ Phys.\ }}
\newcommand{\AP}{{\em Ann.\ Phys.\ }}
\newcommand{\PL}{{\em Phys.\ Lett.\ }}
\newcommand{\PR}{{\em Phys.\ Rev.\ }}
\newcommand{\PRL}{{\em Phys.\ Rev.\ Lett.\ }}
\newcommand{\PRP}{{\em Phys.\ Rep.\ }}
\newcommand{\CMP}{{\em Comm.\ Math.\ Phys.\ }}
\newcommand{\MPL}{{\em Mod.\ Phys.\ Lett.\ }}
\newcommand{\IJMP}{{\em Int.\ J.\ Mod.\ Phys.\ }}

\pagestyle{plain}
\setcounter{page}{1}

\baselineskip16pt

\begin{titlepage}

\begin{flushright}
MIT-CTP-3430\\
hep-th/0311017
\end{flushright}
\vspace{13 mm}

\begin{center}

{\Large \bf D-Branes, Tachyons, and String Field Theory}\\
\vspace{3mm}
Lectures presented at TASI 2001, Boulder, Colorado

\end{center}

\vspace{7 mm}

\begin{center}

Washington Taylor and Barton Zwiebach\\

\vspace{3mm}
{\small \sl Center for Theoretical Physics} \\
{\small \sl MIT, Bldg.  6} \\
{\small \sl Cambridge, MA 02139, U.S.A.} \\
{\small \tt wati {\rm at} mit.edu},
{\small \tt zwiebach  {\rm at} mitlns.mit.edu}\\
\end{center}

\vspace{8 mm}

\begin{abstract}
In these notes we provide a pedagogical introduction to the subject of
tachyon condensation in Witten's cubic bosonic open string field
theory.  We use both the low-energy Yang-Mills description and the
language of string field theory to explain the problem of tachyon
condensation on unstable D-branes.  We give a self-contained
introduction to open string field theory using both conformal field
theory and overlap integrals.  Our main subjects are the Sen
conjectures on tachyon condensation in open string field theory and
the evidence that supports these conjectures.  We conclude with a
discussion of vacuum string field theory and projectors of the
star-algebra of open string fields. We comment on the possible role of
string field theory in the construction of a nonperturbative
formulation of string theory that captures all possible string
backgrounds.
\end{abstract}

\vspace{1cm}
\begin{flushleft}
October 2003
\end{flushleft}
\end{titlepage}
\newpage


\section{Introduction}
\label{sec:introduction}

The last seven years have been a very exciting time for string theory.
A~new understanding of nonperturbative objects in
           string theory, such
as D-branes, has led to exciting new developments that relate string
theory to physical systems such as  black holes and
supersymmetric gauge field theories.  It has also led to the discovery of
unexpected relationships between Yang-Mills theories and
quantum theories of gravity such as closed superstring theories and
M-theory.
Finally, the analysis of unstable D-branes has elucidated
the long-standing mysteries associated with the
open  string tachyon.

The study of unstable D-branes and tachyons has also led to the
realization that string field theory contains significant
non-perturbative information.  This has been somewhat of a surprise.
Certain forms of string field theory were known since the early
1990's, but there was no concrete evidence that they could be used to
give a non-perturbative definition of string theory.  The study of
tachyon condensation, however, has changed our perspective.  These
lecture notes give an introduction to string field theory and review
recent work in which unstable D-branes and their associated tachyons
are described using string field theory.  As we will discuss here,
this work suggests that open string field theory, or some successor of
it, may give a complete definition of string theory in which all
possible backgrounds can be obtained from a single set of degrees of
freedom.  Such a formulation appears to be necessary to address
questions related to vacuum selection and string cosmology.

In the rest of this section we will review briefly the
current status of string theory as a
whole, and summarize the goals of this set of lectures.  In
section~\ref{sec:D-branes}
we review some basic aspects of D-branes.  In
section~\ref{sec:tachyon-D-branes},
we describe a particular D-brane configuration which exhibits a tachyonic
instability.  This tachyon can be seen in the low-energy
Yang-Mills description of the D-brane system.  We also
describe a set of conjectures made by Sen in 1999, which stated
that the tachyonic instability of the open bosonic string
is the  instability
of the space-filling D25-brane.
Sen suggested that open string field theory could be used to give an
analytic description of this instability.  In section~\ref{sec:SFT}
we give an introduction to
Witten's bosonic open string field theory (OSFT).
Section~\ref{sec:CFT} gives a more detailed analytic description of this theory
using the language of conformal field theory.
Section~\ref{sec:overlaps} describes
the  string field
theory using the oscillator approach and overlap integrals.
            The two approaches to OSFT described in these
two sections give complementary ways of
analyzing problems in string field theory.  In section 7 we
summarize evidence from string field theory for Sen's conjectures.  In
section 8 we describe  ``vacuum string
field theory,'' a new version of open string field theory which arises when one
attempts  to directly formulate the theory around
the classically stable vacuum
where the D-brane has disappeared.
This section also  discusses
important structures in string field theory, such as projectors of the
star algebra
of open string fields.  Section 9 contains concluding remarks.

Much new work has been done in this area since these lectures were
presented at TASI in 2001.  Except for some references to
more recent developments
which are related to the topics covered,
these lecture notes primarily cover work
done before summer of 2001.  Previous articles reviewing related work
include those of Ohmori~\cite{Ohmori}, de Smet~\cite{deSmet}, Aref'eva
{\it et al.}~\cite{abgkm},  Bonora {\it et al.}~\cite{Bonora}, and
Taylor~\cite{Taylor-Valdivia}.
There are a number of major related areas which we do not cover
significantly or at all in these lectures.  We do not have any
substantial discussion on the dynamic process of tachyon decay; there
has been quite a bit of work on this subject~\cite{rolling-tachyon}
since the time of these lectures in 2001.
We do not discuss the Moyal approach to SFT taken recently by Bars and
collaborators~\cite{Bars-original,Bars-all,Bars}; this work is an interesting
alternative to the level-truncation method primarily used here.  We also do not
discuss in any detail the alternative boundary string field theory
(BSFT) approach to
OSFT.  The BSFT approach is well suited to derive certain concrete results
regarding the tachyon vacuum~\cite{BSFT}---for example, using this
approach the energy
of the tachyon vacuum can be computed exactly.  On the other hand,
BSFT is not a
completely well-defined framework, as massive string fields cannot yet be
consistently incorporated into the theory.

\subsection{The status of string theory: a brief review}
\label{sec:status}

To understand the significance of developments over the last seven
years, it is useful to recall the status of string theory  in early 1995.
At that time
it was clearly understood that there were five distinct ways in which a
supersymmetric string  could be
quantized to give a microscopic definition  of a theory of quantum
gravity in ten dimensions.  Each of these quantum string theories
gives a set of rules for calculating scattering amplitudes of string
states; these states describe gravitational quanta and other massless
and massive particles moving in a ten-dimensional spacetime.  The
five  superstring theories are known as the type IIA, IIB, I, heterotic
$SO(32)$, and heterotic $E_8 \times E_8$  theories.  While
these string theories give perturbative descriptions of quantum
gravity, there was little understanding in 1995 of nonperturbative aspects of
these theories.

In the years between 1995 and 2002, several new ideas dramatically
transformed our understanding of string theory.  We now
briefly summarize these ideas and mention some aspects of these
developments relevant to the main topic of these lectures.
\vspace*{0.07in}

\noindent
{\bf Dualities:} The five different perturbative formulations of
superstring theory are all related to one another through duality
symmetries~\cite{Hull-Townsend,Witten-various}, whereby the degrees of
freedom in one theory can be described through a duality
transformation in terms of the degrees of freedom of another theory.
Some of these duality symmetries are nonperturbative, in the sense
that the string coupling $g$ in one theory is related to the inverse
string coupling $1/g$ in the dual theory.  The web of dualities that relate
the different theories gives a picture in which, rather than
describing five distinct  fundamental theories, each
superstring theory appears to be a particular
perturbative limit of a single, still unknown,
underlying theoretical structure.
\vspace*{0.07in}

\noindent
{\bf M-theory:} In addition to the five perturbative string theories,
the web of dualities also seems to include a limit which describes a
quantum theory of gravity in eleven dimensions.  This new theory has
been dubbed ``M-theory''.  Although no covariant definition for
M-theory has been given, this theory can be related to type IIA and
heterotic $E_8 \times E_8$ string theories through compactification on
a circle $S^1$ and the space $S^1/Z_2$,
respectively~\cite{Townsend-11,Witten-various,Horava-Witten}.  In
the relation to  type IIA, for example, the compactification radius
$R_{11}$ of M-theory is equal to the product $ g_sl_s$ of the string coupling
$g_s$ and the string length $l_s$.  Thus, M-theory in flat space,
which arises in the limit $R_{11} \rightarrow \infty$, can be thought
of as the strong coupling limit of type IIA string theory.  The field theory
limit of M-theory is eleven-dimensional supergravity.  It is also
suspected that M-theory may be formulated as a quantum theory of
membranes in eleven dimensions~\cite{Townsend-11}.
\vspace*{0.07in}

\noindent
{\bf Branes:} In addition to strings, all five superstring
theories, as well as M-theory, contain extended objects
of various dimensionalities
known as ``branes''.  M-theory has M2-branes and
M5-branes, which have two and five dimensions of spatial extent, respectively.
(A string is a one-brane, since it has one spatial dimension.)
The different superstring theories each have different sets
of (stable) D-branes,
         special branes that are defined
by Dirichlet-type boundary conditions on strings.
           In particular, the IIA/IIB superstring theories contain
(stable) D-branes of all even/odd dimensions.
Each superstring theory also has  a fundamental string and a
Neveu-Schwarz five-brane.  The branes of one theory can be
related to the branes of another through the duality transformations mentioned
above.  Using an appropriate sequence of dualities, any brane can be mapped
to  any other brane, including the string itself.  This suggests that
none of these
objects are really any more fundamental than any others; this idea is known as
``brane democracy''.
\vspace*{0.07in}

\noindent
{\bf M(atrix) theory and AdS/CFT:}  It is a remarkable consequence
of the above developments that for certain
asymptotic space-time backgrounds, M-theory and
string theory can be completely
described through  supersymmetric quantum mechanics and field
theories related to the low-energy description of systems of
branes.  The M(atrix) model of M-theory is a simple supersymmetric
matrix quantum mechanics,  and it is believed to capture (in light-cone
coordinates) all of the
physics of M-theory in asymptotically flat spacetime.
In the AdS/CFT correspondence, certain maximally supersymmetric Yang-Mills
theories can be used to describe  closed superstring  theories in
asymptotic  spacetime backgrounds
that are the product of anti-de Sitter space and a sphere.
It is believed that the
Yang-Mills theories and the matrix model of M-theory, each
give true nonperturbative descriptions of quantum gravity in the
corresponding spacetime geometry. For reviews of
M(atrix) theory and AdS/CFT, see Taylor~\cite{WT-RMP}
and Aharony{\em et.al}~\cite{agmoo}.

\vspace*{0.07in}

\noindent
{\bf Unstable D-branes and open string tachyons:} This is in large
part the subject of these lectures.  The most recent chapter in our
new understanding of nonperturbative effects in string theory has been
the incorporation of unstable branes and open string tachyons into the
overall framework of the theory.  It has turned out that an
understanding of unstable D-branes is necessary to properly describe
all D-branes.  This is natural from the point of view of K-theory,
where brane configurations which are equivalent under the annihilation
of unstable branes are identified~\cite{k-theory}.  The
long-mysterious tachyon instability of open string theory has finally
been given a physical interpretation: it is the instability of the
D-brane that supports the existence of open strings.  The instability
disappears in the tachyon vacuum, in which the D-brane decays.
Moreover, the belief that D-branes are solitonic solutions of string
theory has been confirmed: starting with the appropriate tachyonic
field theory of unstable space-filling branes, one can describe lower
dimensional D-branes as solitonic solutions.
Lower dimensional D-branes are thereby essentially obtained as
solitons of the tachyon field theory, so,  in some sense,
lower-dimensional D-branes
can be thought of as being made of tachyons!
It has also been shown that the physics of unstable D-branes is
captured by string field theory, thus making it a candidate for a
non-perturbative formulation of string theory capable of describing
changes of the string background.

\vspace*{0.1in}

The set of ideas just summarized have greatly increased our
understanding of nonperturbative aspects of string theory.  In
particular M(atrix) theory and the AdS/CFT correspondences
provide nonperturbative
definitions of M-theory and string theory in certain asymptotic space-time
backgrounds which can be used, in principle,  to calculate any local result
in quantum gravity.  Through string field theory we have a possibly
nonperturbative definition of the theory that appears to capture many open
string theory backgrounds.
The existing
formulations of string field theory are not  manifestly
background independent because
a background must be selected to write the theory.
Nevertheless,  as we discuss in
these lectures, the theory describes multiple distinct backgrounds in
terms of a common set of variables,
so it embodies, at least partially, physical background independence.
It remains to
be seen if the theory
incorporates full physical
background independence; this requires an ability
to describe all possible open string backgrounds,
as well as all possible closed string backgrounds.

\subsection{The goal of these lectures}

The goal of these lectures is to describe progress towards a
nonperturbative  formulation of string theory that implements
the physics of background independence.
Open string field theory, as applied to tachyon condensation
and related matters, has shown itself capable of describing
non-perturbative objects in string theory,  and it has demonstrated
an ability to represent various open string backgrounds.

A completely background independent formulation of string theory
may be needed
to address fundamental questions such as:
What is string theory/M-theory?  How is the vacuum of string theory
selected?  ({\it i.e.}, Why can the observable low-energy universe be
accurately described by the standard model of particle physics in four
space-time dimensions with an apparently small but nonzero positive
cosmological constant?), and other questions of a cosmological nature.
Obviously, aspiring to address these questions is an ambitious
undertaking, but we believe that attaining a better understanding of
string field theory is a
useful step in this direction.
More concretely, in these lectures we will describe recent progress on
open string field theory. It may be useful here to recall some basic
aspects of open and closed strings and the relationship between them.

Closed strings, which are topologically equivalent to a circle $S^1$,
give rise upon quantization to a massless set of
states associated
with the graviton $g_{\mu \nu}$, the dilaton $\varphi$, and the
antisymmetric two-form $B_{\mu \nu}$, as well as an infinite family of
massive states.
For the supersymmetric closed string, further
massless fields
appear within  
the graviton supermultiplet---these are the Ramond-Ramond $p$-form fields
$A^{(p)}_{\mu_1
\cdots \mu_p}$ and the gravitini $\psi_{\mu \alpha}$.  Thus, the
quantum theory of closed strings is naturally associated with a theory
of gravity in space-time.  On the other hand, open strings, which are
topologically equivalent to an interval $[0, \pi]$, give rise under
quantization to a massless gauge field $A_\mu$ in space-time.  The
supersymmetric open string also has a massless gaugino field
$\psi_\alpha$.
It is now understood that
the endpoints of  open strings
must lie
on a Dirichlet $p$-brane (D$p$-brane),
and that the massless open string fields describe the fluctuations of
the D-brane and the gauge field living on the world-volume of the D-brane.

It may seem, therefore, that open and closed strings are quite
distinct, and describe disjoint aspects of the physics in a fixed
background space-time that contains some family of D-branes.  At tree
level, the closed strings indeed describe gravitational physics in the
bulk space-time, while the open strings describe the D-brane dynamics.
At the quantum level, however, the physics of open and closed strings
are deeply connected.  Indeed, historically open strings were
discovered first through the form of their
scattering amplitudes~\cite{Veneziano}.  Looking at one-loop processes for open
strings led to the first discovery of closed strings, which appeared as {\em
poles} in nonplanar one-loop open string diagrams~\cite{NGS,Lovelace}.  The
fact that open string diagrams naturally contain closed string
intermediate states indicates that in some sense all closed string
interactions are implicitly defined by the open
string diagrams.  This connection underlies many of the important
recent developments in string theory.  In particular, the M(atrix)
theory and AdS/CFT correspondences between gauge theories and quantum
gravity are essentially limits in which closed string physics in a
fixed space-time background is captured by
the Yang-Mills limit of an open string theory on a family
of branes (D0-branes for M(atrix) theory, D3-branes for the CFT that describes
AdS${}_5\times S^5$, etc.)

Since quantum gravity theories in certain fixed space-time
backgrounds can be described by  field theory limits of open strings,
we may ask if a global change of the space-time background
can be described as well.
If M(atrix) theory or AdS/CFT allowed for this description, it would
indicate that
these models may have background-independent generalizations.
Unfortunately,  such background changes involve the generally intractable
addition of an infinite number of nonrenormalizable interactions to the field
theories in question.
One tractable  situation arises
for the addition of a constant background $B_{\mu\nu}$ field in
space-time (perhaps because this closed string background is gauge
equivalent to the open string background of a D-brane
with a magnetic field).  In the
associated Yang-Mills theory,
this change
in the background field corresponds to replacing products of open string
fields with a noncommutative star-product.  The resulting theory is a
noncommutative Yang-Mills theory.  Such noncommutative theories are
the only well-understood example of a situation where adding an
infinite number of apparently nonrenormalizable terms to a field
theory action leads to a sensible modification of quantum field
theory (for a review of noncommutative field theory and its connection
to string theory, see Douglas and Nekrasov~\cite{Douglas-Nekrasov}).

String field theory is a nonperturbative formulation  of
string theory in which the infinite family of
fields associated with string excitations are described by a
space-time field theory action.
For open strings on a D-brane configuration, this field theory
contains Yang-Mills
fields and an entire hierarchy of massive string
fields.  Integrating out all the massive fields from the string field
theory action
gives rise to a nonabelian Born-Infeld action for the
D-branes, which includes an
infinite set of higher-order terms that arise from string theory
corrections to the
simple Yang-Mills action.  Like the case of noncommutative field
theory discussed
above, the new terms appearing in this action are apparently nonrenormalizable,
but the combination of terms must work together to form a sensible theory.

In the 1980's, a great deal of work was done to formulate string
field theory for open and closed, bosonic and supersymmetric string
theories.  All of this work was based on the BRST approach to
string 
quantization~\cite{Kato:1982im,Siegel:1984xd,Siegel:1985tw,Banks:1985ff}.
For the open bosonic
string Witten~\cite{Witten-SFT} constructed an extremely elegant string
field theory based on the Chern-Simons action.  This cubic open
string field theory
(OSFT) is the primary focus of the work described in these lectures.
Although this theory can be described in a simple abstract language,
practical computations rapidly become
complicated.  The formulation of bosonic closed string field theory
was completed in the early
1990s~\cite{Zwiebach:1992ie,Saadi:tb,Kugo:1989tk,Kaku:zw}.  This theory is the
natural counterpart of Witten's open string field theory, but it is
more technically
challenging because of its nonpolynomiality.
A nonpolynomial string field theory is also required to describe
in a non-singular fashion open and closed string fields~\cite{Zwiebach:1997fe}.
For open superstrings, a cubic formulation~\cite{Witten:1986qs} 
encountered some
difficulties~\cite{Wendt,Greensite-Klinkhamer} (for which there are some
proposed resolutions~\cite{Arefeva:1989cp,Preitschopf:fc}), but the
nonpolynomial formulation of Berkovits~\cite{Berkovits} appears to be fully
consistent. Despite a substantial amount of work in string field
theory in the early 90's,  little
insight was gained at the time concerning non-perturbative physics.  Work on
this subject stalled out until open string field theory  was used to test the
tachyon conjectures beginning in 1999~\cite{Sen-Zwiebach}.

\smallskip
One simple feature of the 26-dimensional bosonic string has been
problematic since the early days of string theory: both the open and
closed bosonic strings have tachyons in their spectra, indicating that
the usual perturbative vacua of these
theories are unstable.  In 1999, Ashoke Sen had a remarkable insight
into the nature of the open bosonic string
tachyon~\cite{Sen-universality}.  He observed that the open bosonic
string theory (the so-called Veneziano model) represents open strings that end
on a space-filling D25-brane. He pointed out that this D-brane is unstable,
as it does not carry any conserved charge, and he suggested that the open
string tachyon is in fact the unstable mode of the D25-brane.  This led
him to conjecture that open string field theory could be used to precisely
determine a new vacuum for the open string, namely one in which the D25-brane
is annihilated through condensation of the tachyonic unstable mode.  Sen made
several precise conjectures regarding the details of the string field theory
description of this new open string vacuum.  As we describe in these lectures,
there is now overwhelming evidence that Sen's picture is correct, demonstrating
that string field theory accurately describes the nonperturbative
physics of D-branes.  This new nonperturbative application of string
field theory has sparked a new wave of work on open
string field theory, revealing many remarkable new structures.
In particular, string field theory now provides a concrete framework
in which disconnected string backgrounds can emerge from the equations
of motion of a single underlying theory.  Although so far this can
only be shown explicitly in the open string context, this work paves
the way for a deeper understanding of background-independence in
quantum theories of gravity.


\section{D-branes}
\label{sec:D-branes}

In this section we briefly review some basic features of D-branes.
The concepts developed here will be useful to describe tachyonic
D-brane configurations in the following section.  For more detailed
reviews of D-branes, see the reviews of
Polchinski~\cite{Polchinski-TASI}, and of
Taylor~\cite{WT-Trieste}.

\subsection{D-branes and Ramond-Ramond charges}

D-branes can be understood from many points of view.  In these
lectures we primarily focus on the viewpoint motivated by the recent
work on tachyon condensation, which is that that D-branes are solitons
in string field theory.  The original realization of the importance of
D-branes in string theory stemmed from Polchinski's realization that
D-branes could be described in two complementary fashions:
         {\it a}) as extended extremal black brane solutions of
supergravity that carry conserved charges, and {\it b}) as hypersurfaces on
which strings have Dirichlet boundary conditions.
We now discuss these
two viewpoints briefly.
\vspace*{0.03in}

\noindent
{\it a}) The ten-dimensional type IIA and IIB supergravity theories
each have a set of $(p + 1)$-form fields $A^{(p + 1)}_{\mu_1 \cdots
\mu_{(p+ 1)}}$ in the supergraviton multiplet, with $p$ even/odd for
type IIA/IIB supergravity.  These are the Ramond-Ramond (RR) fields in the
massless superstring spectrum.  For each of these $(p + 1)$-form
fields, there is a solution of the supergravity field equations that
is invariant under $(p + 1)$-dimensional Lorentz transformations, and which has
the form of an extremal black hole
solution in the
$9-p$ spatial  directions that are not affected
by these Lorentz transformations~\cite{dkl}.
These ``black
$p$-brane'' solutions carry charge under the RR fields $A^{(p + 1)}$,
and are BPS states in the supergravity theory that preserve half the
supersymmetry of the theory.
These solutions represent the gravitational
and gauge backgrounds created by the branes, in a way similar to that in which
the  Schwarzschild solution represents the gravitational background of
a point mass, or the Coulomb field represents the electric field of
a point charge.
\vspace*{0.03in}

\noindent
{\it b})
In type IIA and IIB string theory, it is possible to consider open
strings with Dirichlet boundary conditions on some number $9-p$ of
the spatial coordinates $x^\mu (\sigma)$.  The locus of points defined
by such Dirichlet boundary conditions defines a $(p + 1)$-dimensional
hypersurface $\Sigma_{p + 1}$ in the ten-dimensional spacetime.
When $p$ is even/odd in
type IIA/IIB string theory, the spectrum of the resulting quantum open
string theory contains a massless set of fields $A_\alpha, \alpha = 0,
1, \ldots, p$ and $X^{a}, a = p + 1, \ldots, 9$.
These fields can be associated with a gauge field living on the
hypersurface $\Sigma_{p + 1}$, and a set of  degrees of
freedom describing the transverse fluctuations of this hypersurface in
spacetime, respectively.  Thus, the quantum fluctuations of the open string
describe a fluctuating $(p + 1)$-dimensional hypersurface in spacetime --- a
Dirichlet-brane, or ``D-brane''.

The remarkable insight of Polchinski~\cite{Polchinski}  in 1995
was the observation that the stable Dirichlet-branes
of superstring theory carry Ramond-Ramond
charges, and therefore should be described in the low-energy
supergravity limit of string theory by precisely the black $p$-branes
discussed in {\it a}).  This connection between the string and
supergravity descriptions of these nonperturbative objects paved the
way to a dramatic series of new developments in string theory,
including connections between string theory and supersymmetric gauge
theories, string constructions of black holes, and new approaches to
string phenomenology.  The
bosonic D-branes on which we concentrate
attention in these lectures do not carry conserved charges, and thus
are not associated with supergravity solutions as in {\it a)}; rather,
these D-branes can be described through open bosonic strings with some
Dirichlet boundary conditions as in {\it b)}.

\subsection{Born-Infeld and super Yang-Mills D-brane actions}

In this subsection we briefly review the low-energy super Yang-Mills
description of the dynamics of one or more D-branes.  As discussed in
the previous subsection, the massless open string modes on a
D$p$-brane in type IIA or IIB superstring theory describe a $(p +
1)$-component gauge field $A_\alpha$, $9-p$ transverse scalar fields
$X^a$, and a set of massless fermionic gaugino fields.  The scalar
fields $X^a$ describe small fluctuations of the D-brane around a flat
hypersurface.  If the D-brane geometry is sufficiently far from flat,
it is useful to describe the D-brane configuration by a general
embedding $X^\mu (\xi)$, where $\xi^\alpha$ are $p + 1$ coordinates on
the D$p$-brane world-volume $\Sigma_{(p + 1)}$, and $X^\mu$ are ten
functions giving a map from $\Sigma_{(p + 1)}$ into the space-time
manifold ${\bf R}^{ 9, 1}$.  Just as the Einstein equations which govern
the geometry of spacetime arise from the condition that the one-loop
contribution to the closed
string beta function vanishes,
a set of equations of motion for a general D$p$-brane geometry and associated
world-volume gauge field can be derived from a calculation of the
one-loop open string beta function~\cite{Leigh}.  These equations of
motion arise from the classical Born-Infeld action:
\begin{equation}
S = - T_p  \int d^{p + 1} \xi
\;e^{-\varphi} \;\sqrt{-\det (G_{\alpha \beta} + B_{\alpha \beta} + 2
\pi \alpha'
F_{\alpha \beta})  }   + S_{{\rm CS}}+{\rm fermions}
\label{eq:DBI}
\end{equation}
where $G$, $B$,  and $\varphi$ are the pullbacks of the ten-dimensional metric,
antisymmetric tensor, and dilaton to the D-brane world-volume, while $F$ is the
field strength of the world-volume $U(1)$ gauge field $A_{\alpha}$.
$S_{\rm CS}$ represents a set of Chern-Simons terms which will be
discussed in the following subsection.
This action can be verified by a perturbative string
calculation~\cite{Polchinski-TASI}, which also gives a precise
expression for the brane tension
\begin{equation}
\tau_p =\frac{T_p}{g_s} =
            \frac{1}{g_s\sqrt{\alpha'}}  \frac{1}{ (2 \pi \sqrt{\alpha'})^{p}}
\end{equation}
where $g_s = e^{\langle \varphi \rangle}$ is the closed string
coupling, equal to
the exponential
of the dilaton expectation value.

A particular limit of the Born-Infeld action
(\ref{eq:DBI}) is useful to describe many low-energy aspects of
D-brane dynamics.  Take the background space-time $G_{\mu \nu}=
\eta_{\mu \nu}$ to be flat, and all other supergravity fields ($B_{\mu
\nu}, A^{(p + 1)}_{\mu_1 \cdots \mu_{p + 1}}$) to vanish.  We then
assume that the D-brane is approximately flat, and is close to the
hypersurface $X^a = 0, a > p$, so that we may make the static gauge
choice $X^\alpha = \xi^\alpha$.  We furthermore take the low-energy
limit
in which
$\partial_\alpha X^a$ and $2 \pi \alpha' F_{\alpha \beta}$ are small
and of the same order. The action (\ref{eq:DBI})
can then be expanded as
\begin{equation}
S =-\tau_pV_p  -\frac{1}{4g_{{\rm YM}}^2}
            \int d^{p + 1} \xi
\left(F_{\alpha \beta} F^{\alpha \beta} +\frac{2}{(2 \pi \alpha')^2}
\partial_\alpha X^a \partial^\alpha X^a\right) +\cdots
\label{eq:action-expansion}
\end{equation}
where $V_p$ is the $p$-brane world-volume and the coupling $g_{{\rm
YM}}$ is given by
\begin{equation}
g_{{\rm YM}}^2 = \frac{1}{4 \pi^2 \alpha'^2 \tau_p}
= \frac{g_s}{\sqrt{\alpha'}}  (2 \pi \sqrt{\alpha'})^{p-2}\,.
\label{eq:ym-coupling}
\end{equation}
Ignoring  fermionic terms,
the second term in  the right-hand side of
(\ref{eq:action-expansion})
is simply the reduction to $(p + 1)$ dimensions of the ten-dimensional
${\mathcal N} = 1 $ super Yang-Mills action:
\begin{equation}
S = \frac{1}{g_{{\rm YM}}^2}  \int d^{10}\xi \; \left(
            -\frac{1}{4} F_{\mu \nu}F^{\mu \nu}
+ \frac{i}{2}  \bar{\psi} \Gamma^\mu \partial_{\mu} \psi \right)
\label{eq:SYM1}
\end{equation}
where for $\alpha, \beta \leq p$, $F_{\alpha \beta}$ is the
world-volume $U(1)$ field strength, and for $a > p, \alpha \leq p$, $
F_{\alpha a} \rightarrow \partial_\alpha X^a/(2\pi \alpha')$.

When multiple D$p$-branes are present, the D-brane action is modified
in a fairly simple fashion~\cite{Witten-multiple}.  Consider a system
of $N$ coincident D-branes.  For every pair of branes $\{i, j\}$ there
is a set of massless fields
\begin{equation}
(A_\alpha)_i^{\; j},  \; \; \;(X^a)_i^{\; j}\,,
\label{eq:nonabelian-fields}
\end{equation}
associated with
strings stretching from the $i$th brane to the $j$th brane; the
indices $i, j$ are known as Chan-Paton indices.  Treating the fields
(\ref{eq:nonabelian-fields}) as
$N$-by-$N$ matrices, and letting  Tr  denote
the trace operation for such matrices, the multiple
brane analogue of the Born-Infeld action (\ref{eq:DBI}) takes the
schematic form
\begin{equation}
S \sim \int {\rm Tr}\; \sqrt{-\det \left( G + B +  2\pi \alpha' F \right)}\,.
\label{eq:NBI}
\end{equation}
In order to properly define this nonabelian analog of the Born-Infeld
action (NBI), it is necessary to resolve the ordering ambiguities in
(\ref{eq:NBI}).  Since the spacetime coordinates $X^a$ associated with
the D-brane positions in space-time become themselves matrix-valued,
even evaluating the pullbacks $G_{\alpha \beta}, B_{\alpha \beta}$
involves resolving ordering issues.  Much work has been done recently
to resolve these ordering ambiguities~\cite{ordering-NBI}
but there is still no known
definition of the nonabelian Born-Infeld theory
(\ref{eq:NBI}) which is valid to all orders.

The nonabelian Born-Infeld action (\ref{eq:NBI}) becomes much simpler
when, once again,  the background space-time is
assumed to be flat and we take the low-energy
limit , leading to
the nonabelian $U(N)$ super Yang-Mills action in $p
+ 1$ dimensions.  This action is the  reduction to $p + 1$
dimensions of the
ten-dimensional
$U(N)$ super Yang-Mills action (analogous to (\ref{eq:SYM1})).
In this reduction,
            for $\alpha, \beta \leq p$, $F_{\alpha \beta}$ is the
world-volume $U(N)$ field strength,
and for $a >p,\alpha\leq p$,
$ F_{\alpha a} \rightarrow \partial_\alpha X^a$, where now $ A_\alpha,
X^a,$ and $F_{\alpha \beta}$ are $N \times N$ matrices.
Since the derivatives $\partial_a$ are set to zero in the dimensional
reduction, we
furthermore have, for $a, b > p$, $F_{ab} \rightarrow -i[X^a, X^b]$.

The low-energy description of a system of $N$ coincident flat D-branes
is thus given by $U(N)$ super Yang-Mills theory in the appropriate
dimension.  This connection between D-brane actions in string theory
and super Yang-Mills theory has led to many new developments,
including new insights into supersymmetric field theories, the
M(atrix) theory and AdS/CFT correspondences, and brane world
scenarios.

\subsection{Branes from branes}

In this subsection we describe a remarkable feature of D-brane
systems:
one or more D-branes of a fixed
dimension can be used to construct additional D-branes of higher or
lower dimension.

In our discussion of the D-brane action (\ref{eq:DBI}),
we mentioned a group of terms $S_{\rm CS}$ which we did not describe
explicitly.  For a single D$p$-brane, these Chern-Simons terms can be
combined into a single expression of the form
\begin{equation}
S_{\rm CS}\sim \int_{\Sigma_{p + 1}}\;{\mathcal A} \;e^{F + B}\,,
\label{eq:Chern-Simons}
\end{equation}
where ${\mathcal A} = \sum_{k}A^{(k)} $ represents a formal sum over all
the Ramond-Ramond fields $A^{(k)}$ of various dimensions.  In this
integral, for each term $A^{(k)}$, the nonvanishing contribution to
(\ref{eq:Chern-Simons}) is given by expanding the exponential of $F +
B$ to order $(p + 1-k)/2$, where the dimension of the resulting form
saturates the dimension of the brane.  For example, on  a D$p$-brane,
there is a coupling of the form
\begin{equation}
\int_{\Sigma_{(p + 1)}} A^{(p-1)} \wedge F\,.
\end{equation}
This coupling implies that the $U(1)$ field strength on the D$p$-brane
couples to the RR field associated with $(p-2)$-branes.  Thus, we can
associate magnetic fields on a D$p$-brane with dissolved
$(p-2)$-branes living on the D$p$-brane.  This result generalizes to a
system of multiple D$p$-branes, in which case a trace is included on
the right-hand side of (\ref{eq:Chern-Simons}).
For example, on $N$ compact
D$p$-branes wrapped on a $p$-torus,
the
flux
\begin{equation}
            \frac{1}{2 \pi}  \int {\rm Tr}\; F_{\alpha \beta},
\label{eq:p-2-charge}
\end{equation}
of the magnetic field over a
two-cycle on the torus  is quantized
and measures the number of units of D$(p-2)$-brane charge on
the D$p$-branes that threads the cycle integrated over. Thus, these
branes are encoded in the field strength $F_{\alpha \beta}$.
The object in (\ref{eq:p-2-charge}) is the relevant component of the first
Chern class of the
$U(N)$ bundle described by the gauge field on the $N$ branes.
Similarly,
\begin{equation}
            \frac{1}{8 \pi^2}  \int {\rm Tr}\; F\wedge F
\end{equation}
encodes D$(p -4)$-brane charge on the D$p$-branes, etc..

Just as lower-dimensional branes can be described in terms of the
degrees of freedom associated with a system of $N$ D$p$-branes through
the field strength $F_{\alpha \beta}$, higher-dimensional branes can
be described by a system of $N$ D$p$-branes in terms of the
commutators of the matrix-valued scalar fields $X^a$.  Just as
$\frac{1}{2 \pi} F$ measures $(p-2)$-brane charge, the matrix
\begin{equation}
2 \pi i
[X^a, X^b]
\label{eq:up-charge}
\end{equation}
measures $ (p + 2)$-brane charge~\cite{WT-Trieste,Mark-Wati-4,Myers}.  The
charge (\ref{eq:up-charge}) should be interpreted as a form of local charge
density.
Just as the $N$ positions of the D$p$-branes are replaced by matrices
in the nonabelian theory, so the locations of the charges become matrix-valued.
The
trace of (\ref{eq:up-charge}) vanishes for finite sized matrices
because the net D$p$-brane charge of a
finite-size brane configuration in flat spacetime vanishes.  Higher
multipole moments of the brane charge, however, have a natural
definition in terms of traces of the charge matrix multiplied by
powers of the scalars $X^a$, and generically are nonvanishing.

A simple example of the mechanism by which a system of multiple
D$p$-branes form a higher-dimensional brane is given by the matrix
sphere.  If we take a system of D0-branes with scalar matrices $X^a$
given by
\begin{equation}
X^a = \frac{2r}{ N}  J^a, \;\;\;\;\; a = 1, 2, 3
\label{eq:matrix-sphere}
\end{equation}
where $J^a$ are the generators of $SU(2)$ in the $N$-dimensional
representation, then we have a configuration corresponding to the
``matrix sphere''.  This is a D2-brane of spherical geometry living on
the locus of points satisfying $x^2 + y^2 + z^2 = r^2$.  The ``local''
D2-brane charge of this brane is given by (\ref{eq:up-charge});
here, for example, the D2-brane charge in the $x$-$y$ plane is
proportional to the matrix $X^3$ ($z$), as one would expect from the
geometry of a spherical brane.
The D2-brane configuration given by (\ref{eq:matrix-sphere}) is
rotationally invariant (up to a gauge transformation).  The
restriction of the brane to the desired locus of
points  can be seen from the relation $(X^1)^2 + (X^2)^2 + (X^3)^2 =
r^2\identity+{\mathcal O} (N^{-2})$.

\subsection{T-duality}
\label{sec:T-duality}

We conclude our discussion of D-branes with a brief description of
T-duality.  T-duality is a perturbative and nonperturbative
symmetry  which relates
the type IIA and type IIB string theories.  This duality symmetry was in
fact crucial in the original discovery of D-branes~\cite{Polchinski}.
A more detailed discussion of T-duality can be found in the textbook
by Polchinski~\cite{Polchinski-string}.
Using T-duality, we construct an explicit example of a brane within a
brane encoded in super Yang-Mills theory, illustrating the ideas of
the previous subsection.  This example will be used in the following
section to construct an analogous configuration with a tachyon.

Consider type IIA string theory on a spacetime of the form $M^9 \times
S^1$ where $M^9$ is a generic 9-manifold of Lorentz signature, and
$S^1$ is a circle of radius $R$.  T-duality is the statement that this
theory is precisely equivalent, even
at the perturbative level,
to type IIB
string theory on the spacetime $M^9 \times (S^1)'$, where
$(S^1)'$ is a circle of radius $R' = \alpha'/R$.

T-duality symmetry is most easily understood
in the case of closed strings, where
it amounts to an exchange of winding and momentum modes of the string.
The string winding modes on $S^1$ have energy
$R |m|/\alpha'$, where
$m$ is the winding number.  The T-dual momentum modes on $(S^1)'$ have
energy $|n|/R'$, where $n$ is the momentum quantum number.
These two sets of values coincide when $m$ and $n$ run over all possible
integers.  It is in fact
straightforward to check that the full spectrum of closed string states is
unchanged under T-duality.
For the case of open strings,
T-duality maps an open string with Neumann boundary conditions on
$S^1$ to an open string with Dirichlet boundary conditions on
$(S^1)'$, and vice versa.  Thus, a D$p$-brane which is wrapped around
the circle $S^1$ is mapped under T-duality to a D$(p-1)$-brane which
is localized to a point on the circle $(S^1)'$.  Under T-duality the
low-energy $(p + 1)$-dimensional Yang-Mills theory on the $p$-brane is
replaced by a $p$-dimensional Yang-Mills theory on the dual
$(p-1)$-brane.  Mathematically, the covariant derivative operator in
the direction $S^1$ is replaced under T-duality with an adjoint scalar
field $X^a$.  Formally, this adjoint scalar field is an infinite size
matrix~\cite{WT-T-duality}, which contains information about the open
strings wrapped an arbitrary number of times around the compact
direction $(S^1)'$.

We can summarize the relevant mappings under T-duality in the
following table

\begin{center}
\begin{tabular}{rcr}
IIA/$S^1$ & $\leftrightarrow $ & IIB/$(S^1)'$\\
$R$ & $\leftrightarrow $ & $R' =\alpha'/R $\\
Dirichlet/Neumann b.c.'s & $\leftrightarrow $ & Neumann/Dirichlet b.c.'s\\
$p$-brane & $\leftrightarrow $ & $(p\pm 1)$-brane\\
$2 \pi \alpha' (i \partial_a + A_a)$ & $\leftrightarrow $ & $X^a $
\end{tabular}
\end{center}
\vspace*{0.2in}

The phenomena by which field strengths in one brane describe lower- or
higher-dimensional branes can be easily understood using T-duality.
The following simple example may help to clarify this connection.
(For a more detailed discussion using this point of view, see
Taylor~\cite{WT-Trieste}.)
constant magnetic

\begin{figure}[!ht]
\leavevmode
\begin{center}
\epsfxsize = 12 cm \epsfbox{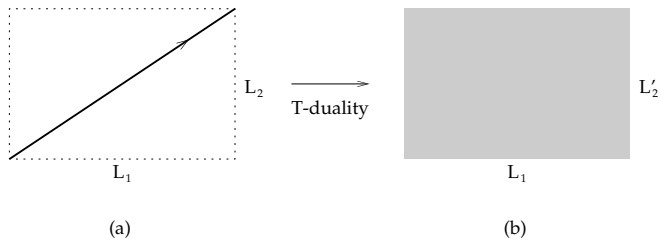}
\end{center}
\caption[x]{\footnotesize T-duality takes a diagonal D1-brane  on a
             two-torus (a) to a D2-brane on the dual torus with 
constant magnetic
             flux encoding an embedded D0-brane (b).}
\label{f:T-duality}
\end{figure}

Consider a D1-brane wrapped diagonally on a two-torus $T^2$ with sides
of length $L_1 = L$ and $L_2 = 2 \pi R$.
(Figure~\ref{f:T-duality}(a)).   This configuration is described in
terms of the world-volume Yang-Mills theory on a D1-brane stretched in
the $L_1$ direction through a transverse scalar field
\begin{equation}
X^2 = 2 \pi R \xi_1/L\,.
\end{equation}
To be technically precise, this scalar field should be treated as an
$\infty \times \infty$ matrix~\cite{WT-T-duality} whose $(n, m)$ entry
is associated with strings that connect the $n$th and $m$th images of the
D1-brane on the covering space of $S^1$.  The diagonal elements
$X^2_{n,n}$ of this infinite matrix are given by $2 \pi R (\xi_1 +
n L)/L$, while all off-diagonal elements vanish.  While the resulting
matrix-valued function of $\xi_1$ is not periodic, it is periodic up
to a gauge transformation
\begin{equation}
X^2 (L) = V X^2 (0) V^{-1}
\label{eq:boundary-1}
\end{equation}
where $V$ is the shift matrix with nonzero elements $V_{n, n + 1} = 1$.

Under T-duality
in the $x^2$ direction
the infinite matrix $X^2_{nm}$ becomes the Fourier mode representation of a
gauge field on a dual D2-brane:
\begin{equation}
A_2 = \frac{1}{ R' L}  \xi_1\,.
       \label{eq:a2}
\end{equation}
The magnetic flux associated with this gauge field is
\begin{equation}
F_{12} = \frac{1}{ R' L}\,,
\end{equation}
so that
\begin{equation}
\frac{1}{2 \pi} \int F_{12} \; d \xi^1 \, d \xi^2 = 1\,.
\label{eq:0-charge}
\end{equation}
Note that the boundary condition (\ref{eq:boundary-1}) on the
infinite matrix $X^2$ transforms under T-duality to the boundary
condition on the gauge field
\begin{eqnarray}
A_2 (L, x_2) & = &
e^{2 \pi i \xi_2/L_2'}
\left(A_2 (0, x_2)  + i \partial_2 \right)
e^{-2 \pi i \xi_2/L_2'}\\
& = &
A_2 (0, x_2)
       + \frac{2 \pi}{ L_2'}, \nonumber
\end{eqnarray}
which (\ref{eq:a2}) clearly satisfies.
The off-diagonal elements of the shift matrix $V$ in
(\ref{eq:boundary-1}) describe winding modes which correspond after
T-duality to the first Fourier mode $e^{2 \pi i \xi_2/L_2'}$.  The
boundary condition on the gauge fields in the $\xi_2$ direction is
trivial, which simplifies the T-duality map; a similar construction
can be done with a nontrivial boundary condition in both directions,
although the configuration looks more complicated in the D1-brane
picture.

This construction gives a simple Yang-Mills description of the mapping
of D-brane charges under T-duality: the initial configuration
described above has charges associated with a single D1-brane wrapped
around each of the directions of the 2-torus: D$1_1 +$ D$1_2$.  Under
T-duality, these D1-branes are mapped to a D2-brane and a D0-brane
respectively: D$2_{12} +$ D$0$.  The flux integral (\ref{eq:0-charge})
is the representation in the D2-brane world-volume Yang-Mills theory
of the charge associated with a D0-brane which has been uniformly
distributed over the surface of the D2-brane, just as in
(\ref{eq:p-2-charge}).


\section{Tachyons and D-branes}
\label{sec:tachyon-D-branes}

We now turn to the subject of tachyons.  Certain D-brane
configurations are unstable, both in supersymmetric and
nonsupersymmetric string theories.  This instability is manifested as
a tachyon,  that is, as
a state with $M^2 < 0$ in the spectrum of open strings that end on the
D-brane.  We will explicitly describe the tachyonic mode in the case
of the open bosonic string in Section \ref{sec:bosonic-string}; this
open bosonic string tachyon will be the focal point of most of the
developments described in these notes.  In this section we list some
elementary D-brane configurations where tachyons arise, and we
describe a particular situation in which the tachyon can be seen in
the low-energy Yang-Mills description of the D-branes.  This
Yang-Mills background with a tachyon provides a simple field-theory
model of a system analogous to the more complicated string field
theory tachyon we describe in the later part of these notes.  This
simpler model may be useful to keep in mind in the later analysis.

\subsection{D-brane configurations with tachyonic instabilities}
\label{sec:D-brane-tachyons}

Some simple examples of unstable D-brane configurations where the open
string contains a tachyon include the following:
\vspace*{0.1in}

{\bf Brane-antibrane:} A pair of parallel D$p$-branes with opposite
orientation in type IIA or IIB string theory which are separated by a
distance $d \ll l_s$ give rise to a tachyon in the spectrum of open
strings stretched between the branes~\cite{Banks-Susskind}.
The difference in orientation of the branes means that the two branes are
really a brane and antibrane, carrying equal but opposite RR charges.
Since the net RR charge is 0, the brane and antibrane can annihilate,
leaving an uncharged vacuum configuration.
\vspace*{0.05in}

{\bf Wrong-dimension branes:} In type IIA/IIB string theory, a
D$p$-brane of even/odd spatial dimension $p$ is a stable BPS state
with  nonzero RR charge.  On the other hand, a D$p$-brane of the
{\it wrong} dimension ({\it i.e.,} odd/even for IIA/IIB) carries no
charges under the classical IIA/IIB supergravity fields, and has a
tachyon in the open string spectrum.  Such a brane can decay into
the vacuum without violating charge conservation.
\vspace*{0.05in}

{\bf Bosonic D-branes:} Like the wrong-dimension branes of IIA/IIB
string theory, a D$p$-brane of any dimension in the bosonic string
theory carries no conserved charge and has a tachyon in the open
string spectrum.  Again, such a brane can decay into the vacuum
without violating charge conservation.

\subsection{Example: tachyon in low-energy field theory of two D-branes}
\label{sec:example-SYM}

In order to illustrate the physical
behavior of  tachyonic configurations, we consider in this subsection a
simple example~\cite{Hashimoto-Taylor,gns} where a brane-antibrane tachyon
can be seen in the context of the low-energy Yang-Mills theory.

\begin{figure}[!ht]
\leavevmode
\begin{center}
\epsfxsize = 12 cm \epsfbox{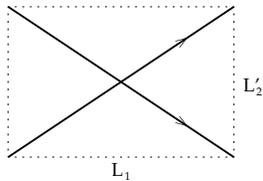}
\end{center}
\caption[x]{\footnotesize A pair of crossed D1-branes, T-dual to a
             pair of D2-branes with uniformly embedded D0- and anti-D0-branes.}
\label{f:crossed}
\end{figure}

The system we want to consider is a simple generalization of the (D2 +
D0)-brane configuration we described using Yang-Mills theory in
section 2.4.  Consider a pair of D2-branes wrapped on a two-torus, one
of which has a D0-brane embedded in it as a constant positive magnetic
flux, and the other of which has an anti-D0-brane within it described
by a constant negative magnetic flux.  We take the two dimensions of
the torus to be $L_1, L_2$.  Following the discussion of Section 2.4,
this configuration is equivalent under T-duality in the $L_2$
direction to a pair of crossed D1-branes (see Figure~\ref{f:crossed}).

The Born-Infeld energy of this configuration is
\begin{eqnarray}
E_{{\rm BI}}  & = & 2 \sqrt{(\tau_2L_1 L_2)^2 + \tau_0^2}  \nonumber\\
            & = & \frac{1}{g}  \left[
\frac{2L_1 L_2}{\sqrt{2 \pi}}  + \frac{(2 \pi)^{3/2}}{L_1 L_2}  +
            \cdots \right]\,, \label{eq:energy-bi}
\end{eqnarray}
in units where $2 \pi \alpha' = 1$.
This can be computed either directly from the Born-Infeld action on
the D2-branes (the abelian theory can be used since the matrices are
diagonal), or by simply using the Pythagorean theorem in the T-dual
D1-brane picture.
The second term in the last line
corresponds to the Yang-Mills approximation.  In this approximation
(dropping the D2-brane energy) the energy is
\begin{equation}
E_{{\rm YM}} = \frac{\tau_2}{4}  \int {\rm Tr}\; F_{\alpha \beta}
F^{\alpha \beta} = \frac{1}{4 \sqrt{2 \pi} g}  \int {\rm Tr}\;
F_{\alpha \beta}
F^{\alpha \beta}\,.
\end{equation}

We are interested in studying this configuration in the Yang-Mills
approximation, in which we have a $U(2)$ theory on $T^2$ with field
strength
\begin{equation}
F_{12} = \left(\begin{array}{cc}
\frac{2 \pi}{ L_1 L_2}  & 0\\
0 &-\frac{2 \pi}{ L_1 L_2}
\end{array}\right)
= \frac{2 \pi}{ L_1 L_2}  \tau_3 \,.
\end{equation}
This field strength can be realized as the curvature of a linear gauge
field
\begin{equation}
A_1 = 0, \;\;\;\;\;
A_2 =\frac{2 \pi}{ L_1 L_2}  \xi\tau_3\,,
\label{eq:unstable-background}
\end{equation}
which satisfies the boundary conditions
\begin{equation}
A_j (L, \xi_2) = \Omega (i \partial_j+ A_j (0, \xi_2)) \Omega^{-1} \,,
\label{eq:bc-tachyon}
\end{equation}
where
\begin{equation}
\Omega = e^{2 \pi i ( \xi_1/L_2) \tau_3} \,.
\end{equation}

\begin{figure}[!ht]
\leavevmode
\begin{center}
\epsfxsize = 12 cm \epsfbox{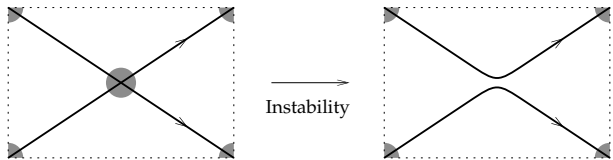}
\end{center}
\caption[x]{\footnotesize The brane-antibrane instability of a
             D0-D$\bar{0}$ system embedded in two D2-branes, as seen in the
             T-dual D1-brane picture.}
\label{f:instability}
\end{figure}

It is easy to check that this configuration indeed satisfies
\begin{equation}
E_{{\rm YM}} = \frac{1}{2g}  \frac{(2 \pi)^{3/2}}{L_1 L_2}  {\rm Tr}\;
\tau_3^2 = \frac{1}{g}  \frac{(2 \pi)^{3/2}}{L_1 L_2}\,,
\label{eq:unstable-energy}
\end{equation}
as desired from (\ref{eq:energy-bi}).
Since
\begin{equation}
{\rm Tr}\; F_{\alpha \beta} = 0,
\end{equation}
the gauge field we are considering is in the same topological
equivalence class as $F = 0$.  This corresponds to the fact that the
D0-brane and anti-D0-brane can annihilate.  To understand the
appearance of the tachyon, we can consider the spectrum of excitations
$\delta A_\alpha$ around the background
(\ref{eq:unstable-background})~\cite{Hashimoto-Taylor}.  The
eigenvectors of the quadratic mass terms in this background are
described by torus  theta functions which satisfy boundary
conditions related to (\ref{eq:bc-tachyon}).  There are precisely two
elements in the spectrum with the negative eigenvalue $-4 \pi/L_1
L_2$.  These theta functions
            ~\cite{Hashimoto-Taylor} are  tachyonic modes of the theory which
are associated with the annihilation of the positive and negative
fluxes that encode the D0- and anti-D0-brane.  These tachyonic modes are
perhaps easiest to understand in the dual configuration, where they
provide a direction of instability in which the two crossed D1-branes
reconnect as in Figure~\ref{f:instability}.

It is also interesting to note that in
the T-dual picture the
tachyonic modes of the gauge field have support which is localized near
the two  intersection points and take the off-diagonal form
\begin{equation}
\delta A_t \sim \left(\begin{array}{cc}
0 & \star\\
\star & 0
\end{array} \right) \,,
\end{equation}
which naturally encodes our geometric
understanding that the tachyonic modes  reconnect the two D1-branes near
each  intersection point.

The full Yang-Mills action around the background
(\ref{eq:unstable-background}) can be written as a quartic function of
the mass eigenstates around this background.  Written in terms of
these modes, there are nontrivial cubic and quartic terms which couple
the tachyonic modes to all the massive modes in the system.  If we
integrate out the massive modes, we know from the topological
reasoning above that an effective potential arises for the tachyonic
mode $A_t$, with a maximum value of (\ref{eq:unstable-energy}) and a
minimum value of 0.  This system is highly analogous to the bosonic
open string tachyon we will discuss in the remainder of these
lectures.  Our current understanding of the bosonic string through
bosonic string field theory is analogous to that of someone who only
knows the Yang-Mills theory around the background
(\ref{eq:unstable-background}) in terms of a complicated quartic
action for an infinite family of modes.  Without knowledge of the
topological structure of the theory, and given only a list of the
coefficients in the quartic action, such an individual would have to
systematically calculate the tachyon effective potential by explicitly
integrating out all the massive modes one by one.  This would give a
numerical approximation to the minimum of the effective potential,
which could be made arbitrarily good by raising the mass of the cutoff
at which the effective action is computed.  It may be helpful to keep
this example
        in mind in the following sections, where an
analogous tachyonic system is considered in string field theory.  For
further discussion of this unstable configuration in Yang-Mills
theory, see the
references~\cite{Hashimoto-Taylor,gns,Morosov,Hashimoto-Nagaoka}.


\subsection{The Sen conjectures}
\label{conjectures}

The existence of the tachyonic mode in the open bosonic string
indicates that the standard choice of perturbative vacuum for this
theory is unstable.  In the early days of the subject, there were some
calculations suggesting that this tachyon could condense, leading to a
more stable vacuum~\cite{Bardakci-tachyon}.  Kostelecky and Samuel
argued early on that the stable vacuum could be identified in string
field theory in a systematic way ~\cite{ks-open}, however there was no
clear physical picture for the significance of this stable vacuum.  In
1999, Ashoke Sen reconsidered the problem of tachyons in string field
theory.  Sen suggested that the open bosonic string should really be
thought of as living on a D25-brane, and hence that the perturbative
vacuum for this string theory should have a nonzero vacuum energy
associated with the tension of this D25-brane.  He suggested that the
tachyon is simply the instability mode of the D25-brane, which carries
no conserved charge and hence is not expected to be stable, as
discussed in section~3.1.  More precisely, Sen conjectured that the
following three statements are true~\cite{Sen-universality}:
\begin{enumerate}
\item The tachyon potential has  a locally stable minimum, whose energy
density  $\mathcal{E}$, measured with respect to that of the
unstable critical
point,  is equal to minus the  tension of the D25-brane:
\begin{equation}
\mathcal{E}  = -T_{25}\,.
\end{equation}
\item Lower-dimensional D-branes are solitonic solutions
of the string theory on the background of a D25-brane.

\item The locally stable vacuum of the system is the closed string vacuum.
In this vacuum the D25-brane is absent and no conventional
open string excitations exist.
\end{enumerate}

It was also suggested by Sen that open string field theory
was a natural setup to test the above conjectures.  He sharpened the
first conjecture by suggesting that Witten's OSFT should precisely
reproduce the tension of the D25-brane, which he expressed
in terms of the open string coupling constant $g$ which appears
in the formulation of open string field theory:
\begin{equation}
T_{25}= \frac{1}{2 \pi^2 g^2} \,.
\end{equation}
We will give the instructive derivation of this result in
section~\ref{sec:evidence}.

Our first encounter with the tachyon conjectures will happen in
section~\ref{sec:CFT}, where we calculate the first nontrivial term in
the tachyon potential, find a minimum, and discover that, even with
this rough approximation, the calculated $\mathcal{E}$ gives about
70\% of the expected answer. In Section 7 of these lectures we
systematically explore the evidence for these conjectures in Witten's
OSFT.  First, however, we need to develop the technical tools to do
specific calculations in string field theory.


\section{Witten's cubic open string field theory}
\label{sec:SFT}

In this  and in the following two sections we give a detailed
description
of Witten's open string field theory~\cite{Witten-SFT}.  This
section contains a general introduction to this string field theory.
Subsection \ref{sec:bosonic-string} reviews the quantization of the
open bosonic string in 26 dimensions and sets notation.  Subsection
\ref{sec:Witten-SFT} gives a heuristic introduction to open string field
theory, which follows Witten's original paper.
In subsection \ref{sec:algebraic-structure} we discuss the
algebraic structure of open string field theory
which  emerges naturally in the context of conformal field
theory. This discussion also develops the properties of the twist
operator $\Omega$ which reverses the orientation of open strings.

The work in the present section prepares the ground for sections
\ref{sec:CFT} and \ref{sec:overlaps}, in which
precise definitions of the open bosonic SFT are given
using conformal field theory and the mode decomposition of overlap
equations.
For further background on Witten's OSFT see the reviews of LeClair,
Peskin and Preitschopf~\cite{lpp}, of Thorn~\cite{Thorn}, and of
Gaberdiel and Zwiebach~\cite{Gaberdiel-Zwiebach}.


\subsection{The bosonic open string}
\label{sec:bosonic-string}

In this subsection we review the quantization of the open bosonic
string.  For further details see the textbooks by Green, Schwarz, and
Witten~\cite{gsw} and by Polchinski~\cite{Polchinski-string}.
The bosonic open string can be quantized using the BRST
approach starting from the action
\begin{equation}
S = -\frac{1}{4 \pi \alpha'}  \int \sqrt{-\gamma} \gamma^{ab}
\partial_a X^\mu \partial_b X_\mu,
\label{eq:string-action}
\end{equation}
where $\gamma$ is an auxiliary dynamical metric on the world-sheet.
This action can be gauge-fixed to conformal gauge $\gamma_{ab} \sim
\delta_{ab}$,  introducing at the same time
ghost and antighost fields $c^{\pm} (\sigma), b_{\pm \pm} (\sigma)$.  The
gauge-fixed action is
\begin{equation}
S = -\frac{1}{4 \pi \alpha'}  \int
\partial_a X^\mu \partial^a X_\mu
+ \frac{1}{ \pi}  \int \left( b_{++} \partial_-c^+ + b_{--} \partial_+
c^-\right)\,.
\label{eq:gauge-fixed-action}
\end{equation}

The matter fields $X^\mu$ can be expanded in modes using
\begin{equation}
X^\mu (\sigma, \tau) = x_0^\mu + 2 p^\mu \tau +
i\sqrt{2} \sum_{n \neq 0}
\frac{1}{n}  \alpha^\mu_n \cos (n \sigma) e^{-in \tau}\,,
\label{eq:mode-decomposition}
\end{equation}
where we have fixed $l_s = \sqrt{2 \alpha'} = \sqrt{2}$, so that
$\alpha' = 1$.  In the quantum theory, $x_0^\mu$ and $p^\mu$ obey the
canonical commutation relations
\begin{equation}
[x^\mu_0, p^\nu] = i \eta^{\mu \nu}\,.
\end{equation}
The $\alpha^\mu_n$'s with negative/positive
values of $n$ become raising/lowering operators for the oscillator
modes on the string.  They satisfy the Hermiticity conditions
       $(\alpha^\mu_n)^{\dagger} = \alpha^\mu_{-n}$
       and the
commutation relations
\begin{equation}
[\alpha^\mu_m, \alpha^\nu_n] = m \eta^{\mu \nu} \delta_{m + n, 0}\,.
\end{equation}
We will often use the canonically normalized oscillators:
\begin{equation}
a^\mu_n = \frac{1}{ \sqrt{n}}  \alpha^{\mu}_n\,, \quad  n\geq 1 \,,
\end{equation}
which obey the commutation relations
\begin{equation}
[a^\mu_m, a^{\nu\dagger}_n] =  \eta^{\mu \nu} \delta_{m,n}\,.
\end{equation}
We will also frequently use position modes $x_n$ for $n \neq 0$ and
lowering and raising operators $a_0, a^{\dagger}_0$ for the zero
modes.  These are related to the modes in
(\ref{eq:mode-decomposition}) through (dropping space-time indices)
\begin{eqnarray}
x_n & = &  \frac{i}{ \sqrt{2n}} (a_n- a^{\dagger}_n) \\
x_0 & = &  \frac{i}{2}  (a_0-a_0^{\dagger}) \nonumber
\end{eqnarray}

The ghost and antighost fields can be decomposed into modes through
\begin{eqnarray}
c^{\pm} (\sigma) & = &  \sum_{n}c_n e^{\mp in \sigma} \\
b_{\pm \pm} (\sigma) & = &  \sum_{n}b_n e^{\mp in \sigma} \,.
\nonumber
\end{eqnarray}
The ghost and antighost modes satisfy the anticommutation relations
\begin{eqnarray}
\{c_n, b_m\} & = &  \delta_{n + m, 0}\\
\{c_n, c_m\}  & = & \{b_n, b_m\} = 0\,. \nonumber
\end{eqnarray}

A general state in the open string Fock space can be written in the
form
\begin{equation}
\alpha^{\mu_1}_{-n_1} \cdots \alpha^{\mu_i}_{-n_i} \;
c_{-m_1} \cdots c_{-m_j} \;
b_{-p_1} \cdots b_{-p_l} \; | 0; k \rangle
\end{equation}
where
\begin{equation}
n_i  \geq 1\,, \quad  m_i \geq  -1\,, \quad \hbox{and} \quad p_i \geq 2\,,
\end{equation}since $| 0; k\rangle$ is
annihilated by
\begin{eqnarray}
b_n | 0; k \rangle  & = &  0, \; \; \; n \geq -1\, \nonumber\\
c_n | 0; k \rangle  & = &  0, \; \; \; n \geq 2\,, \\
\alpha^\mu_{n} | 0; k \rangle  & = &  0, \; \; \; n \geq 1 \,.\nonumber
\end{eqnarray}
The state $|0;k\rangle$ is a  momentum eigenstate:
\begin{equation}
p^\mu | 0; k \rangle = k^\mu | 0; k \rangle\,.
\end{equation}
The zero-momentum state $| 0;  0 \rangle$ is the SL(2,R) invariant
vacuum; we  will often write it simply as $| 0 \rangle$.  This vacuum is
defined to have ghost number 0,  and it is normalized by the equation
\begin{equation}
\label{conventionalnormalization}
\langle 0; k | c_{-1} c_0c_1 | 0; k' \rangle =
(2 \pi)^{26}
\delta (k-k')
\end{equation}
For string field theory we will also find it
convenient to work with the vacua of ghost number one and two:
\begin{eqnarray}
G = 1: & \hspace*{0.1in} &  | 0_1  \rangle = c_1| 0 \rangle\\
G = 2: & \hspace*{0.1in} &  | 0_2  \rangle = c_0 c_1| 0 \rangle\,. \nonumber
\end{eqnarray}
In the notation of Polchinski~\cite{Polchinski-string}, these two vacua
are written as
\begin{eqnarray}
| 0_1 \rangle & = & | 0 \rangle_m \otimes |\! \downarrow \rangle \nonumber\\
| 0_2 \rangle & = & | 0 \rangle_m \otimes |\! \uparrow \rangle  \,,
\end{eqnarray}
where $| 0 \rangle_m$ is the matter vacuum and $|\!\downarrow\rangle,
|\!\uparrow\rangle$ are the ghost vacua annihilated by $b_0, c_0$.

The BRST operator of this theory is given by
\begin{equation}
Q_B = \sum_{n = -\infty}^{\infty}  c_nL_{-n}^{({\rm m})}
+ \sum_{n, m = -\infty}^{ \infty}  \frac{(m-n)}{2}
:c_mc_nb_{-m-n}:-c_0
\label{eq:BRST}
\end{equation}
where the matter Virasoro operators are given by
\begin{equation}
L_q^{({\rm m})} = \left\{
\begin{array}{ll}
\frac{1}{2}\sum_{n}\alpha^\mu_{q-n} \alpha_{\mu \; n}, & q \neq 0\\[1.0ex]
p^2 + \sum_{n = 1}^{ \infty}  \alpha^\mu_{-n} \alpha_{\mu
\; n}\,.
\end{array}
\right.
\end{equation}


\subsection{Witten's cubic bosonic SFT}
\label{sec:Witten-SFT}

The discussion of the previous subsection leads to a systematic
quantization of the open bosonic string in the conformal field theory
framework.  Using this approach it is possible, in principle, to
calculate an arbitrary perturbative on-shell scattering amplitude for
physical string states.  To study tachyon condensation in string
theory, however, we require a nonperturbative, off-shell formalism for
the theory--- a string field theory.

A very simple form for the
off-shell open bosonic string field theory action was proposed by
Witten in 1986:~\cite{Witten-SFT}
\begin{equation}
S = -\frac{1}{2}\int \Psi \star Q \Psi -\frac{g}{3}  \int \Psi \star
\Psi \star \Psi\,.
\label{eq:SFT-action}
\end{equation}
This action has the general form of a Chern-Simons theory on a
3-manifold, although for string field theory there is no explicit
interpretation of the integration in terms of a concrete 3-manifold.
In Eq.~(\ref{eq:SFT-action}), $g$ is interpreted as the (open) string
coupling constant.  The field $\Psi$ is a string field, which takes
values in a graded algebra ${\mathcal A}$.  Associated with the algebra
${\mathcal A}$ there is a star product
\begin{equation}
\star:{\mathcal A} \otimes{\mathcal A} \rightarrow{\mathcal A}, \;\;\;\;\;
\end{equation}
under which the degree $G$ is additive ($G_{\Psi \star \Phi} = G_\Psi
+ G_\Phi$).  There is also a BRST operator
\begin{equation}
Q:{\mathcal A} \rightarrow{\mathcal A}, \;\;\;\;\;
\end{equation}
of degree one ($G_{Q \Psi} = 1 + G_\Psi$).  String fields can be
integrated using
\begin{equation}
\int:{\mathcal A} \rightarrow {\bf C}\,.
\end{equation}
This integral vanishes for all $\Psi$ with degree $G_\Psi \neq 3$.
Thus, the action (\ref{eq:SFT-action}) is only nonvanishing for a
string field $\Psi$ of degree 1.

The elements $Q, \star, \int$ that define the string field theory are
assumed to satisfy the following axioms:
\vspace*{0.15in}

\noindent {\bf (a)} Nilpotency of $Q$: $\;Q^2 \Psi = 0, \;\; \; \forall \Psi
\in{\mathcal A}$.
\vspace*{0.08in}

\noindent {\bf (b)} $\int Q\Psi = 0, \; \; \; \forall \Psi \in{\mathcal A}$.
\vspace*{0.08in}

\noindent {\bf (c)} Derivation property of $Q$:\\
\hspace*{0.4in}$\;Q (\Psi \star \Phi) = (Q \Psi) \star \Phi +
(-1)^{G_\Psi} \Psi \star (Q \Phi), \; \; \forall \Psi, \Phi \in{\mathcal A}$.
\vspace*{0.08in}

\noindent {\bf (d)} Cyclicity:  $\;\int \Psi \star \Phi = (-1)^{G_\Psi
G_\Phi} \int \Phi \star \Psi, \; \; \; \forall \Psi, \Phi \in{\mathcal A}$.
\vspace*{0.08in}

\noindent {\bf (e)}  Associativity:  $(\Phi \star \Psi) \star \Xi =
\Phi \star (\Psi \star \Xi), \; \; \;
\forall \Phi, \Psi, \Xi \in{\mathcal A}$.
\vspace*{0.15in}

When these axioms are satisfied, the action (\ref{eq:SFT-action}) is
invariant under the gauge transformations
\begin{equation}
\delta \Psi = Q \Lambda + \Psi\star \Lambda - \Lambda \star \Psi\,,
             \label{eq:SFT-gauge}
\end{equation}
for any gauge parameter $\Lambda \in{\mathcal A}$ with degree 0.

When the string coupling $g$ is taken to vanish, the equation of
motion for the theory defined by (\ref{eq:SFT-action}) simply becomes
$Q \Psi = 0$, and the gauge transformations (\ref{eq:SFT-gauge})
simply become
\begin{equation}
\delta \Psi = Q \Lambda\,.
\end{equation}
Thus, when $g = 0$ this string field theory gives precisely the structure
needed to describe the free bosonic string.  The motivation for
introducing the extra structure in (\ref{eq:SFT-action}) was to
find a simple interacting extension of the free theory, consistent
with the perturbative  expansion of open bosonic string theory.

Witten presented this formal structure and argued
that all the needed axioms are satisfied when ${\mathcal A}$ is taken to
be the space of string fields
\begin{equation}
{\mathcal A} =\{\Psi[x (\sigma); c (\sigma), b
(\sigma)]\}
             \label{eq:string-functionals}
\end{equation}
which can be described as functionals of the matter, ghost and
antighost fields describing an open string in 26 dimensions with $0
\leq \sigma \leq \pi$.  Such a string field can be written as a formal
sum over open string Fock space states with coefficients given by an
infinite family of space-time fields
\begin{equation}
\Psi =
\int d^{26}p \;
\left[ \phi (p)\; | 0_1; p \rangle + A_\mu (p) \; \alpha^\mu_{-1} | 0_1; p
\rangle + \cdots \right]
             \label{eq:field-expansion}
\end{equation}
Each Fock space state is associated with a given string functional,
just as the states of a harmonic oscillator are associated with
wavefunctions of a particle in one dimension.  For example, the matter
ground state $| 0 \rangle_m$ annihilated by $a_n$ for all $n \geq 1$
is associated (up to a constant $C$) with the functional of matter
modes
\begin{equation}
| 0 \rangle_m \rightarrow
C \exp \left( -\frac{1}{2}\sum_{n > 0}^{ \infty}
nx_n^2 \right)\,.
\end{equation}

For Witten's cubic string field theory, the BRST operator $Q$ in
(\ref{eq:SFT-action}) is the usual open string BRST operator $Q_B$,
given in (\ref{eq:BRST}), and the degree associated with a Fock space
state is the
ghost number of that state.
The star product $\star$ acts on a pair
of functionals $\Psi, \Phi$ by gluing the right half of one string to
the left half of the other using a delta function interaction

\begin{center}
\begin{picture}(100,60)(- 50,- 30)
\put(-40,20){\line(1,0){37}}
\put(40,20){\line(-1,0){37}}
\put(-3,20){\line( 0, -1){ 37}}
\put(3,20){\line( 0, -1){ 37}}
\put(-20,2){\makebox(0,0){$\Psi$}}
\put(20, 2){\makebox(0,0){$\Phi$}}
\end{picture}
\end{center}

This star product factorizes into separate matter and ghost parts.
In the matter sector, the star product is given by the formal functional
integral
\begin{eqnarray}
\lefteqn{\left(\Psi \star   \Phi\right) [z(\sigma)]} \label{eq:mult} \\ &
\equiv&
\int
\prod_{{0} \leq \tilde{\tau} \leq {\pi\over 2}} dy(\tilde{\tau}) \; dx
(\pi -\tilde{\tau})
\prod_{{\pi\over 2} \leq
\tau \leq \pi}
\delta[x(\tau)-y(\pi-\tau)]
\;   \Psi [x(\tau)]  \Phi [y(\tau)]\, ,\nonumber\\
& &
\hspace*{1.2in}x(\tau)  = z(\tau) \quad {\rm for} \quad {0} \leq \tau \leq
{\pi\over 2}\, ,
\nonumber\\
& &
\hspace*{1.2in}y(\tau)  = z(\tau)\quad {\rm for} \quad   {\pi\over 2} \leq
\tau \leq \pi\, .
\nonumber
\end{eqnarray}
Similarly, the integral over a string field factorizes into matter and
ghost parts, and in the matter sector is given by
\begin{equation}
\int \Psi = \int \prod_{0 \leq \sigma \leq \pi} dx (\sigma) \;
\prod_{0 \leq
\tau \leq \frac{\pi}{2} }
\delta[x(\tau)-x(\pi-\tau)] \;\Psi[x (\tau)]\,.
\label{eq:integral-p}
\end{equation}
This corresponds to gluing the left and right halves of the string
together with a delta function interaction

\begin{center}
\begin{picture}(100,60)(- 50,- 30)
\put(-3,20){\line(1,0){6}}
\put(-3,20){\line( 0, -1){ 37}}
\put(3,20){\line( 0, -1){ 37}}
\put(10,-3){\makebox(0,0){$\Psi$}}
\end{picture}
\end{center}

The ghost sector of the theory is defined in a similar fashion, but
has an anomaly due to the curvature of the Riemann surface that  describes
the three-string vertex.  The ghost sector can be described either in
terms of fermionic ghost fields $c (\sigma), b (\sigma)$ or through
bosonization in terms of a single bosonic scalar field $\phi_g
(\sigma)$.  From the functional point of view of Eqs.~(\ref{eq:mult},
\ref{eq:integral-p}), it is easiest to describe the ghost sector in the
bosonized language.  In this language, the ghost fields $b (\sigma)$
and $c (\sigma)$ are replaced by the scalar field $\phi_g (\sigma)$, and
the star product in the ghost
sector is given by (\ref{eq:mult}) with an extra insertion of $\exp
(3i \phi_g (\pi/2)/2)$ inside the integral.  Similarly, the integration
of a string field in the ghost sector is given by (\ref{eq:integral-p})
with an insertion of $\exp (-3i \phi_g (\pi/2)/2)$ inside the integral.
Witten first described the cubic string field theory using
bosonized ghosts.  While this approach is useful for some purposes, we
will use fermionic ghost fields in the remainder of these lecture notes.
With the fermionic ghosts, there is no need for insertions at the
string midpoint.

The expressions (\ref{eq:mult}, \ref{eq:integral-p}) may seem rather
formal, as they are written in terms of functional integrals.  These
expressions, however, can be given precise meaning when described in
terms of creation and annihilation operators acting on the string Fock
space.  In Sections \ref{sec:CFT} and
\ref{sec:overlaps} we give a more precise definition of the string
field theory action using conformal field theory and the countable
mode decomposition of the string.


\subsection{Algebraic structure of OSFT}
\label{sec:algebraic-structure}

Here we discuss an approach to the algebraic structure
of OSFT that is inspired by conformal field theory.
This approach can be used to write rather general
string actions, including those whose interactions are not based
on delta-function overlaps.  In this language the string action
takes the form
\begin{equation} \label{e1}
S (\Phi) = \,-\, {1\over g^2}\,\,\bigg[\, {1\over 2} \langle \,\Phi
\,,\, Q\,
\Phi
\rangle + {1\over 3}\langle \,\Phi \,,\, \Phi *
\Phi \rangle \bigg]\,.
\end{equation}
Here $g$ is the open string coupling constant, the string field
$\Phi$ is a state in the total matter plus ghost CFT, $Q$ is the
kinetic operator, $*$ denotes a multiplication or star-product,
and $\langle \cdot \,, \cdot \rangle$ is a bilinear inner product
on the state space of the CFT.
We will discuss the relationship
between the action (\ref{e1}) and the form of the action
(\ref{eq:SFT-action}) used in the previous subsection shortly.

The kinetic operator $Q$  satisfies the following identities
\begin{eqnarray} \label{l1e1}
&& Q^2 A = 0, \nonumber \\
&& Q (A * B) = (QA) * B + (-1)^{A} A * (QB)\,, \\
&& \langle \, Q A , B \,\rangle = - (-)^A \langle A , Q B \rangle
\,.\nonumber
\end{eqnarray}
The first equation is the nilpotency condition, the second states
that $Q$ is a derivation of the star product, and the third states
that $Q$ is self-adjoint.
There are also identities involving the inner product and the star operation
\begin{eqnarray}
\label{l1e2}
&& \langle A, B \rangle = (-)^{AB} \langle  B , A \rangle\,, \nonumber \\
&& \langle \, A \,, B * C \, \rangle = \, \langle A* B \,,\, C \,
\rangle \, \\
&& A * (B * C) = (A*B) * C \,.\nonumber
\end{eqnarray}
In the sign factors, the exponents $A, B, \cdots$ denote the
Grassmanality
of the state, and should be read as $(-)^A \equiv (-)^{\epsilon (A)}$
where $\epsilon (A) = 0 \, (\hbox{mod}~ 2)$ when  $A$ is Grassmann even,
and
$\epsilon (A) = 1 \, (\hbox{mod}~ 2)$ when $A$ Grassmann odd.
The first property above is a symmetry condition, the second indicates
that the inner product has a cyclicity property analogous to the similar
property of the trace operation.  Finally, the last equation is the
statement that the star product is associative.

Finally, we also
have that the star operation is an {\it even} product of degree zero
(as before, we
identify degree with ghost number).  In plain english, this means that both the
grassmanality and the ghost number of the star
product of two string fields is obtained from those of the string
fields without any additional offset:
\begin{eqnarray}
\label{l1e3}
&& \epsilon (A * B ) = \epsilon (A) + \epsilon (B) \,,\nonumber \\
&& \hbox{gh} (A* B ) = \hbox{gh} (A) + \hbox{gh} (B) \,.
\end{eqnarray}
In this language $Q$ is an odd operator of degree one:
\begin{eqnarray}
\label{l1e3q}
&& \epsilon (QA  ) = \epsilon (A) +1\,, \nonumber \\
&& \hbox{gh} (QA ) = \hbox{gh} (A) + 1 \,.
\end{eqnarray}
In  the conventions we shall work
             the SL(2,R) vacuum $|0\rangle$ is assigned ghost
number zero.   The  Grassmanality $\epsilon (A)$
of a string field $A$ is an integer mod~2.
In open string field theory, Grassmanality and ghost number (degree)
are related because Grassmann odd operators carry odd units of
ghost number.

The algebraic structure discussed here is very similar, but
not identical to that in section~\ref{sec:Witten-SFT}.
The string field $\Phi$ and the action (\ref{e1}) can be related to
the string field $\Psi$ and the action (\ref{eq:SFT-action}) of the
previous section by taking
\begin{equation}
\Phi = g \Psi
\end{equation}
and by relating the inner product used here to the integral used in
(\ref{eq:SFT-action}) through
\begin{equation}
\label{identifystructures}
\langle A \,, B\rangle  = \int A \star B \,.
\end{equation}
The first two conditions in (\ref{l1e1}) are then clearly equivalent to
properties (a) and (c) of section~\ref{sec:Witten-SFT}.  You can also
readily see that the first two properties in (\ref{l1e2}) hold given
properties (d) and (e).  Property (b), however, does not have a
counterpart in this formalism.  A counterpart exists if we assume the
existence of a suitable identity string field $\mathcal{I}$, as we
will discuss at the end of this subsection.

Throughout these lectures we will go back and forth between the CFT
notation with string field $\Phi$ and action (\ref{e1}) and the
oscillator description with string field $\Psi$ and action
(\ref{eq:SFT-action}) (which we rewrite more explicitly as
(\ref{eq:action-Fock-1}) in section \ref{sec:overlaps}).  While we
could have chosen to use one notation and neglect the other, both
formalisms are used extensively in the literature, and some results
are more easily expressed in one notation than the other.  When in
doubt, the reader should return to the previous paragraph to see how
the two notations are related.

\medskip
Let us now deduce some basic properties of the string field,
in particular its ghost number and its Grassmanality. The Grassmanality
of $\Phi$ can be deduced from the condition that the kinetic term
of the string action must be non-vanishing. Using the above properties
we have
\begin{equation}
\label{deducephi}
\langle \Phi, Q \Phi \rangle = (-1)^{\Phi (1+ \Phi)}
\langle Q \Phi, \Phi \rangle  = \langle Q \Phi, \Phi \rangle
= - (-1)^\Phi \langle \Phi, Q \Phi \rangle \,.
\end{equation}
It is clear that the string field $\Phi$ must be Grassmann odd.
At this point we must use some CFT knowledge to decide on
the Grassmanality of the SL(2,R) vacuum and on the ghost number
of the string field.  For bosonic strings we have that zero momentum
tachyon states are of the form  $t c_1 |0\rangle$, where $c_1$ is a
ghost field oscillator.  Since this oscillator is Grassmann odd, and
the string field is also Grassmann odd, we must declare the SL(2,R)
vacuum to be Grassmann even. Thus
\begin{equation}
\label{propzero}
|0\rangle \,\, \hbox{is a Grassmann even state of ghost number zero}\,.
\end{equation}
Since the $c_1$ oscillator carries ghost number one, we also deduce
that the open string field must have ghost number one.
\begin{equation}
\label{propsf}
|\Phi\rangle \,\, \hbox{is a Grassmann odd state of ghost number one}\,.
\end{equation}

Equations (\ref{l1e1}), (\ref{l1e2}),
(\ref{l1e3}),  (\ref{l1e3q}), and (\ref{propsf}) guarantee that the
string field action is
invariant under the gauge transformations:
\begin{equation} \label{l1e5}
\delta \Phi = Q \Lambda + \Phi * \Lambda - \Lambda * \Phi\, ,
\end{equation}
for any Grassmann-even ghost-number zero state $\Lambda$.
Moreover, variation of the action gives the field equation
\begin{equation}
Q \Phi + \Phi * \Phi = 0\,.
\end{equation}

\medskip
\noindent
{\it Exercise} Verify that the string action in (\ref{e1}) is gauge
invariant under the transformations (\ref{l1e5}).

\bigskip

It is convenient to use the above structures to define a multilinear object
that given three string fields
yields a number:
\begin{equation}
\langle\, A\,,\, B\,,\, C \rangle \equiv \langle \, A \,, \, B * C \,
\rangle
\end{equation}The middle equation in (\ref{l1e2}) implies the {\it cyclicity}
of the multilinear form. A small calculation immediately gives:
\begin{equation}
\langle\, A\,,\, B\,,\, C \rangle  = (-)^{A(B+C)}\langle\, B\,,\, C\,,\,
A
\rangle
\end{equation}
A basic consistency check of the signs above is that the cubic
term
$\langle\, \Phi\,,\, \Phi\,,\, \Phi \rangle$ in the action
(\ref{l1e1}) is strictly
cyclic  for odd $\Phi$, and therefore does not vanish.

\bigskip
Open string theory has additional algebraic structure that
sometimes plays a crucial role.  One such structure arises from
the twist operation, which reverses the parametrization of a
string.  From the algebraic viewpoint this is summarized by the
existence of an operator $\Omega$ that  satisfies the following
properties:
\begin{eqnarray}
\label{l1e4}
&& \Omega ( QA) =  Q (\Omega A) \nonumber\\
&&\langle \, \Omega A \,, \, \Omega B \, \rangle = \langle\, A\,, \, B
\,\rangle  \\
&&\Omega \, (A* B) =  (-)^{AB + 1} \,  \Omega (B) * \Omega (A)\nonumber\,.
\end{eqnarray}
The first property means that the BRST operator has zero twist, or
does not change the twist property of the states it acts on.
The second property states that the bilinear form is twist
invariant. The third property is crucial. Up to signs, twisting
the star product of string fields amounts to multiplying the
twisted states in {\it opposite order}. This change of order
is a simple consequence of the basic multiplication rule where
the second half of the first string must be glued to the first
half of the second one.  The sign factor is also important.
For the string field $\Phi$, which is grassmann odd, it gives
\begin{equation}
\label{l1e59}
\Omega ( \Phi * \Phi ) = +  \, (\Omega \Phi) * (\Omega \Phi)
\end{equation}
with the plus sign.  This result, together with the first
two equations in (\ref{l1e4}) immediately implies that the
string field action in (\ref{l1e1}) is twist invariant:
\begin{equation}
\label{l1e6}
S (\Omega \Phi) = S (\Phi) \,.
\end{equation}
This invariance under twist transformations allows one to
construct new string theories by truncating the spectrum to
the subset of states that are twist even.  Moreover, in solving
the string field equations it will be possible to find consistent
solutions by restricting oneself to the twist even subspace of
the string field.

\medskip
\noindent
{\it Exercise}.  Letting $\Omega_A$ denote the $\Omega$ eigenvalue of $A$,
show that
\begin{equation}
\langle A, B, C\rangle = \Omega_A\Omega_B\Omega_C (-1)^{AB + BC+CA +1}
\langle C, B, A\rangle \,.
\end{equation}

\medskip
\noindent
{\it Exercise}.  Let $\Omega A_\pm = \pm A$ and $\epsilon (A_\pm) =1$.
Show that
\begin{equation}
\label{fortwistproperty}
\langle A_+, A_+, A_-\rangle = 0\,.
\end{equation}

\medskip
\noindent
{\it Exercise}.  We will see later
             that the star product of the vacuum with itself is the vacuum
plus Virasoro descendents:
\begin{equation}
|0\rangle * |0\rangle = |0\rangle  + \cdots
\end{equation}
Show that this implies that the vacuum is twist odd:
\begin{equation}
\Omega |0\rangle  = -|0\rangle .
\end{equation}

The star algebra may have
an identity element ${\mathcal I}$.
If  ${\mathcal I}$ exists, it  is presumed to satisfy
\begin{equation}
{\mathcal I}* A = A *  {\mathcal I} = A\,,
\end{equation}
for all states $A$.
Some properties of ${\mathcal I}$
are immediately deduced from the above definition:
\begin{equation}
{\mathcal I} \,\,\hbox{is Grassmann even, ghost number zero, twist odd
string field.}
\end{equation}
The twist odd property follows from the twist property of products
\begin{equation}
\Omega ( {\mathcal I} *A) = (-1)^{0\cdot A + 1} (\Omega A) * (\Omega
{\mathcal I})= -
             (\Omega A) * (\Omega {\mathcal I})\,.
\end{equation}
Since the left hand side is also just $(\Omega A)$ it must follow that
\begin{equation}
             \Omega {\mathcal I} = - {\mathcal I}\,.
\end{equation}
This is consistent with the fact that the SL(2,R) vacuum is also
twist odd. Indeed the identity string field is just the vacuum plus
Virasoro descendents of the vacuum, as we shall see in
Section~\ref{subsec:slivers}.

Any derivation $D$ of the star algebra should
annihilate the identity:
\begin{equation}
D ( {\mathcal I}* A) = (D{\mathcal I}) * A + {\mathcal I} * DA =
(D{\mathcal I}) * A +  DA\,.
\end{equation}
Since the left hand side also equals $DA$, one concludes that
$ (D{\mathcal I}) * A=0$ for all $A$, and thus the expectation that
$D{\mathcal I} =0$.
In the star algebra of open strings
an identity state has been identified~\cite{Gross-Jevicki-12,efhm,Schnabl} that
satisfies the expected properties for most well-behaved states.

Finally,
if the identity string field is annihilated by the derivation $Q$, then
\begin{equation}
\langle Q \Psi \,, \mathcal{I} \rangle = - (-)^\Psi \langle \Psi, Q
\mathcal{I}\rangle =0\,.
\end{equation}
The identification (\ref{identifystructures}) then yields
\begin{equation}
0= \int Q\Psi \star \mathcal{I} = \int Q\Psi \,,
\end{equation}
which is property (b) in the axiomatic formulation of OSFT
discussed in section~\ref{sec:Witten-SFT}.


\section{String field theory: conformal field theory approach}
\label{sec:CFT}

A direct conformal field theory evaluation of the string action
is perhaps the most economical way to proceed in the case of
simple computations.  We will explain this definition of the
action, using at the same time the example of the
action restricted to only the tachyon field
to illustrate the definitions.  The string action, written before
in (\ref{e1}) is given by
\begin{equation} \label{e1p}
S (\Phi) = \,-\, {1\over g^2}\,\,\bigg[\, {1\over 2} \langle \,\Phi
\,,\, Q\,
\Phi
\rangle + {1\over 3}\langle \,\Phi \,,\, \Phi *
\Phi \rangle \bigg]\,.
\end{equation}
This OSFT action can be used to describe the spacetime field theory
of any D-brane.  For example, for a Dp-brane we would have an
underlying conformal field theory of a  field $X^0$
and $p$ fields $X^i$
with Neumann boundary
conditions, and $(25-p)$ fields $X^a$
        with Dirichlet boundary conditions.
In our computations, we will assume that the brane has unit volume,
in which case the mass $M$ of the brane coincides with its tension.
One
can show that, in units where $\alpha'=1$,
\begin{equation}
\label{tf1}
M = {1\over 2\pi^2}  {1\over g^2}\,.
\end{equation}
We will prove this result in section~\ref{sec:tension}.

We will evaluate the OSFT action by truncating the string field
down to the zero momentum tachyon.
The systematic approximation of
the full theory by successive level truncation is described in  detail
in Section \ref{sec:vacuum}.
In the level expansion this
zero momentum tachyon is assigned level zero.  The tachyon
vertex operator is $e^{ipX(z)} c(z)$ and the associated state is
$c_1 |0;p\rangle$.  The zero momentum tachyon state is
$c_1 |0;0\rangle$ or in simpler notation $c_1 |0\rangle$.
Since we have
\begin{equation}
L_0 c_1 |0\rangle = - c_1 |0\rangle\,,
\end{equation}
the level $\ell$ of a state is related to the $L_0$ eigenvalue as
\begin{equation}
\ell = L_0 + 1\,.
\end{equation}
The string field truncated to the zero momentum tachyon is written
as
\begin{equation}
|T\rangle = t \, c_1 |0\rangle \,,
\end{equation}
where the variable $t$ denotes the expectation value of the
tachyon field, and it is a spacetime constant.
The variable $t$ is related to the tachyon field $\phi$ in the
expansion (\ref{eq:field-expansion}) through
\begin{equation}
t = g \phi (0) \,.
\end{equation}
As we alternate between notation $\Psi$ and $\Phi$ for the string
field, we will use $\phi$ and $t$ for the zero-momentum tachyon.
After truncating to just the tachyon degree of freedom $t$
the tachyon potential $V(t)$ is just minus $S(|T\rangle)$ and thus
\begin{equation}
V(t) = - S(|T\rangle) =  M (2\pi^2) \Bigl( {1\over 2} \langle T, Q T\rangle
+ {1\over 3} \langle T, T , T \rangle \Bigr)\,.
\end{equation}
In fact, it is convenient to define the ratio
\begin{equation}
\label{ftest}
f(t) \equiv {V(t)\over M}  =  (2\pi^2) \Bigl( {1\over 2}
\langle T, Q T\rangle + {1\over 3} \langle T, T , T \rangle \Bigr)\,.
\end{equation}
The function $f(t)$ is a rescaled version of the tachyon potential.
By construction,  $f(t)$ has a quadratic term and a cubic term, so $f(t=0)=0$.
The Sen conjecture requires that $f(t)$ have a critical point at $t=t^*$ that
satisfies
\begin{equation}
f(t^*) = -1 \,.
\end{equation}
This is indeed equivalent to
saying that the energy difference between
the D-brane vacuum and the stable vacuum equals the energy $M$
of the D-brane.  It suggests strongly that the stable vacuum is a
vacuum without a D-brane.   It is perhaps useful to remark that
$V(t)$ as obtained directly from the OSFT action does not convey the
true gravitational picture where absolute vacuum energies are important.
The vacuum with the D-brane, namely at $t=0$ has a positive cosmological
constant, or vacuum energy.  This is in fact the D-brane energy.  As the
theory rolls to the stable vacuum, the vacuum energy goes to zero.
Thus the tachyon potential $V(t)$ is missing an additive constant, which
becomes important when coupling to gravity (which we will not consider
in the present lectures).  Such a constant term at least morally belongs
in a more general OSFT action where the disk partition function
would
naturally appear as a field independent contribution to the string action.
This disk partition function calculated with the boundary condition
appropriate to the  D-brane is in fact proportional to the D-brane energy.

\subsection{Kinetic term computations}

Let us begin the computation of the string action truncated to
the tachyon by evaluating $\langle T , QT\rangle$.
To this end we need to use the normalization condition
\begin{equation}
\label{ourashokenorm}
\langle 0| c_{-1} c_0 c_1 |0\rangle = 1\,,
\end{equation}
which is appropriate if we  compactify all coordinates (including
time) into circles of unit circumference.  Indeed, comparing with
(\ref{conventionalnormalization}), we see that the right-hand side
of (\ref{ourashokenorm}) should have a $(2\pi)^{26} \delta (0)$, which
is equivalent to the full spacetime volume $V$.  In our full compactification,
$V=1$.  The compactification of time is only a formal
trick that facilitates computations but is not strictly necessary.

\smallskip
\noindent
{\it Exercise:}  Given $c(z) = \sum_n {c_n\over z^{n-1}}$ show that
\begin{equation}
\label{ghcorr}
\langle 0| c(z_1) c(z_2) c(z_3) |0\rangle  = (z_1-z_2) (z_1 - z_3)
(z_2 - z_3)\,.
\end{equation}

\medskip
Now that we must compute precisely we should make clear the CFT
definition of the inner product

\noindent
{\it Definition: }  $ \langle A , B\rangle = \langle bpz(A) | B\rangle$.
Here $bpz: {\mathcal H} \to {\mathcal H}^*$  is BPZ conjugation,
which we review
next.

Given a primary field $\phi(z)$ of dimension $d$, it has
a mode expansion
\begin{equation}\
\label{oscill}
\phi(z) = \sum_n {\phi_n\over z^{n+d}}
\quad \to \quad \phi_n = \oint {dz\over 2\pi i} z^{n+d-1} \phi(z)\,.
\end{equation}
We define
\begin{equation}
bpz(\phi_n) \equiv  \oint {dt\over 2\pi i} t^{n+d-1} \phi(t)\,, \qquad
\hbox{with} \quad t = -{1\over z} \,.
\end{equation}
Note that this simply defines the BPZ conjugation of the oscillator with
the same formula as the oscillator itself (\ref{oscill}) but referred to
a coordinate at $z=\infty$. This integral is evaluated by using the
transformation law
\begin{equation}
\phi(t) (dt)^d = \phi(z) (dz)^d\,.
\end{equation}
We therefore get
\begin{equation}
bpz(\phi_n) \equiv  - \oint {dz\over 2\pi i}{1\over z^2}
            \Bigl( - {1\over z} \Bigr)^{n+d-1} \phi(z) (z^2)^d\,.
\end{equation}
The minus sign in front arises from a reversal of contour of integration
(a contour circling $t=0$ clockwise circles $z=0$ counterclockwise).
Moreover the transformation law was used to reexpress $\phi(t)$ in
terms of the field $\phi(z)$ whose mode expansion is given.  Simplifying
the integral one finds
\begin{equation}
bpz(\phi_n) =  (-1)^{n+d}  \oint {dz\over 2\pi i}
            z^{-n+d-1} \phi(z) \, = (-1)^{n+d} \phi_{-n}\,.
\end{equation}
In summary, we have shown that
\begin{equation}
bpz(\phi_n) =  (-1)^{n+d} \phi_{-n}\,.
\end{equation}
This equation defines BPZ conjugation when we supplement it with
the rule
\begin{equation}
bpz \Bigl( \phi_{n_1}  \cdots \phi_{n_p}|0\rangle \Bigr)
= \langle 0 |  bpz(\phi_{n_1}) \cdots  bpz(\phi_{n_p})\,.
\end{equation}
This formula is correct as stated also when the oscillators
are anticommuting. The only condition for its validity is that
the various modes with mode numbers of the same sign must
commute (or anticommute).  Otherwise BPZ conjugation produces
a sequence of oscillators in {\em reverse} order.

A nontrivial example of the above rules arises when we calculate
the BPZ conjugates of the modes $L_n$ of the stress tensor.  Although the
stress tensor $T(z)$ is not a primary field, it transforms as a primary under
SL(2,C) transformations, and therefore it does transform as a dimension two
primary under the inversion needed in the definition of BPZ.  Thus we have
\begin{equation}
bpz(L_n) = (-1)^n L_{-n}\,,
\end{equation}
and for a string of oscillators we must write
\begin{equation}
bpz \Bigl( L_{n_1}  \cdots L_{n_p}|0\rangle \Bigr)
= \langle 0 |  bpz(L_{n_p}) \cdots  bpz(L_{n_1})\,.
\end{equation}

\medskip
Since  $c(z)$ has dimension minus one,
$bpz(c_1) = (-1)^{1+1} c_{-1} = c_{-1}$, so
$bpz (c_1 |0\rangle) = \langle 0| c_{-1}$.  With this
we have
\begin{equation}
\langle T, Q T \rangle = t^2 \langle 0| c_{-1} Q c_1 |0\rangle\,.
\end{equation}
Because of the form of the inner product only the term $c_0L_0$
in $Q$ can contribute and we have
\begin{equation}
\label{tachkin}
\langle T, Q T \rangle = t^2 \langle 0| c_{-1} c_0 L_0 c_1 |0\rangle
= - t^2  \langle 0| c_{-1} c_0 c_1 |0\rangle = - t^2\,.
\end{equation}
This completes the computation of the quadratic term in the tachyon
potential.  The negative sign obtained is the expected one, showing
the instability of the $t=0$ field configuration.

\subsection{Interaction term computation}
\medskip
To compute the interaction of three tachyons we must explain
how the three vertex is defined in CFT language.  Consider three
states $A,B,$ and $C$ and their associated vertex operators
${\mathcal O}_A$, ${\mathcal O}_B$, and ${\mathcal O}_C$. We define
\begin{equation}
\label{defvertex}
\langle A, B, C \rangle \equiv \Bigl\langle f_1^D \circ {\mathcal O}_A(0),
f_2^D \circ {\mathcal O}_B(0), \, f_3^D \circ {\mathcal O}_C(0)\Bigr\rangle_D
\end{equation}
Here the right hand side denotes the CFT correlator of the conformal
transforms of the vertex operators
${\mathcal O}_A$, ${\mathcal O}_B$, and ${\mathcal O}_C$. The conformal
transforms are specified by the functions $f_i$ as we explain now.
Let there be three canonical coordinates $\xi_i$, with $i=1,2,3$.
The three functions $f_i(\xi_i)$ define maps from the upper half
disks $\Im (\xi_i) \geq 0, |\xi_i| \leq 1$ into a disk $D$, with the points
$\xi_i=0$ being taken into  points in the boundary of the disk $D$.
The meaning of the conformal map of operators is that:
$f_i\circ {\mathcal O}_A (0)$ {\it is the operator} ${\mathcal O}_A(\xi_i=0)$
{\it expressed in terms of local operators at} $f_i(\xi_i=0)$.
The disk $D$ may have the form of a unit disk, or can be the
(conformally equivalent) upper half plane, or any other arbitrary
form.  Of course, the unit disk and the upper half plane are
especially convenient for explicit computations.

\begin{figure}[!ht]
\leavevmode
\begin{center}
\epsfxsize = 11 cm \epsfbox{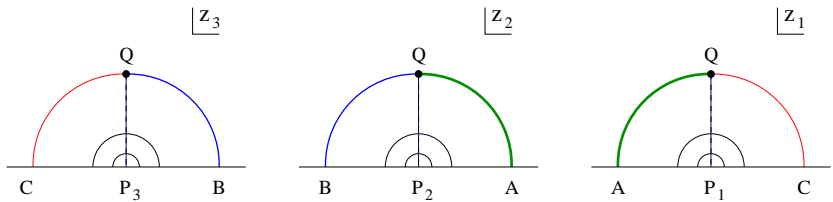}
\end{center}
\caption[]{\footnotesize Representation of the cubic vertex as the gluing of
3 half--disks.} \label{pictorialfig}
\end{figure}

For the SFT
at hand, the picture
is given in  Fig.{\ref{pictorialfig}}.
The worldsheets of the three strings
are represented as the unit half--disks
$\{ |\xi_i| \leq 1, \Im \, \xi \geq 0 \}$, $i=1,2,3$,
            in three copies of the complex plane.
The boundaries $|\xi_i| =1$ in the respective upper half-disks
are {\it the} strings.  Thus the point $\xi_i=i$ is the string
midpoint.  The interaction defining the vertex is
built by gluing the three half-disks to form a single disk.  This is done
by the half-string identifications:
\begin{eqnarray}
\label{wittengluing}
\xi_1 \xi_2  =  -1\,, && \quad {\rm for} \; |\xi_1| =1, \,~~ \Re ( \xi_1 )\leq
0\,, \nonumber \\
\xi_2 \xi_3   =  -1\,,  && \quad {\rm for} \; |\xi_2| =1, \,~~ \Re (\xi_2) \leq
0\,, \\ \xi_3 \xi_1   =  -1\,,  && \quad {\rm for} \; |\xi_3| =1,
\,~~ \Re (\xi_3)
\leq 0\,. \nonumber
\end{eqnarray}
Note that the common interaction point $Q$,
is indeed  $\xi_i=i$ (for
$i=1,2,3$), namely the mid--point of each open string $|\xi_i| =1, \,
\Im (\xi_i)
\geq 0$.  The left-half of the first string is glued with the
right-half of the second string, and the same is repeated cyclically.
This construction defines a specific `three--punctured disk', a
genus zero Riemann surface with a boundary, three marked points
(punctures) on this boundary, and a choice of local coordinates
$\xi_i$ around each puncture.

\begin{figure}[!ht]
\leavevmode
\begin{center}
\epsfxsize = 6 cm \epsfbox{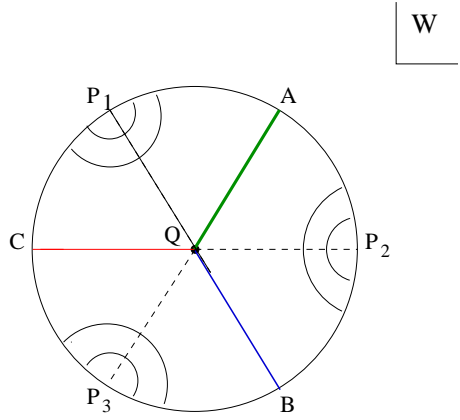}
\end{center}
\caption[]{\footnotesize Representation of the cubic vertex as a 3--punctured
unit disk.} \label{3wedgesfig}
\end{figure}

The calculation of the functions $f^D_i(\xi)$ require a choice of
disk $D$.  We begin with the case when the disk $D$ is simply
chosen to be the interior of the unit disk
$|w| < 1$,  as shown in Fig. \ref{3wedgesfig}.
In this case the functions $f_i^{D_w} \equiv f_i$ must map each half-disk
to a $120^\circ$ wedge of this unit disk.
To construct the explicit maps that
send
$\xi_i$ to the
$w$ plane, one notices that the SL(2,C) transformation
\begin{equation}
\label{hdefi}
h(z)= \frac{1+i\xi}{1-i\xi}\,,
\end{equation}
maps the unit upper--half disk $\{ |\xi| \leq 1, \Im \xi \geq 0 \}$ to the
`right' half--disk $\{ |h| \leq 1, \Re \, h\geq 0 \}$, with
$z=0$ going to $h(0) =1$. Thus the functions
\begin{eqnarray}
\label{unitvertex}
f_1(\xi_1) &  =&   e^{\frac{2 \pi i}{3}}\left(
\frac{1+i\xi_1}{1-i\xi_1} \right)^{\frac{2}{3}}\,, \nonumber \\
            f_2(\xi_2) &  =&  \left(
\frac{1+i\xi_2}{1-i\xi_2} \right)^{\frac{2}{3}}\,, \\
            f(\xi_3) &  =& e^{-\frac{2 \pi i}{3}} \left(
\frac{1+i\xi_3}{1-i\xi_3} \right)^{\frac{2}{3}}\,,\nonumber
\end{eqnarray}
will send the  three half-disks to three wedges in the $w$
plane of Fig. \ref{3wedgesfig}, with punctures at $e^{\frac{2 \pi
i}{3}}$, $1$, and $e^{-\frac{2 \pi i}{3}}$ respectively.
This specification of the functions $f_i(\xi_i)$
gives the definition of the cubic vertex.
In this representation cyclicity ({\it i.e.}, $\langle \Phi_1, \Phi_2 ,
\Phi_3 \rangle =
\langle \Phi_2, \Phi_3 , \Phi_1 \rangle $) is manifest by construction.
By SL(2,C) invariance, there are many other possible representations
that give exactly the same off--shell amplitudes.

\medskip

\begin{figure}[!ht]
\leavevmode
\begin{center}
\epsfxsize = 10 cm \epsfbox{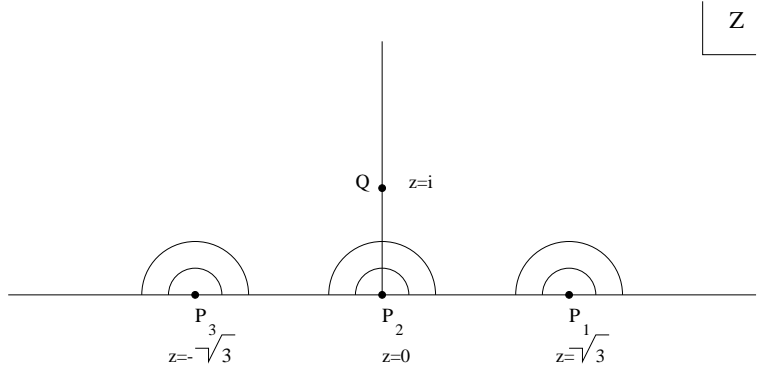}
\end{center}
\caption[]{\footnotesize Representation of the cubic vertex as the upper--half
plane with 3 punctures on the real axis.}\label{uhpfig}
\end{figure}

A useful choice is obtained by  mapping the interacting $w$ disk
symmetrically to the upper half $z$-plane $H$.  This is the convention
that we shall mostly be using.
We can therefore
define the functions
$f_i^H$
by composition of the earlier maps $f_i$ (that send the half-disks to
the $w$ unit disk) with the  map $h^{-1}(w) =-i
\,\frac{w-1}{w+1}$ that takes this unit disk
to the upper--half--plane, with the three punctures on the real
axis (Fig. \ref{uhpfig}),
\begin{eqnarray}
\label{uhpvertex}
            f_1^H (\xi_1)& \equiv & h^{-1}\circ f_1 (\xi_1)=
S ( f_3^H (\xi_1))  \cr\cr
&=& \sqrt {3}+{\frac {8}{3}}\,\xi_1
+{\frac {16}{9}}\,\sqrt {3}\,{\xi_1}^{2}+{\frac
{248}{81}}\,{\xi_1}^{3}+O\left ({\xi_1}^{4}\right )\,. \cr\cr
            f_2^H (\xi_2) & \equiv & h^{-1}\circ f_2 (\xi_2)  =S(f_1^H(\xi_2))=
\tan
\left(\frac{2}{3} \, \arctan(\xi_2) \right)  \cr\cr
&=&
{\frac {2}{3}}\, \xi_2-{\frac {10}{81}}\, {\xi_2}^{3}
+O\left
({z_2}^{5}\right )\,.
            \cr\cr
f_3^H (\xi_3) & \equiv & h^{-1}\circ f_3\, (\xi_3) =
S (f_2^H (\xi_3)) \cr\cr
&=&
  -\sqrt {3}+{\frac {8}{3}}\, \xi_3-{\frac {16}{9}}\,\sqrt
{3}\, {\xi_3}^{2}+{\frac {248}{81}}\, {\xi_3}^{3} +O({\xi_3}^{4})  \,.
\end{eqnarray}
The three punctures are at
            $ f_1^H(0)=+\sqrt{3}, f_2^H(0)=0, f_3^H(0)=-\sqrt{3},$
and the SL(2,R) map $S(z) = \frac{z - \sqrt{3}}{1+\sqrt{3}z}$
cycles them
(thus $S\circ S\circ S(z) =z$).

\medskip
This completes the definition
of the string field theory action. When the disk $D$ is presented as
a unit disk the functions $f_i$ in (\ref{defvertex}) are the functions
given in equation (\ref{unitvertex}).  When the disk $D$ is presented
as the upper half  plane $H$ the relevant functions in (\ref{defvertex})
are the functions $f_i^H$ given in (\ref{uhpvertex}) above.

\medskip\noindent
{\it Exercise:}  Verify explicitly by a {\it by-hand} calculation
that the first two terms in the expansion of $f_1^H$ and $f_3^H$,
as well as the first term in $f_2^H$ are correct.

\medskip
Let us now return to the computation of the tachyon action.
For our string field $|T\rangle = tc_1|0\rangle$ the interaction
term $\langle T, T, T\rangle$ will be given by
\begin{equation}
\langle T, T, T\rangle  = t^3 \langle c_1, c_1, c_1\rangle\,.
\end{equation}
Since the vertex operator associated to $c_1|0\rangle$ is
$c(z)$,  using
(\ref{defvertex}) we write:
\begin{eqnarray}
\langle T, T, T\rangle = t^3\langle  f_1^H\circ c (0), f_2^H\circ c
(0),  f_3^H\circ c (0) \rangle_{H}
\end{eqnarray}
Since the field $c(z)$ is a primary of dimension minus one, we
have
\begin{equation}
{c(z)\over dz} = {c(\xi)\over d\xi} \quad \to c(\xi) = {c(z)\over {dz\over
d\xi}}
\end{equation}
Therefore
\begin{eqnarray}
f\circ c(0) \equiv c(\xi=0) = {c(f(0)) \over f'(0)}.
\end{eqnarray}
Using equations (\ref{uhpvertex}) to read the values of
$f_1^H(0)$ and ${df_1^H\over d\xi }(0)$ we therefore get, for
example,
\begin{equation}
f_1^H\circ c(0)  = {c(f_1^H(0)) \over {f_1^H}'(0)}  =
{c(\sqrt{3})\over {8/3}}\,.
\end{equation}
The other two insertions are dealt with similarly, and we find
\begin{eqnarray}
\langle T, T, T \rangle &=& t^3 \Bigl\langle\,\, { c (\sqrt{3})\over
{8/ 3}}\,\,, { c
(0)\over {2 / 3}}, { c (-\sqrt{3})\over {8/ 3}}
\Bigr\rangle_H \cr & =& {3^3 \over 2^7}
\langle c(\sqrt{3}) c(0) c(-\sqrt{3}) \rangle_H =
{3^4 \sqrt{3} \over 2^6}\,,
\end{eqnarray}
where in the last step we made use of (\ref{ghcorr}). Our  answer is
therefore
\begin{eqnarray}
\label{tachver}
\langle T, T, T \rangle = {81\sqrt{3}\over 64} t^3\equiv t^3 K^3\,,
\quad \langle c_1, c_1, c_1 \rangle = {81\sqrt{3}\over 64} = K^3\,.
\end{eqnarray}
This completes the calculation of an interaction term.

\subsection{A first test of the tachyon conjecture}\label{firsttestofit}

Having obtained the kinetic term of the tachyon truncated
action in (\ref{tachkin}) and the cubic term in (\ref{tachver}) we are now in
a position  to evaluate the function $f(t)$ in (\ref{ftest}):
\begin{equation}
f(t) = 2\pi^2 \Bigl( - {1\over 2} t^2 + {1\over 3} K^3 t^3\Bigr)\,.
\label{eq:CFT-cubic-potential}
\end{equation}
We must now find the (locally) stable critical point $t=t^*$ of this
potential and evaluate the value of $f(t^*)$.  It is clear the answer
will not be the precise one $f=-1$, since we have truncated
the string field  dramatically.
Nevertheless, we hope
to get an answer that is reasonably close, if level expansion is
supposed to make sense.

\medskip
The
equation of motion is
\begin{equation}
  -t^* + t^{*2} K^3 = 0 \quad \to \quad t^* = {1\over K^3}\,,
\end{equation}
and substituting back we find
\begin{equation}
f(t^*) = -{1\over 3} {\pi^2\over K^6} = -\pi^2 {2^{12}\over 3^{10}}
= - \pi^2 {4096\over 59049} \simeq -0.684.
\end{equation}
Thus is this simplest approximation, where we only kept the tachyon
zero mode we have found that the critical point cancels about 70\%
of the D-brane energy.
In section (\ref{sec:vacuum}) we discuss the
extension of this calculation to include massive string modes.

\subsection{String vertex in the CFT approach: Neumannology}
\label{sec:more-CFT}

When doing explicit computations in OSFT we need to consider
interactions of fields other than the tachyon. The explicit
computation of the previous section becomes a lot more involved
for massive fields, and it is useful to find an automated procedure
to deal with such calculations.  One such procedure is based on
conformal field theory conservation
laws.  This is a very effective method, but we will not review it
here since its
explanation in Rastelli and Zwiebach~\cite{Rastelli-Zwiebach} is 
self-contained.
Another approach uses the explicit Fock representations
of the string vertex.  This will be our subject of interest here.
We will provide a self-contained derivation of the Neumann
coefficients that define the three string vertex both in the matter
and in the ghost sector.  In fact, our construction will be
general and applies to three string interactions other than the
one used in OSFT.  We will determine the full structure of
the three string vertex, except for the matter zero modes.

In the Fock space representation of the vertex, we must find
a state $\langle V_3| \in {\mathcal H}^* \otimes  {\mathcal
H}^*\otimes  {\mathcal H}^*$
such that for any Fock space states $A, B$ and $C$ one finds that
\begin{equation}
\label{introvert}
\langle A, B, C\rangle \equiv  \langle V_3 | A\rangle_{(1)}
            | B\rangle_{(2)}  | B\rangle_{(3)}\,.
\end{equation}
Since we provided in (\ref{defvertex}) a definition of the left hand
side of the above equation, the vertex $\langle V_3|$ is implicitly
defined.  Our procedure will be general in that the functions
$f_r (\xi)$ that  map the canonical half-disks to the upper half plane
will be kept arbitrary.
There is a natural ansatz for the vertex:
\begin{eqnarray}
\label{ansatzv}
\langle V_3| &=& {\mathcal N}  (\langle 0| c_{-1} c_0 )^{(3)} )
(\langle 0| c_{-1} c_0 )^{(2)})  (\langle 0| c_{-1} c_0 )^{(1)}) \\
&& \exp \Bigl( - {1\over 2} \sum_{r,s} \sum_{n,m\geq 1} \,
\alpha_m^{(r)} \, N^{rs}_{mn} \, \alpha_n^{(s)} \Bigr)
            \exp \Bigl( \sum_{r,s} \sum_{m\geq 0\atop n\geq 1} \,
b_m^{(r)} \, X^{rs}_{mn} \, c_n^{(s)} \Bigr) \,.\nonumber
\end{eqnarray}
Here ${\mathcal N}$ is a normalization factor, which will be determined
shortly. In fact, its determination is essentially the tachyon computation
of the previous section.  Moreover, note that the nontrivial oscillator
dependence in the matter sector  is in the form of an exponential of
a quadratic
form. This is a
general result that follows from the free field property of the  matter
CFT.  Having just a quadratic form is possible also for the ghost sector,
but it requires a careful choice of vacua. This is because there is
a sum rule regarding ghost number-- if the vacua are not chosen
conveniently, extra linear ghost factors are necessary in the vertex.
Since the vertex state $\langle V_3|$  is a bra we use out-vacua,
in particular the vacua
$\langle0|c_{-1} c_0$.  This is
quite convenient because the ghost number conservation
law is satisfied when each of the states $A,B$ and $C$ in
(\ref{introvert}) is of ghost number one.  Indeed in each of the three
state spaces we must have a total ghost number of three--
two are supplied by the out-vacuum, and one by the in-state.
This clearly allows the nontrivial ghost dependence of the vertex
to be just a pure exponential with zero ghost number.
A final point concerns the sum restrictions over the ghost
oscillators.  These simply arise because only oscillators that
do not kill the vacua $\langle 0| c_{-1} c_0$ should appear in
the exponential. Thus for the antighost oscillators $b_m$
            we find $m\geq 0$
and for the ghost oscillators $c_n$ we find $n\geq 1$

The normalization factor ${\mathcal N}$ can be determined by
finding the overlap of the vertex with three zero momentum
tachyons $c_1|0\rangle$. In this case we have
\begin{equation}
\langle c_1, c_1, c_1\rangle = \langle V_3 | c_1\rangle_{(1)}
            | c_1\rangle_{(2)}  | c_1\rangle_{(3)} = {\mathcal N}\,,
\end{equation}
since all oscillators in the exponentials kill the zero momentum
tachyon.  In (\ref{tachver}) we found the value of this constant
for the case of the OSFT vertex
\begin{equation}
{\mathcal N} = K^3 = \frac{3^{9/2}}{2^6}  \,.
\end{equation}
The calculation in the general
case is not any more complicated and it is a good exercise!

\medskip
\noindent
{\it Exercise: }  Show that for arbitrary functions $f_i(\xi)$,
$i=1,2,3$,  that map half-disks to the UHP, the  constant $\mathcal{N}$ in the
        vertex (\ref{ansatzv}) is given by:
\begin{equation}
\label{normv}
{\mathcal N} = {(f_1(0) - f_2(0)) (f_1(0) - f_3(0)) (f_2(0) - f_3(0)) \over
f_1'(0) f_1'(0) f_1'(0) }\,.
\end{equation}

\medskip
\noindent
Our goal now is to find explicit expressions for the Neumann
coefficients $N^{rs}_{mn}$ and $X^{rs}_{mn}$ in terms of the
functions $f_i$ that define the vertex.

We begin with the matter sector,
where the following conventions are used
\begin{equation}
\label{modex}
i\partial X(z) = \sum{\alpha_n\over z^{n+1}} \,, \quad
\alpha_n = \oint {dz\over 2\pi i}  \, z^n \, i \partial X\,,
\end{equation}
\begin{equation}
\label{opex}
\langle i\partial X (z)\, \, i\partial X(w) \rangle = {1\over (z-w)^2}\,,\quad
[\alpha_n ,
\alpha_m]
            = n \delta_{m+n,0}\,.
\end{equation}
To find the matter Neumann coefficients we evaluate
\begin{equation}
\label{expmatt}
M = \langle V_3 | \, \, R\left( i \partial X^{(r)} (z) \,\,  i
\partial X^{(s)} (w)
\right)\, c_1^{(1)}|0\rangle_{(1)}\,
c_1^{(2)} | 0\rangle_{(2)} \,  c_1^{(3)}|0\rangle_{(3)}
\end{equation}
in
two different ways.  In here $R( \, \dots\,)$ denotes radial ordering,
necessary when $r=s$.   For our first
computation we use the mode expansion (\ref{modex}) of the conformal
fields to find that
\begin{equation}M = \langle V_3 | \Bigl( \sum_{m,n}  {1\over
z^{-m+1}} {1\over w^{-n+1} }
           \, \alpha_{-m}^{(r)} \,
\alpha_{-n}^{(s)}\,  + {\delta^{rs}\over (z-w)^2} \Bigr)c_1^{(1)}
|0\rangle_{(1)}\,
c_1^{(2)} | 0\rangle_{(2)} \,  c_1^{(3)}|0\rangle_{(3)}
            \end{equation}and the oscillator form  (\ref{ansatzv}) of
the vertex to obtain
\begin{equation}\label{firsteval}
M = -{\mathcal N} \sum_{m,n} z^{m-1}w^{n-1}  \, m n N^{rs}_{mn} +
{\mathcal N} {\delta^{rs}\over (z-w)^2} \,.
\end{equation}In the second evaluation we first rewrite $M$ as
\begin{equation}
\label{rewritem}
M = \langle V_3 | \, \,  i \partial X^{(r)} (z) \,\,  i \partial X^{(s)} (w) \,
c^{(1)}(0)c^{(2)}(0)c^{(3)}(0) |0\rangle_{(1)}\,
            | 0\rangle_{(2)} \,  |0\rangle_{(3)}\,,
\end{equation}
and  reinterpret as a correlator,
in the spirit of (\ref{defvertex}):
\begin{eqnarray}
M = \Bigl\langle f_r\circ i \partial X (z)
\,  f_s \circ  i \partial X (w) \,  f_1 \circ c(0) \, f_2 \circ c(0)
\, f_3 \circ
c(0)\Bigr\rangle\,.
\end{eqnarray}
The ghost part of this correlator gives the factor ${\mathcal N}$.  The matter
part, using $i\partial X(z) = i\partial X (f(z)) {df\over dz}$, and
(\ref{opex}) finally  gives
\begin{equation}\label{seceval}
M= {\mathcal N}  \, f'_r(z) \, f'_s(w)  \Bigl\langle
\,i\partial X (f_r(z))\,\,  i\partial X (f_s(w))\,\Bigr\rangle
= {\mathcal N} {\, f'_r(z) \, f'_s(w)\over ( f_r(z) - f_s(w))^2}\,.
\end{equation}Equating the results (\ref{firsteval}) and
(\ref{seceval}) of the two
evaluations of $M$ we obtain:
\begin{equation}
\sum_{m,n} z^{m-1}w^{n-1}  \, m n N^{rs}_{mn}  - {\delta^{rs}\over (z-w)^2}= -
            {\, f'_r(z) \, f'_s(w)\over ( f_r(z) - f_s(w))^2}\,.
\end{equation}
It is now simple to pick up the coefficients $N^{rs}_{mn}$ by
contour integration over small circles surrounding $z=0$ and
$w=0$.  The second term on the left-hand side gives no contribution,
and one finally finds
\begin{equation}
\label{neumannmatter}
N^{rs}_{mn} = -{1\over  mn} \oint_0{dz\over 2\pi i}{1\over z^m}
\oint_0{dw\over 2\pi i}{1\over w^n} \, {\, f'_r(z) \, f'_s(w)\over ( f_r(z) -
f_s(w))^2}\,.
\end{equation}
This is the desired expression for the Neumann coefficients of the
matter sector.  They can be used for  an  arbitrary  vertex.
The above contour integrals are straightforward to compute and
they can be easily done by a computer in a series expansion. In terms
of residues the expression above is equivalent to
\begin{equation}
\label{neuresidue}
N^{rs}_{mn} = -{1\over  mn} \hbox{Res}_{z=0} \hbox{Res}_{w=0}
\Bigl[ {1\over z^m}
{1\over w^n} \, {\, f'_r(z) \, f'_s(w)\over ( f_r(z) -
f_s(w))^2}\Bigr]
\end{equation}

\medskip
\noindent
{\it Exercise:} Show that the contour integrals in
(\ref{neumannmatter}) can be evaluated in any order.
Do this both for the case when $r\not= s$ and for the
case when $r=s$.

\medskip
We now turn to the calculation of the
ghost Neumann coefficients $X^{rs}_{mn}$.
For this we need mode expansions and two point functions for
the ghost CFT:
\begin{equation}
c(z) = \sum_n {c_n\over z^{n-1}} \,, \quad b(z) = \sum_n {b_n\over z^{n+2}}
\,, \quad \langle c(z) b(w) \rangle = {1\over z -w}\,.
\end{equation}
The strategy is once more based on the computation of a certain
expression in two different ways. Indeed, we consider the overlap
\begin{equation}
G = \langle V_3 | \, R\left( b^{(s)} (z) \,\,  c^{(r)} (w) \right)\,  |
\,c_1^{(1)}|0\rangle_{(1)}\,
c_1^{(2)} | 0\rangle_{(2)} \,  c_1^{(3)}|0\rangle_{(3)}
\end{equation}
and first evaluate it by using the mode expansion of the antighost
and ghost fields, and then the explicit expression for the vertex in
(\ref{ansatzv}). In this way we find
\begin{eqnarray}
\label{fghost}
G &=& \langle V_3 |\Bigl( \sum_{m,n}  {1\over z^{-n+2}}  {1\over w^{-m-1} }
             \, b_{-n}^{(s)} \, c_{-m}^{(r)}  +{w\over
z(z-w)}\Bigr) \,\,c_1^{(1)}|0\rangle_{(1)}\, c_1^{(2)} | 0\rangle_{(2)} \,
c_1^{(3)}|0\rangle_{(3)}
\nonumber  \\ &=& {\mathcal N} \sum_{m,n} z^{-n+2}w^{-m-1}  \,   X^{rs}_{mn}
        + \mathcal{N} {w\over
z(z-w)}\,.
            \end{eqnarray}
In the second computation $G$ is interpreted as a correlator and we
have
\begin{eqnarray}
G &=& \Bigl\langle f_s\circ b(z)
\,\, f_r\circ c (w)\,\, f_1\circ  c(0) \,
            f_2\circ  c(0) \, f_3\circ  c(0)
\Bigr\rangle \\
&=& {(f'_s (z))^2\over f'_r(w) } {1\over f'_1(0)
f'_2(0) f'_3(0)}  \Bigl\langle  b(f_s(z))
\,\,  c (f_r(w))\,\,  c(f_1(0)) \,c(f_2(0)) \,c(f_3(0))
\Bigr\rangle\,, \nonumber
            \end{eqnarray}
where we used the standard conformal maps of the relevant
operators, all of which are primary. The final correlator is in the upper half
plane and all the field arguments refer to the coordinates in the
upper half plane.  The correlator can be calculated by using OPE's, but it is
simpler to use the singularity structure and derive the normalization from a
special configuration. Note, for example, that  there must be zeroes when any
pair of
$c$ fields approach each other. In particular,  this will include a
factor $(f_1(0) - f_2(0)) (f_1(0) - f_3(0)) (f_2(0) - f_3(0))$
as in ${\mathcal N}$ (see (\ref{normv})). We will also have poles
when the antighost approaches any ghost.  These considerations
imply that
\begin{equation}\label{sghost}
G ={\mathcal N} {(f'_s (z))^2\over f'_r(w) } {1\over f_s(z) - f_r(w)}
{\prod_{I=1}^3 (f_r(w) -f_I(0)) \over \prod_{J=1}^3 (f_s(z) - f_J(0))}\,.
\end{equation}We can now equate the results
obtained in  (\ref{fghost}) and (\ref{sghost}).  Picking up
the coefficients via contour integration,  and noting that the
second term on the right-hand side of  (\ref{fghost}) does not
contribute for the relevant values of $m$ and $n$,  we find
\begin{equation}
X^{rs}_{mn} = \oint{dz\over 2\pi i}{1\over z^{n-1}}
\oint{dw\over 2\pi i}{1\over w^{m+2} } \, {(f'_s (z))^2\over f'_r(w) } {1\over
f_s(z) - f_r(w)}{\prod_{I=1}^3 (f_r(w) -f_I(0)) \over \prod_{J=1}^3 (f_s(z) -
f_J(0)}\,.
\end{equation}
This is the general result for the ghost Neumann coefficients.
Again, for any vertex they are easily calculated by power series
expansions and picking up residues.  For particular vertices one
can simplify somewhat the above expressions and find interesting
relations.  In fact, a fair amount of work can be done for the OSFT
vertex in simplifying the above results.  One can show that the
matrices $N$ and $X$ are related, and while no closed form expressions
are known for the coefficients, they can be generated quite
efficiently from the simpler expressions.

Since any specific Neumann coefficient can
be calculated exactly with
a finite number of operations, the  exact computation
of $\langle A, B, C\rangle$
for any three Fock space states $A$, $B$, and $C$  requires a finite number
of operations, as well.


\section{SFT action: oscillator approach}
\label{sec:overlaps}

In this section, we give a more detailed discussion of Witten's open
bosonic string field theory from the oscillator point of view.
The main goal of this section is to explicitly formulate the OSFT
action in the string Fock space, where the action
(\ref{eq:SFT-action}) takes the form
\begin{equation}
S = -\frac{1}{2}\langle V_2 | \Psi, Q \Psi \rangle
  -\frac{g}{3}   \langle V_3 | \Psi, \Psi, \Psi \rangle\,.
\label{eq:action-Fock-1}
\end{equation}
In this expression, $\langle V_2 |$ and $\langle V_3 |$ are elements
of the two-fold and three-fold product of the dual Fock space $({\mathcal
         H}^*)^2$ and  $({\mathcal  H}^*)^3$, respectively.  These
objects defined
in terms of the string Fock space give a rigorous definition to the
abstract action (\ref{eq:SFT-action}) through the replacement
\begin{eqnarray*}
\langle V_2 | A, B \rangle &  \rightarrow &  \int A\star B\\
\langle V_3 | A, B, C \rangle & \rightarrow &  \int A\star B\star C \,.
\end{eqnarray*}

Subsection \ref{sec:warmup} is a warmup, in which we review some basic
features of the simple harmonic oscillator and discuss squeezed
states.  In subsection \ref{sec:split} we relate modes on the full
string to modes on half strings, giving formulae needed to compute the
three-string vertex.
In subsection \ref{sec:v2} we
derive the two-string vertex in oscillator form, and in subsection
\ref{sec:v3} we give an explicit formula for the three-string vertex.
In subsection \ref{sec:calculating} we put these pieces together and
discuss the calculation of the full SFT action.


\subsection{Squeezed states and the simple harmonic oscillator}
\label{sec:warmup}

Let us consider a simple harmonic oscillator with annihilation
operator
\begin{equation}
a = -i \left( \sqrt{\frac{\alpha}{ 2}}x + \frac{1}{ \sqrt{2 \alpha}}
             \partial_x   \right)
\end{equation}
where $\alpha$ is an arbitrary constant.  The oscillator ground state
is associated with the wavefunction
\begin{equation}
| 0 \rangle \rightarrow
       \left( \frac{\alpha}{ \pi}  \right)^{1/4} e^{-\alpha x^2/2}\,.
\end{equation}
In the harmonic oscillator basis $|n \rangle$, the Dirac position basis
states $| x \rangle$ have a squeezed state form
\begin{equation}
| x \rangle = \left( \frac{\alpha}{ \pi}  \right)^{1/4}
\exp\left(-\frac{\alpha}{ 2}x^2
  -i \sqrt{2 \alpha} a^{\dagger} x +\frac{1}{2} (a^{\dagger})^2 \right)
| 0 \rangle\,.
\end{equation}

A general wavefunction is associated with a state through the
correspondence
\begin{equation}
f (x) \rightarrow \int_{-\infty}^\infty dx \;f (x) | x \rangle \,.
\label{eq:function-state}
\end{equation}
In particular, we have
\begin{eqnarray}
\delta (x) & \rightarrow &
\left( \frac{\alpha}{ \pi}  \right)^{1/4}
\exp\left(\frac{1}{2} (a^{\dagger})^2 \right)
| 0 \rangle \,,\label{eq:squeezed-d1}
\\
1 & \rightarrow &  \int dx \; | x \rangle =
\left( \frac{4\pi}{ \alpha}  \right)^{1/4}
\exp\left(-\frac{1}{2} (a^{\dagger})^2 \right)
| 0 \rangle\,.\nonumber
\end{eqnarray}
This shows that the delta and constant functions both have squeezed
state representations in terms of the harmonic oscillator basis.
The norm of a squeezed state
\begin{equation}
| s \rangle =\exp\left(\frac{1}{2} s (a^{\dagger})^2 \right)
| 0 \rangle
\end{equation}
is given by
\begin{equation}
\langle s| s \rangle =
\frac{1}{ \sqrt{1-s^2}}\,.
\end{equation}
The states (\ref{eq:squeezed-d1}) are non-normalizable,   but since they
have $s=\pm 1$, they are right on the border of normalizability.
The states
(\ref{eq:squeezed-d1}) can be used to calculate, just like we do with
the Dirac basis states $| x\rangle$, which lie outside
the single-particle Hilbert space.

It will be useful for us to generalize the foregoing considerations in
several ways.  A particularly simple generalization arises when we
consider a pair of degrees of freedom $x, y$ described by a two-harmonic
oscillator Fock space basis.  In such a basis, repeating the preceding
analysis leads us to a function-state correspondence for the delta
functions relating $x, y$ of the form
\begin{equation}
\delta (x \pm y) \rightarrow
\exp\left(\pm a^{\dagger}_{(x)}a^{\dagger}_{(y)}\right)
\left( | 0 \rangle_x \otimes | 0 \rangle_y \right)\,.
\label{eq:squeezed-two}
\end{equation}
Note that this result is independent of $\alpha$; like the $\delta$
function, the resulting state is again non-normalizable.
we will find these squeezed state expressions very useful in
describing the two- and three-string vertices of Witten's open string
field theory.  It is worth pointing out here that there are several
ways of deriving (\ref{eq:squeezed-two}).  The most straightforward
way is to carry out a two-dimensional Gaussian integral analogous to
(\ref{eq:squeezed-d1}).  We can also derive (\ref{eq:squeezed-two})
indirectly, however (at least up to an overall constant) from the
following argument.  From the general result that delta functions give
squeezed states, we expect that up to an overall constant
\begin{eqnarray}
\lefteqn{\delta (x \pm y) \rightarrow} \\
    & | D_{\pm} \rangle = &
\exp\left(\pm\frac{1}{2}\left[
A a^{\dagger}_{(x)}a^{\dagger}_{(x)} +
2 B a^{\dagger}_{(x)}a^{\dagger}_{(y)}+
C a^{\dagger}_{(y)}a^{\dagger}_{(y)} \right]
\right)
\left( | 0 \rangle_x \otimes | 0 \rangle_y \right)\,. \nonumber
\end{eqnarray}
The state associated with the delta function must satisfy the
constraints
\begin{eqnarray}
(x \pm y) | D_{\pm} \rangle & = &  0\\ \label{eq:delta-constraints}
(p_x \mp p_y)  | D_{\pm} \rangle& = &  0 \,.  \nonumber
\end{eqnarray}
Rewriting $x, p_x$ in terms of $a_{(x)}, a^{\dagger}_{(x)}$ and
similarly for $y, p_y$,
these conditions impose the constraints
\begin{eqnarray}
\left[(A \pm B -1) a_{(x)}^{\dagger}+( B \pm C \pm 1)
             a_{( y)}^{\dagger} \right] | D_{\pm} \rangle& = &  0\\
\left[( A \mp B + 1) a_{(x)}^{\dagger}+( B \mp C \mp 1)
            a_{(y)}^{\dagger} \right]
| D_{\pm} \rangle & = &  0 \,, \nonumber
\end{eqnarray}
from which it follows that $A = C = 0$ and $B = \pm 1$, reproducing
(\ref{eq:squeezed-two}) up to an overall constant.  We will use this
indirect method, following Gross and Jevicki, to derive the
three-string vertex in subsection \ref{sec:v3}.

\subsection{Half-string modes}
\label{sec:split}

For many computations it is useful to think of the string as being
``split'' into a left half and a right half.  Formally, the string
field can be expressed as a functional $\Psi[L, R]$, where $L, R$ describe
the left and right parts of the string.  This is a very appealing
idea, since it leads to a simple picture of the  star product in terms of
matrix multiplication
\begin{equation}
(\Psi \star \Phi)[L, R] = \int{\mathcal D} A \; \Psi[L, A] \Phi[A, R]  \,.
\end{equation}
While there has been quite a bit of work aimed at making this ``split
string'' formalism precise~\cite{split,Gross-Taylor-I,Gross-Taylor-II},
the technical details in this approach become quite complicated when one
attempts to precisely deal with the string midpoint where the left and right
parts of the string attach.  In particular, the BRST operator $Q_B$ becomes
rather awkward in this formulation.

Nonetheless, some of the structure of the star product encoded in the
three-string vertex is easiest to understand using the half-string
formalism, and many formulae related to the 3-string vertex are most
easily expressed in terms of the linear map from full-string modes
to half-string modes.  In this subsection we discuss this  linear map,
encoded in a matrix $X$,
which we use in subsection \ref{sec:v3} to give an explicit formulae
for the three-string vertex.

Recall that the
matter fields are expanded in modes through
\begin{equation}
x (\sigma) = x_0 + \sqrt{2} \sum_{n = 1}^{ \infty}  x_n \cos n \sigma\,.
            \label{eq:mode-decomposition-2}
\end{equation}
(We suppress Lorentz indices in most of this section for clarity.)
We are interested in considering an analogous expansion of the left
and right halves of the string.  We expand in odd modes with Neumann
boundary conditions at the ends of the string, and Dirichlet boundary
conditions at the string midpoint:
\begin{eqnarray}
l (\sigma) & = &  x (\sigma) =
\sqrt{2} \sum_{k = 0}^{ \infty} l_{2k + 1}
\cos (2k + 1) \sigma,\quad \sigma < \pi/2 \\
r (\sigma) & = &  x (\pi -\sigma) =
\sqrt{2} \sum_{k = 0}^{ \infty} r_{2k + 1}
\cos (2k + 1) \sigma,\quad  \sigma < \pi/2  \,.  \nonumber
\end{eqnarray}
Note that there are subtleties associated with the midpoint in this
expansion.  For example, while we have taken $l (\pi/2)$ to formally
vanish, by choosing coefficients like $l_{2k + 1} = (-1)^k2 \sqrt{2}
a/(2k + 1) \pi$ we have  $l (\sigma) = a, \forall \sigma < \pi/2$, so
$\lim_{\sigma \rightarrow \pi/2_-}= a$.  These subtleties become
important when dealing with the full theory, but are not important in
the calculation we carry out below of the three-string vertex.

Let us define the quantities
\begin{eqnarray}
X_{2k+1,2n}=X_{2n,2k+1} & = & { 4(-1)^{k+n}(2k+1)\over
\pi\left({(2k+1)^2-4n^2}\right)} \;\;\;  \quad  (n \neq 0)\, , \label{eq:X}\\
\quad X_{ 2k+1,0} =X_{0,2k+1} & = &
{  2 \sqrt{2}(-1)^{k}\over \pi{(2k+1)}}\, .\nonumber
\end{eqnarray}
The matrix
\begin{equation}
X =
\begin{pmatrix} 
0 & X_{2k+1,2n} \\
                    X_{2n, 2k+1} & 0
\end{pmatrix}
\equiv \left(\begin{array}{cc}
0 & X_{oe}\\
X_{eo} & 0
\end{array}\right) \,
\label{eq:matX}
\end{equation}
where $e, o$ refer to the set of even and odd indices respectively,
is manifestly symmetric and turns out to be orthogonal: \ \  $X=X^T=X^{-1}$.
Performing a Fourier decomposition we can relate the full-string and
half-string modes through
\begin{eqnarray}
                   x_{2n+1} &= &\half\left(l_{2n+1}-r_{2n+1}\right)  \, ,
\label{eq:xrel}\\
x_{2n}
             & = &  \half
\sum_{k=0}^{\infty}X_{2n,2k+1}\left(l_{2k+1}+r_{2k+1}\right)\,,
\nonumber
\end{eqnarray}
We can invert (\ref{eq:xrel}) to derive
\begin{eqnarray}
l_{2k+1}   &=
&x_{2k+1}+\sum_{n=0}^{\infty}X_{2k+1,2n}
x_{2n}  \, ,
\label{eq:half-full}\\
r_{2k+1} & = &
  -x_{2k+1}+\sum_{n=0}^{\infty}X_{2k+1,2n}
x_{2n}  \, .
\nonumber
\end{eqnarray}
One must be careful with the order of summation
in sequences of coefficients which do not go to zero
faster than $1/n$:  different orders of summation
may give different results.  Fortunately, such associativity
anomalies are not relevant for the calculations we do here.


\subsection{The two-string vertex $\langle V_2 |$}
\label{sec:v2}

We can immediately apply the oscillator formulae from subsection
\ref{sec:warmup} to calculate the two-string vertex.  Using the mode
decomposition (\ref{eq:mode-decomposition-2}), we associate the string
field functional $\Psi[x (\sigma)]$ with a function $\Psi (\{x_n\})$
of the infinite family of string oscillator mode amplitudes.  The
overlap integral combining (\ref{eq:integral-p}) and (\ref{eq:mult})
can then be expressed in modes as
\begin{equation}
\int \Psi \star \Phi = \int \prod_{n = 0}^{ \infty}  dx_ndy_n
\; \delta (x_n-(-1)^ny_n) \Psi (\{x_n\}) \Phi (\{y_n\})\,.
\end{equation}
Geometrically this just encodes the overlap condition $x (\sigma) = y
(\pi -\sigma)$ described through

\begin{center}
\centering
\begin{picture}(100,40)(- 50,- 20)
\put( 20,2){\vector( -1,0){40}}
\put( -20,-2){\vector( 1,0){40}}
\put(0,-10){\makebox(0,0){$\Psi$}}
\put(0, 10){\makebox(0,0){$\Phi$}}
\end{picture}
\end{center}
It follows from (\ref{eq:squeezed-two}) that we can write the
two-string vertex as the squeezed state
\begin{equation}
\langle V_2 |_{{\rm matter}} =
\left(\langle 0 | \otimes \langle 0 |\right)
\exp\Bigl( \sum_{n, m = 0}^{ \infty}
  -a_n^{(1)}  C_{nm} a_m^{ (2)} \Bigr)\,,
\label{eq:v2o}
\end{equation}
where $C_{nm} = \delta_{nm} (-1)^n$ is an infinite-size matrix
connecting the oscillator modes of the two single-string Fock spaces,
and the sum is taken over all oscillator modes, including zero.  In the
expression (\ref{eq:v2o}), we have used the formalism in which $| 0
\rangle$ is the vacuum annihilated by $a_0$.  To translate this
expression into a momentum basis, we use only $n, m > 0$, and replace
\begin{equation}
\left(\langle 0 | \otimes \langle 0 |\right)
\exp\left(-a^{(1)}_0a^{(2)}_0\right) \rightarrow
\int d^{26} p
\left( \langle 0; p | \otimes \langle 0; -p | \right)\,.
\end{equation}

The extension of this analysis to ghosts is straightforward.  For the
ghost and antighost respectively, the overlap conditions corresponding
with
$x_1 (\sigma) = x_2 (\pi -\sigma)$ are ~\cite{Gross-Jevicki-12}
$c_1 (\sigma) = -c_2 (\pi -\sigma)$ and
$b_1 (\sigma) = b_2 (\pi -\sigma)$.  This leads to the overall formula
for the two-string vertex
\begin{eqnarray}
\langle V_2 |  & = & \int d^{26} p
\left( \langle 0; p | \otimes \langle 0; -p | \right)
(c^{(1)}_0 + c^{(2)}_0) \label{eq:v2}\\
            &  & \hspace{0.3in} \times
\exp\left(
  -\sum_{n = 1}^{ \infty} (-1)^n
[a^{(1)}_n a^{(2)}_n+c^{(1)}_n b^{(2)}_n+c^{(2)}_n b^{(1)}_n] \right)\,.
\nonumber
\end{eqnarray}
This expression for the two-string vertex can also be derived directly
from the conformal field theory approach, computing the two-point
function of an arbitrary pair of states on the disk.


\subsection{The three-string vertex $| V_3 \rangle$}
\label{sec:v3}

The three-string vertex, which is associated with the three-string
overlap
\begin{center}
\centering
\begin{picture}(100,60)(- 50,- 30)
\put( 20,14){\line( -2,-1){20}}
\put(0, 4){\vector( -2,1){20}}
\put(-20,9){\line( 2, -1){18}}
\put( -2, 0 ){\vector( 0, -1){ 20}}
\put( 2,-20){\line( 0, 1){ 20}}
\put(2,0){\vector( 2,1){18}}
\put(-15,-10){\makebox(0,0){$\Psi_2$}}
\put(0, 15){\makebox(0,0){$\Psi_1$}}
\put(15, -10){\makebox(0,0){$\Psi_3$}}
\end{picture}
\end{center}
can be computed in a very similar fashion to the two-string vertex
above.  The details of the calculation, however, are significantly
more complicated.  In this subsection we follow the original approach
of Gross and Jevicki~\cite{Gross-Jevicki-12}; similar approaches were
taken by other authors~\cite{cst,Samuel}.  The method used by Gross and
Jevicki is essentially the method used in (\ref{eq:squeezed-two}) to
write a delta function of two variables in oscillator form by imposing
the constraints (\ref{eq:delta-constraints}) on a general squeezed
state.  The 3-string vertex can also be computed by explicitly
performing~\cite{Ohta,Shelton} the relevant Gaussian
integrals.\footnote{Another approach to the cubic vertex has been
explored extensively.  By diagonalizing the Neumann matrices, the star
product takes the form of a continuous Moyal 
product~\cite{Bars-original,Moyal}.  This simplifies the vertex but
complicates the propagator.  For a recent discussion of this work,
applications of this approach, and
further references, see the review of Bars~\cite{Bars}.}

{}From the general structure of the overlap conditions it is clear that,
like the two-string vertex, the
three-string vertex takes the  form of a squeezed state:
\begin{eqnarray}
\lefteqn{ | V_3 \rangle = \kappa\int d^{26} p^{(1)}d^{26} p^{(2)}d^{26}
             p^{(3)}} \nonumber \\
& &
\times\exp\Bigl(
  -\frac{1}{2}\sum_{r, s = 1}^{3}
[a^{(r)}_m V^{rs}_{mn} a^{(s)}_n+
2 a^{(r)}_m V^{rs}_{m0} p^{(s)}+
p^{(r)} V^{rs}_{00}  p^{(s)}+
c^{(r)}_m X^{rs}_{mn} b^{(s)}_n] \Bigr)\,, \nonumber\\
& &
\times
\delta (p^{(1)} + p^{(2)} + p^{(3)}) c_0^{(1)}c_0^{(2)}c_0^{(3)}
\left( | 0; p^{(1)} \rangle \otimes
| 0; p^{(2)} \rangle \otimes | 0; p^{(3)} \rangle \right)
\label{eq:v3}
\end{eqnarray}
where $\kappa ={\mathcal N} = K^3 = 3^{9/2}/2^6$, and where the Neumann
coefficients $V^{rs}_{mn}$ and  $X^{rs}_{mn}$ are constants.  Writing the
momentum basis states in oscillator form
\begin{equation}
| p \rangle = \frac{1}{ ( \pi)^{1/4}}
\exp \left[-\frac{1}{2}p^2
+ \sqrt{2}a_0^{\dagger} p -\frac{1}{2}(a^{\dagger}_0)^2 \right]
| 0 \rangle,
\label{eq:p-basis}
\end{equation}
we can write matter part of the 3-string vertex as
\begin{equation}
\label{eq:vertex-0}
| V_3 \rangle   =
\left( \frac{2 \; \pi^{1/4}}{  \sqrt{3} (1 + V_{00})}  \right)^{26}
\exp \Bigl(-\frac{1}{2}\sum_{r, s \leq 3}
\;  \sum_{m, n \geq 0} V'^{rs}_{m n} (a^{(r)\dagger}_m \cdot
a^{(s) \dagger}_n)\Bigr)
\left( | 0 \rangle \otimes | 0 \rangle \otimes
| 0  \rangle \right) \,,
\end{equation}
where $V_{00} = V^{rr}_{00}$ and
\begin{eqnarray}
V'^{rs}_{mn} & = &  V^{rs}_{mn}
  -\frac{ 1}{1 + V_{00}} \sum_{t}V^{rt}_{m0}V^{ts}_{0n}  \nonumber\\
V'^{rs}_{m0} & = &  V'^{sr}_{0m} =
\frac{\sqrt{2}}{1 + V_{00}}
V^{rs}_{m0}   \label{eq:v-relations}\\
V'^{rs}_{00} & = &
\frac{2}{3 (1 + V_{00})}
+ \delta^{rs} \left( 1-2/(1 + V_{00}) \right)\,.
\nonumber
\end{eqnarray}

We now want to determine the coefficients $V'^{rs}_{mn}$ by using
overlap conditions analogous to (\ref{eq:delta-constraints}).  It is
convenient to use a ${\bf Z}_3$ Fourier decomposition of the three
string modes $x^{(i)}(\sigma)$
\begin{eqnarray}
Q & = &  \frac{1}{ \sqrt{3}} (x^{(1)} + \omega x^{(2)} + \omega^2
x^{(3)})
\label{eq:Fourier-q}\\
Q^{(3)} & = &  \frac{1}{ \sqrt{3}} (x^{(1)} +  x^{(2)} +
            x^{(3)})   \nonumber
\end{eqnarray}
where $\omega = \exp\left(2 \pi i/3\right)$.
The definitions (\ref{eq:Fourier-q}) can be used to define $Q
(\sigma)$ in terms of $x^{(i)}(\sigma)$ as well as to define $Q_n$ in
terms of $x^{(i)}_n$ (and similarly for  $Q^{(3)}$); henceforth by $Q$
we denote the collection of full-string modes $Q_n$ (and similarly for
$Q^{(3)}$).
We can relate the
full-string modes $Q, Q^{(3)}$ to half-string modes $L, R, L^{(3)},
R^{(3)}$ through the equations (\ref{eq:half-full}).
In terms of these
variables, the overlap conditions are
\begin{eqnarray}
L-\omega R & = &  0,\\
L^{(3)} -R^{(3)}& = &  0 \,. \nonumber
\end{eqnarray}
In terms of the even and odd full-string modes (which, using the same
notation as in section \ref{sec:split}, we denote by $Q_{e, o}$)
these conditions are
expressed as
\begin{equation}
Q_o -i \sqrt{3} X_{oe} Q_e = 0\,,
\label{eq:condition1}
\end{equation}
and
\begin{equation}
Q^{(3)}_o = 0 \,.
\label{eq:condition2}
\end{equation}
Multiplying by $X_{eo}$, (\ref{eq:condition1}) can be rewritten as
\begin{equation}
\frac{i}{ \sqrt{3}}  X_{eo} Q_o + Q_e = 0 \,.
\end{equation}
This can be combined with (\ref{eq:condition1}) and written in the
simpler form
\begin{equation}
(1-Y) Q = 0,
\label{eq:o1}
\end{equation}
where
\begin{equation}
Y = -\frac{1}{2}C + \frac{\sqrt{3}}{2}  X \,.
\end{equation}
Note that $Y^2 = 1$.
Similarly, we can write (\ref{eq:condition2})
as
\begin{equation}
(1-C) Q^{(3)} = 0 \,.
\label{eq:o2}
\end{equation}
Equations (\ref{eq:o1}) and (\ref{eq:o2}) are the essential overlap
equations satisfied by the three-string vertex.  Writing the
three-string vertex as a squeezed state in terms of oscillators $A,
A^{(3)}$ related to the string oscillators $a^{(i)}$ through the
analogue of (\ref{eq:Fourier-q}), we then have
\begin{equation}
            | V_3 \rangle \sim\exp\left( -A^{\dagger} U \bar{A}^{\dagger}
              -\frac{1}{2}(A^{(3)})^{\dagger} C(A^{(3)})^{\dagger}\right)\,,
\label{eq:vu}
\end{equation}
where $U$ satisfies the overlap constraint
(\ref{eq:o1}).
Recall that the string modes are proportional to
\begin{equation}
x \sim E (a-a^{\dagger})
\end{equation}
where
\begin{equation}
E_{mn} = \delta_{mn} \frac{1}{ \sqrt{m}}, \quad m \neq 0, \;\;\;\;\;
E_{00} = 1/\sqrt{2}  \,.
\end{equation}
Thus, from (\ref{eq:o1}) and (\ref{eq:vu}) we see that $U$ must
satisfy the overlap constraint
\begin{equation}
(1-Y) E (1 + U) = 0 \,.
\label{eq:oo1}
\end{equation}
As we discussed in the last part of subsection \ref{sec:warmup},
associated with this constraint there is an analogous constraint on
the derivatives in the perpendicular direction.  Since $Y^2 = 1$, we
have
\begin{eqnarray}
(1 + Y) (1-Y) & = & (1-Y) (1 + Y) = 0 \,.
\end{eqnarray}
Since derivatives with respect to the $x$ modes go as
\begin{equation}
\partial \sim E^{-1} (a + a^{\dagger}),
\end{equation}
we have the additional overlap constraint on $U$
\begin{equation}
            (1 + Y)E^{-1} (1-U) = 0 \,.
\label{eq:oo2}
\end{equation}
Equations (\ref{eq:oo1}) and (\ref{eq:oo2}) determine $U$ completely,
giving
\begin{equation}
U = (2-EYE^{-1} + E^{-1} YE)[EYE^{-1} + E^{-1} YE]^{-1}.
\label{eq:implicit}
\end{equation}
Unfortunately, the matrix combination in brackets is difficult to
explicitly invert.  This does, however, give a closed form expression
for the three-string vertex (\ref{eq:vertex-0}), where
\begin{eqnarray}
V'^{rr} & = & \frac{1}{3} (C + U + \bar{U})\\
V'^{r, r \pm 1} & = & \frac{1}{6} ( 2C -U-\bar{U}) \pm \frac{i
             \sqrt{3}}{6}  (U- \bar{U}) \,.\nonumber
\end{eqnarray}

While (\ref{eq:implicit}) is difficult to directly compute, given a
formula for $U$ one can check that the formula is correct by checking
the overlap conditions (\ref{eq:oo1}) and (\ref{eq:oo2}).
        Expressions for $V$ and $X$ and hence for $U$
and $V'$ were computed~\cite{Gross-Jevicki-12} by essentially the method used
in the previous section.  Their results for $V$ and $X$ are given as
follows\footnote{Note that in some references, signs and various
factors in $\kappa$ and the Neumann coefficients may be slightly
different.  In some papers, the cubic term in the action is taken to
have an overall factor of $g/6$ instead of $g/3$; this choice of
normalization gives a 3-tachyon amplitude of $g$ instead of $2g$, and
gives a different value for $\kappa$.  Often, the sign in the
exponential of (\ref{eq:v3}) is taken to be positive, which changes
the signs of the coefficients $V^{rs}_{nm}, X^{rs}_{nm}$.  When the
matter Neumann coefficients are defined with respect to the oscillator
modes $\alpha_n$ rather than $a_n$, the matter Neumann coefficients
$V^{rs}_{nm}, V^{rs}_{n0}$ must be divided by $\sqrt{nm}$ and
$\sqrt{n}$.  This is the case for the coefficients $N^{rs}_{nm}$
computed in (\ref{neumannmatter}), which are related to the $V$'s
through $N^{rs}_{nm}= V^{rs}_{nm}/\sqrt{nm}$.  Finally, when $\alpha'$
is taken to be $1/2$, an extra factor of $1/\sqrt{2}$ appears for each
$0$ subscript in the matter Neumann coefficients.  }.  Define $A_n,
B_n$ for $n \geq 0$ through
\begin{eqnarray}
\left( \frac{1 + ix}{1-ix} \right)^{1/3}  & = &
\sum_{n\, {\rm even}} A_n x^n + i
\sum_{m\, {\rm odd}} A_m x^m  \label{eq:ab}\\
\left( \frac{1 + ix}{1-ix} \right)^{2/3}  & = &
\sum_{n\, {\rm even}} B_n x^n + i
\sum_{m\, {\rm odd}} B_m x^m \,.  \nonumber
\end{eqnarray}
These coefficients can be used to define 6-string Neumann coefficients
             $N^{r, \pm s}_{nm}$ through
\begin{eqnarray}
N^{r, \pm r}_{nm} & = &
\left\{\begin{array}{l}
\frac{1}{3 (n \pm m)}  (-1)^n (A_nB_m \pm B_nA_m), \;\;\;\;\;
m + n\, {\rm even}, \;m \neq n\\
0, \;\;\;\;\; m + n\, {\rm odd}
\end{array} \right.\label{eq:n6}\\
N^{r, \pm (r + \sigma)}_{nm} & = &
\left\{\begin{array}{l}
\frac{1}{6 (n \pm \sigma m)}  (-1)^{n + 1} (A_nB_m \pm \sigma B_nA_m),
\;\;\;\;\;
m + n\, {\rm even}, \;m \neq n\\
\sigma
\frac{\sqrt{3}}{6 (n \pm \sigma m)} (A_nB_m \mp \sigma B_nA_m), \;\;\;\;\;
m + n\, {\rm odd}
\end{array}\right].\nonumber
\end{eqnarray}
where in $N^{r, \pm (r + \sigma)}$, $\sigma = \pm 1$, and $r +\sigma$
is taken modulo 3 to be between 1 and 3.  The 3-string matter Neumann
coefficients $V^{rs}_{nm}$ are then given by
\begin{eqnarray}
V^{rs}_{nm} & = &  -\sqrt{mn} (N^{r, s}_{nm} + N^{r, -s}_{nm}),
\;\;\;\;\; m \neq n,\, {\rm and}\, m, n \neq 0 \nonumber\\
V^{rr}_{nn} & = &  -\frac{1}{3}  \left[
2 \sum_{k = 0}^{n}  (-1)^{n-k} A_k^2-(-1)^n-A_n^2 \right], \;\;\;\;\;
n \neq 0 \nonumber\\
V^{r, r + \sigma}_{nn} & = &\frac{1}{2} \left[ (-1)^n-V^{rr}_{nn}
             \right], \;\;\;\;\;  n \neq 0 \label{eq:n3}\\
V^{rs}_{0n}& = & -\sqrt{2n} \left( N^{r, s}_{0n} + N^{r, -s}_{0n}
\right), \;\;\;\;\; n \neq 0\nonumber\\
V^{rr}_{00} & = & \ln (27/16) \nonumber
\end{eqnarray}
The ghost Neumann coefficients $X^{rs}_{m n}, m\geq 0, n >0$ are
given by
\begin{eqnarray}
X^{rr}_{mn} & = &  \left( -N^{r, r}_{nm} + N^{ r, -r}_{nm} \right),
            \;\;\;\;\; n \neq
            m\nonumber\\
X^{r (r \pm 1)}_{mn} & = &  m \left(\pm N^{r, r \mp 1}_{nm} \mp N^{ r, - (r
            \mp 1)}_{nm} \right), \;\;\;\;\; n \neq
            m \label{eq:x3}\\
X^{rr}_{nn} & = &  \frac{1}{3}
\left[ -(-1)^n-A_n^2 + 2 \sum_{k = 0}^{n}  (-1)^{n-k} A_k^2 -2
            (-1)^nA_nB_n \right] \nonumber\\
X^{r (r \pm 1)}_{nn} & = &
  -\frac{1}{2}(-1)^n -\frac{1}{2} X^{rr}_{nn}
\nonumber
\end{eqnarray}

These expressions for the matter and ghost
Neumann coefficients were computed by
Gross and Jevicki~\cite{Gross-Jevicki-12}, and include minor corrections
published later~\cite{RSZ-2}.  It was shown that the resulting
matter matrices $U$ indeed satisfy the overlap conditions
(\ref{eq:oo1}) and (\ref{eq:oo2}).  This shows that the conformal
field theory method and the oscillator method give the same results
for the matter part of the three-string vertex.  The same is true for
the ghost part of the vertex, although we will not go into the details
of this discussion here.

Before leaving the three-string vertex, it is worth noting that
the Neumann coefficients have a number of simple symmetries.  There is
a cyclic symmetry under $r \rightarrow r + 1, s \rightarrow s + 1$,
which corresponds to the obvious geometric symmetry of rotating the
vertex.  The coefficients are also symmetric under the exchange $r
\leftrightarrow s, n \leftrightarrow m$.  Finally, there is a
twist symmetry which, as discussed in section
\ref{sec:algebraic-structure}, is associated with reflection of the
strings
\begin{eqnarray}
V^{rs}_{nm} & = &  (-1)^{n + m}V^{sr}_{nm}\\
X^{rs}_{nm} & = &  (-1)^{n + m}X^{sr}_{nm}\,. \nonumber
\end{eqnarray}
This symmetry follows from the fact that half-strings carrying odd
modes pick up a minus sign under reflection.  Since each string
carrying an odd mode gets two changes of sign, from the two ends of
the string, it is straightforward to see that this symmetry guarantees
that the three-vertex is invariant under reflection, and therefore
satisfies condition (\ref{l1e59}).


\subsection{Calculating the SFT action}
\label{sec:calculating}

Given the action
\begin{equation}
S = -\frac{1}{2}\langle V_2 | \Psi, Q \Psi \rangle
  -\frac{g}{3}   \langle V_3 | \Psi, \Psi, \Psi \rangle\,,
\label{eq:action-Fock}
\end{equation}
and the explicit formulae
(\ref{eq:v2}, \ref{eq:v3}) for the two- and three-string vertices, we
can in principle calculate the string field action term by term for
each of the fields in the string field expansion
\begin{eqnarray}
\Psi  & = & \int d^{26}p \;
\left[ \phi (p)\; | 0_1; p \rangle + A_\mu (p) \; \alpha^\mu_{-1} | 0_1; p
\rangle + \chi ( p) b_{-1} c_0| 0_1; p \rangle  \right. \nonumber\\
            &  & \hspace*{0.9in} \left.
+ B_{\mu \nu} ( p)
\alpha^\mu_{-1} \alpha^\nu_{-1} | 0_1; p \rangle + \cdots
\right]\,. \label{eq:field-expansion-2}
\end{eqnarray}

Since the resulting action has an enormous gauge invariance given by
(\ref{eq:SFT-gauge}), it is often helpful to fix the gauge before
computing the action.  A particularly useful gauge choice is the
Feynman-Siegel gauge
\begin{equation}
b_0 | \Psi \rangle = 0\,.
\label{eq:FS-gauge}
\end{equation}
This is a good gauge choice locally, fixing the linear gauge
transformations $\delta | \Psi \rangle = Q | \Lambda \rangle$.  This
gauge choice is not, however, globally valid; we will return to this
point in subsection \ref{sec:gauge}.  In this gauge, all fields
in the string field expansion which are associated with states that have
an antighost zero-mode $c_0$ are taken to vanish.  For example, the
field $\chi (p)$ in (\ref{eq:field-expansion-2}) vanishes.  In
Feynman-Siegel gauge, the BRST operator takes the simple form
\begin{equation}
Q = c_0L_0 =c_0 (N + p^2 -1)
\label{eq:FS-BRST}
\end{equation}
where $N$ is the total (matter + ghost) oscillator number.

Using (\ref{eq:FS-BRST}), it is straightforward to write the quadratic
terms in the string field action.  They are
\begin{equation}
\frac{1}{2}\langle V_2 | \Psi, Q \Psi \rangle =
\int d^{26} p \; \left\{
\phi (-p) \left[ \frac{p^2 -1}{2}  \right]\phi (p)
+ A_\mu (-p) \left[ \frac{p^2}{2}  \right]A^\mu (p) + \cdots
\right\}\,.
\end{equation}

The cubic part of the action can also be computed term by term,
although the terms are somewhat more complicated.  The leading terms
in the cubic action  are
given by
\begin{eqnarray}
\lefteqn{\frac{1}{3}
\langle V_3 | \Psi, \Psi, \Psi \rangle =
\int d^{26}pd^{26}q \; \frac{\kappa g}{3}  \;
            e^{(\ln 16/27) (p^2 + q^2 + p \cdot q)}
}\label{eq:expanded-action} \\
& &\hspace{1.0in} \times
\left\{\phi (-p) \phi (-q) \phi (p + q) + \frac{16}{9}
A^{\mu} (-p) A_\mu (-q) \phi (p + q)
            \right.
\nonumber
\\
            & & \hspace{1.1in}
\left.
  - \frac{8}{9}
(p^\mu +2q^\mu) (2p^{\nu} + q^{\nu})A^{\mu} (-p) A_\nu (-q) \phi (p +
q)  + \cdots
\right\}\,. \nonumber
\end{eqnarray}
In computing the $\phi^3$ term we have used
\begin{equation}
V^{rs}_{00} = \delta^{rs}
\ln (\frac{27}{16} )\,.
\end{equation}
The $A^2 \phi$ term uses
\begin{equation}
V^{rs}_{11} = -\frac{16}{27} , \; \;r \neq s,
\end{equation}
while the $ (A \cdot p)^2 \phi$ term uses
\begin{equation}
V^{12}_{10} = -V^{13}_{10} = -\frac{2 \sqrt{2}}{3 \sqrt{3}}\,.
\end{equation}
The most striking feature of this action is that for a generic set of
three fields, there is a {\it nonlocal} cubic interaction term that
contains an exponential of a quadratic form in the momenta.  This
means that the target space formulation of string theory has a
dramatically different character from a standard quantum field theory.
{}From the point of view of quantum field theory, string field theory
seems to contain an infinite number of nonrenormalizable interactions.
Just like the simpler case of noncommutative field theories, however,
the magic of string theory seems to combine this infinite set of
interactions into a sensible model.
It has been shown that all on-shell amplitudes computed from the
string field theory action we have discussed here precisely reproduce
the amplitudes given by the usual conformal field theory approach,
including the correct measure on moduli
space~\cite{Giddings-Martinec,gmw,Zwiebach-proof}.
Note, though, that the
bosonic open theory becomes problematic at the quantum
level because of the closed string tachyon, whose instability
is not yet understood.
For the purposes of these lectures,
we will restrict our attention to the classical bosonic open string action.
Open superstring field theory
should be better behaved since the closed string sector has
no tachyon.   There has been significant progress in understanding
tachyon condensation in superstring field theory~\cite{Aref'eva,Berkovits},
even though superstring field theory is less developed than
bosonic string field theory.


\section{Evidence for the Sen conjectures}
\label{sec:evidence}

In this section we review the evidence from Witten's OSFT for Sen's
conjectures.  Subsection \ref{sec:tension} contains a derivation of
the formula for the tension of a bosonic D-brane.  In subsection
\ref{sec:constraints-symmetries}, a general discussion is given of
symmetries in the string field theory action and resulting constraints
on the set of string fields which take nonzero values in the tachyon
vacuum.  Subsection \ref{sec:vacuum} contains a summary of existing
results for the determination of the stable vacuum in Witten's OSFT
(Sen's first conjecture),
including some results which appeared after these lectures were
originally given in 2001.  In subsection \ref{sec:gauge} we discuss
the Feynman-Siegel gauge choice and its limitations.  Subsection
\ref{sec:solitons} summarizes results on lower-dimensional D-branes as
solitons in OSFT (Sen's second conjecture).  Subsection
\ref{subsec:thebackgroundsosft} discusses the general problem of
finding all open string backgrounds within OSFT.  Sen's third
conjecture is discussed in the following section 8, which is completely
devoted to a discussion of the physics in the stable vacuum (vacuum
string field theory).


\subsection{Tension of bosonic D$p$-branes}
\label{sec:tension}

In this section we learn how to relate the open string
coupling constant of string field theory to the mass of
the D-brane described by the open string field theory.
The material presented in this subsection is an unpublished
result due to
Ashoke Sen~\cite{senunpub}, who cited the result
in the paper~\cite{Sen-universality}  on the subject
of universality.  Subsequently, the closely
related computation for the superstring was explained
in detail~\cite{superstring}.  For an alternative
check of the result, see Appendix~A of the paper by Okawa~\cite{Okawa}.

\smallskip
In general, an open string field theory is formulated
using a BCFT (boundary conformal field theory) which
describes some D-brane (or a configuration of D-branes).
In order to describe a D-brane with finite mass, we
consider a compactification of $p$ spatial coordinates
and wrap a D$p$-brane around
along these dimensions.  The string field theory
associated with this D-brane  is written as before:
\begin{equation} \label{ems340909}
S (\Phi) = \,-\, {1\over g^2}\,\,\bigg[\, {1\over 2} \langle \,\Phi
\,,\, Q\,
\Phi
\rangle + {1\over 3}\langle \,\Phi \,,\, \Phi *
\Phi \rangle \bigg]\,.
\end{equation}
The D-brane in question is perceived by the effective
$(25-p)$-dimensional observer as a point particle.
The BCFT includes a Neumann field $X^0$, a set
of Dirichlet fields $X^i$, with $i= 1, \ldots , 25-p$ and some set
of Neumann fields $X^a$, with $a= 25-p+1, \ldots, 25$ that describe
the internal sector of the BCFT.  The string field theory
effectively describes an infinite collection of
fields $\phi_i(t, x^a)$.   These fields do not depend on $x^1, \ldots
, x^{25-p}$
because the corresponding string coordinates are Dirichlet.
Since the coordinates $x^a$ are compact, the fields $\phi_i(t, x^a)$ can be
expanded in Fourier modes. These are a collection of
degrees of freedom that are just time
dependent.  The string field theory action
then reduces to an integral over
time of a time-dependent Lagrangian density.

We will set up the string field theory in such a way that
all dimensions (including time) are compactified on circles
of unit circumference.  In this case, the mass $M$ of the D$p$-brane
coincides with the tension of the D$p$-brane.   The claim is that
\begin{equation}
\label{massbraneorig}
M = {1\over 2\pi^2 \, g^2}\,.
\end{equation}
In this formula and in the following,  we  set
$\alpha'=1$.   In these units the  string tension is $T_0 = 1/(2\pi)$.
When we consider the string field theory of a D25-brane,
        (\ref{massbraneorig}) gives
\begin{equation}
\label{tnebraneorig}
T_{25} = {1\over 2\pi^2 \, g^2}\,.
\end{equation}

We begin our study by considering some special momentum
states of the BCFT:
\begin{equation}
\label{themomstatesorg}
|k_0\rangle \equiv  e^{ (ik_0 X^0(0))}  |0\rangle \,.
\end{equation}
Moreover, we will normalize these states by declaring
\begin{equation}
\langle k_0 | \, c_{-1} c_0 c_1 \, | k_0'\rangle =  \delta_{k_0, k_0'}\,,
\end{equation}
consistent with the discussion below (\ref{ourashokenorm}).
Since the time direction has been made compact via
$t \sim t+1$, the time component $k_0$ of the momentum
is quantized: $k_0 = 2\pi n$, with $n$ integer, and we can use
a Kronecker delta in the above inner product.

\medskip
We will consider the computation of the brane mass in three steps.

\medskip
\noindent
{\em Step 1}:  We consider time-dependent displacements of the D-brane.
We will write down a string field that describes such a displacement
and evaluate the kinetic term of the string action.  This will make it
clear how we can hope to calculate the brane mass.

Let $X^i$ be one of the Dirichlet directions for the D-brane and
assume that $x^i=0$ is the original position of the brane.  Consider now a
displacement field  $\phi^i(t)$ that is expected to be proportional to a
coordinate displacement $x^i(t)$.
We expand the field $\phi^i(t)$ as:
\begin{equation}
\label{thefieldfordiskpvkakgk}
\phi^i(t)  = \sum_{k_0}  e^{ik_0 t}  \, \phi^i(k_0) \,,
\end{equation}
and we use the Fourier components $\phi^i(k_0)$ to assemble
the corresponding string field:
\begin{equation}
\label{corrstringfieldkysla}
|\Phi\rangle = \sum_{k_0} \, \phi^i(k_0)  \, c_1 \alpha_{-1}^i \,
|k_0\rangle\,.
\end{equation}
As you can see, the string  field is built using states of the massless scalar
field that represents translations of the D-brane.
For this string field, the kinetic term $S_2 (\Phi)$ of the string action is
given by
\begin{equation}
\label{fskuyuluay}
S_2(\Phi) = -{1\over g^2} \sum_{k_0, k_0'}  \phi^i(k_0) \phi^i(k_0')
\langle -k_0'| c_{-1} \alpha_1^i \, c_0 L_0 \, c_1 \alpha_{-1}^i |k_0\rangle\,.
\end{equation}
Since $L_0= p^2 + \ldots$~ where the terms indicated by dots vanish
in the present case, $L_0 =
  -k_0^2$ in (\ref{fskuyuluay}) and
\begin{equation}
\label{fskuyuluay89uf}
S_2(\Phi) = {1\over 2g^2} \sum_{k_0}  \phi^i(-k_0) \, k_0^2 \phi^i(k_0)\,.
\end{equation}
Let us now rewrite this string action in terms of the field $\phi^i(t)$
introduced in (\ref{thefieldfordiskpvkakgk}).  A short computation
gives
\begin{equation}
\int_0^1 dt \,  \partial_t \phi^i \, \partial_t \phi^i = \sum_{k_0}
\phi^i(-k_0) \,
k_0^2 \phi^i(k_0)\,.
\end{equation}
Comparing with (\ref{fskuyuluay89uf}) we find that
\begin{equation}
S_2(\Phi)  = {1\over 2 g^2} \int_0^1 dt \,\partial_t \phi^i \,
\partial_t \phi^i\,.
\end{equation}
As we mentioned earlier,  the field $\phi^i(t)$ is expected to be
proportional to the position $x^i(t)$ of the brane (at least for small,
slowly varying displacements), so we can rewrite the above action as
\begin{equation}
S_2(\Phi)  = {1\over  g^2}  \Bigl( {\delta \phi^i\over \delta x^i} \Bigr)^2
\int_0^1  dt\,\, {1\over 2} \partial_t
x^i \, \partial_t x^i
\end{equation}
where the derivatives are evaluated at zero displacement.  Since
$\partial_t x^i$ is the velocity of the D-brane, the above action
represents the contribution from the (non-relativistic) kinetic energy
of a D-brane that has a mass $M$ given
by \footnote{Note that at this point, it is possible to take a shortcut
to get the D-brane mass directly using the fact that SFT at tree level
reproduces Yang-Mills theory~\cite{Coletti-Sigalov-Taylor}, with
$g_{{\rm YM}} = g/\sqrt{2}$ ~\cite{Polchinski,Coletti-Sigalov-Taylor},
where the Yang-Mills
field appears in the string field expansion as $A_\mu (k)
\alpha^{\mu}_{-1} | 0_1; k \rangle$, and where an additional factor of
$2 \pi$
        arises from the T-duality relation from section \ref{sec:T-duality},
       $ X \rightarrow 2 \pi
\alpha' A$.  Thus,
replacing $A^i \rightarrow X^i/2 \pi$ in the Yang-Mills action we have
$1/2g_{{\rm YM}}^2
F_{0i} F^{0i}\rightarrow 1/2 g^2
(\partial_0 x^i/\sqrt{2} \pi)^2$, so $M = 1/2 \pi^2 g^2$.  Although
perhaps more transparent, this is essentially the same calculation as
the original argument of Sen presented here, which we include in full
as it sheds light on the structure of the theory.}
\begin{equation}
\label{theequationforpssmass}
M = {1\over  g^2}  \Bigl( {\delta \phi^i\over \delta x^i} \Bigr)^2\,.
\end{equation}

\medskip
\noindent
{\em Step 2.~}  To find out how $\delta \phi^i$ is related to a true
displacement $\delta x^i$, we add a reference D-brane a
distance $b$ away from our original brane, in the direction~$x^i$.
We will then consider a string stretched between the branes.
We will use the string field action to
compute the change in the mass of  such string when our D-brane is displaced
by some $\delta \phi^i$.  Since the string tension is known, we will
be able to calculate the value of the physical displacement $\delta x^i$.

Given a string of length $L$, its  mass includes a contribution
$T_0 L = L/(2\pi)$,
and the corresponding contribution to the mass-squared is
$L^2/(4\pi^2)$.  If the original stretched string
has length $b$ and its length is then changed to $b +
\delta x^i$, the change $\delta m^2$ of the mass-squared is
\begin{equation}
\label{thechangeinwetnp}
\delta m^2 = {1\over 4\pi^2} \Bigl( (b+ \delta x^i)^2 - b^2 \Bigr)
\simeq  {1\over 2\pi^2} \, b \, \delta x^i \,.
\end{equation}
Let us now consider a time independent displacement, that is, a configuration
with $k_0=0$ (see (\ref{thefieldfordiskpvkakgk})).  We thus set $\delta \phi^i
\equiv \delta
\phi^i (k_0 =0) $ and $\delta \phi^i(k_0\not= 0) =0$.  The
string field associated to this displacement is obtained using
(\ref{corrstringfieldkysla}):
\begin{equation}
\label{thefirsttrysvnkeoii}
|\delta \Phi\rangle = \delta \phi^i \, c_1 \alpha_{-1}^i \, |0\rangle \,.
\end{equation}
We want to learn the effect of this string field perturbation on
the masses of stretched strings.  To do so, we introduce a complex
field $\eta$.  The fields
$\eta$ and $\eta^*$   represent the string that stretches from our
brane (brane one) to the reference brane (brane two)  and the string
that stretches from the reference brane to our brane,
respectively.  The string field that describes these states is written as:
\begin{equation}
\label{stretchedstringvifield}
|\psi\rangle = \eta \, c_1 |k_0, b\rangle \otimes
\begin{pmatrix}
0&1\\ 0&0
\end{pmatrix}
+ \eta^*  \, c_1  |-k_0, -b\rangle \otimes
\begin{pmatrix}
0&0\\ 1&0
\end{pmatrix} \,.
\end{equation}
The matrices included here are Chan-Paton matrices $a_{\alpha\beta}$,
with $\alpha, \beta = 1, 2$.  A value of one for a given $a_{\alpha\beta}$,
with zero for all other entries, is used to represent a string that stretches
from brane $\alpha$ to brane $\beta$.  When we have parallel
D-branes, the string field is matrix-valued.  The string action
includes a trace operation Tr that applies to the matrices, and the
star product
includes matrix multiplication.  The state $| k_0 , b\rangle$ represents
the ground state of a string with momentum $k_0$ that stretches
a distance $b$ in the $x^i$ direction.  It is necessary for our analysis
to determine the CFT vertex operator that corresponds
to this stretched string.  We claim that the operator is
\begin{equation}
\label{theopeforstretchstr}
|k_0, b\rangle \quad \longleftrightarrow \quad e^{i k_0 X^0} \, e^{i
{b\over 2\pi}
(X_L^i - X_R^i) }\,.
\end{equation}
The $k_0$ dependence of the operator is already familiar from
(\ref{themomstatesorg}).  The field multiplying $b$ is formed from
the left-moving and right-moving
parts of the open string coordinate $X^i$, evaluated at the string
endpoint.  This coordinate $X^i$ satisfies Dirichlet boundary conditions, so
at the boundary $X_L^i = - X_R^i$, and we can replace $X_L^i - X_R^i$ by
$2X_L^i$.  We also have the operator products and stress tensor
\begin{equation}
\partial X_L^i (x) \, \partial X_L^i (y) \sim -{1\over 2} {1\over (x-y)^2} \,,
\qquad  T_{X^i} = - \partial X_L^i \partial X_L^i \,.
\end{equation}
These relations allow us to compute the conformal dimension of
an exponential.  One readily finds that $\exp (i\alpha X_L^i)$ has
conformal dimension $\alpha^2/4$.  It follows that
\begin{equation}
\hbox{dimension}~\Bigl(  e^{i {b\over \pi} X_L^i}\Bigr)  =  \Bigl( {b
\over 2\pi}
\Bigr)^2 \,.
\end{equation}
Since conformal dimension is the value of $L_0$, which, in turn, determines
the mass-squared, this
result confirms that the operator in (\ref{theopeforstretchstr}) has
correctly
reproduced
the mass-squared of the stretched string.
For  future use, the vertex operator can be written as
\begin{equation}
\label{theopefohfltchstr}
            e^{i k_0 X^0} \, e^{i {b\over \pi}  X_L^i  }\,.
\end{equation}
The evaluation of the kinetic term for the field in
(\ref{stretchedstringvifield})
is relatively straightforward.  The only terms that survive are the
off-diagonal ones, coupling $\eta$ and $\eta^*$.  There are two such
terms, and their contributions are identical.  The product of the two
matrices give a matrix of trace one, and the overlap $\langle -k_0, -b|k_0,
b\rangle$ is also equal to one.  We then find
\begin{equation}
\label{readmassstrecvnjejj}
g^2 \, S_2 (\eta, \eta^*) =-{1\over 2} \cdot 2 \, \eta^*  \Bigl(
  -k_0^2 + {b^2\over (2\pi)^2} \Bigr) \eta= \eta^* \Bigl( k_0^2 -
{b^2\over (2\pi)^2}
\Bigr) \eta \,.
\end{equation}
In the setup with two branes, the fluctuation (\ref{thefirsttrysvnkeoii}) that
represents the displacement of our brane is fully represented by
\begin{equation}
\label{thefirsttrysvnkmmmdeoii}\
|\delta \Phi\rangle = \delta \phi^i \, c_1 \alpha_{-1}^i \, |0\rangle
\otimes
\begin{pmatrix}1&0\\ 0&0\end{pmatrix} \,.
\end{equation}
With the chosen normalization for $X^i$, the vertex operator
associated with $\alpha^i_{-1} |0\rangle$ is  $i \sqrt{2} \partial X_L$.

\medskip
\noindent
{\em Step 3.~}  We must now include the effects of the interactions
to see how the fluctuation (\ref{thefirsttrysvnkmmmdeoii}) affects
the mass of the stretched string.  Since the mass can be read from
equation (\ref{readmassstrecvnjejj}), we will find a term proportional
to $\eta^* \eta$ that arises from the  interaction and modifies the
value of the mass.

The interaction term takes the form
\begin{equation}
g^2 S_3 (\Phi) = -{1\over 3} \langle \Phi , \Phi, \Phi \rangle \,,
\end{equation}
and the string field is taken to be $|\Phi\rangle = |\psi\rangle
+|\delta
\Phi\rangle$, in order to see the effect of the fluctuation on the
stretched string.
We are looking for the terms of the form $\eta^* \eta \,\delta
\phi^i$, so we have
three different operators to insert at three different punctures.
There are a total
of six possible arrangements, that can be divided into two groups of
three arrangements each.  In each of these groups the cyclic ordering
of the operators is the  same.  The Chan-Paton matrices imply that
one cyclic ordering contributes while the other does not.
Indeed,
\begin{equation}
\begin{pmatrix}0&1\\ 0&0\end{pmatrix}
\begin{pmatrix}0&0\\ 1& 0\end{pmatrix}
\begin{pmatrix}1&0\\ 0&0\end{pmatrix}
= \begin{pmatrix}1&0\\ 0&0\end{pmatrix}
\end{equation}
is a matrix of unit trace, while
\begin{equation}
\begin{pmatrix}0&0\\ 1&0\end{pmatrix}
\begin{pmatrix}0&1\\ 0& 0\end{pmatrix}
\begin{pmatrix}1&0\\ 0&0\end{pmatrix}
= \begin{pmatrix}0&0\\ 0&0\end{pmatrix}
\end{equation}
is a matrix of vanishing trace.  We conclude that the Chan-Paton
matrices contribute a factor of $+3$.   The operators to be inserted
can be chosen to be physical (dimension zero) so we need not worry
about local coordinates at the punctures.  Using punctures at
$\infty$, $-1,$ and $0$ we then find:
\begin{equation}
g^2 S_3 (\Phi) = - \eta^* \eta \delta \phi^i
            \Bigl\langle  e^{-i k_0 X^0-i {b\over \pi}  X_L^i  } c(\infty)
\, \sqrt{2}\, i \, \partial X_L  c(-1) \,
            e^{i k_0 X^0+i {b\over \pi}  X_L^i  } c(0) \Bigr\rangle \,.
\end{equation}
Since the vertex operators are on-shell, and the ghost insertions
are placed at standard positions, the whole correlator gives a factor of
one, except for the contraction between the $\partial X_L (-1)$ and
the finitely located $\exp (i{b\over \pi} X_L^i (0))$:
\begin{equation}
g^2 S_3  = - \eta^* \eta \delta \phi^i\, \sqrt{2} \,i \, {ib\over \pi}\,\Bigl(
  -{1\over 2}\Bigr) {1\over (-1-0)}  =  {b\over \sqrt{2} \, \pi}
\eta^* \eta \delta
\phi^i \,.
\end{equation}
Combining this result with (\ref{readmassstrecvnjejj}) we find
\begin{equation}
\label{readmastkjrecvnjejj}
g^2 \, (S_2 + S_3) = \eta^* \Bigl( k_0^2 - {b^2\over (2\pi)^2}+
{b\over \sqrt{2}}
{\delta \phi^i\over \pi}
\Bigr) \eta \,.
\end{equation}
The last term in parenthesis corresponds to a change in $m^2$.  So, comparing
with (\ref{thechangeinwetnp}) we obtain
\begin{equation}
{1\over 2\pi^2} b\, \delta x^i = -{b\over \sqrt{2}}
{\delta \phi^i\over \pi} \quad \to \quad  {\delta \phi^i\over \delta x^i}
= -{1\over \sqrt{2} \, \pi} \,.
\end{equation}
This is the needed relation between the field  $\delta \phi^i$ that represents
a displacement of the brane and the resulting displacement $\delta x^i$.
The mass of the brane now follows directly from (\ref{theequationforpssmass}):
\begin{equation}
M = {1\over g^2} \Bigl( {1\over \sqrt{2}\, \pi} \Bigr)^2 = {1\over 2\pi^2
g^2}\,.
\end{equation}
This is the result we wanted to establish.


\subsection{Constraints and symmetries}
\label{sec:constraints-symmetries}

It may appear that {\em a priori} all scalar fields in the
spectrum of open strings could acquire a vacuum expectation
value in the tachyonic vacuum.  Nevertheless, there are a
set of considerations that imply that only a subset of these
scalar fields acquire expectation values.  In this section we explore
these ideas.   They are subdivided into the following:

\begin{enumerate}

\item[(1)] Universality conditions.

\item[(2)] Twist conditions.

\item[(3)] Gauge fixing conditions.

\item[(4)] $SU(1,1)$ conditions.

\end{enumerate}

Among these conditions, the third one, which concerns gauge fixing, is
on a somewhat different footing.  The other three conditions apply because
of a simple general argument which we discuss first.

Consider a subdivision of all the scalar fields   into two disjoint set of
fields.  The first set contains the fields $t_0, t_1, t_2, \ldots$
and the second
set contains the fields $u_0, u_1, u_2, \ldots$.  Let us denote by $t_i$
the elements of the first set and by $u_a$ the elements of the second set.
Suppose the string field action $S(t_i, u_a)$ is such that there are no
terms that are linear in $u_a$.  We then claim that it is consistent
to search for a solution of the equations of motion that assumes $u_a=0$
for all $a$.  The reason is easy to explain.  If all terms with $u$
fields contain at
least two of them, the equations of motion for the $u$ fields are composed
of terms all of which contain at least one $u$ field.
As a result, $u_a =0$ satisfies these equations of motion.
In our analysis we will try to construct a set $\{ t_i\}$ with the
smallest possible number of fields,
so that none of the remaining fields couples linearly in the action.
The tachyon field, of course, is
one of the elements of the set $\{t_i\}$.

\medskip
Let's begin by explaining how (1) works.  For this we split the state
space of the BCFT into three groups.  In each of these groups, the
ghost part of the states is the same:  it includes all states of ghost
number one.  The nontrivial part of the argument uses the matter
part of the conformal field theory.   We write
\begin{equation}
\mathcal{H} = \mathcal{H}_1 \oplus  \mathcal{H}_2\oplus \mathcal{H}_3\,,
\end{equation}
where $\mathcal{H}_1,  \mathcal{H}_2,$ and  $\mathcal{H}_3$ will
be disjoint vector subspaces of $\mathcal{H}$ (their intersections give
the zero vector).  We also write
\begin{equation}
\mathcal{H}_i = \mathcal{M}_i \otimes |\mathcal{G}\rangle \,,\quad i=1,2,3\,.
\end{equation}
where $|\mathcal{G}\rangle$ denotes the general state of ghost number one
in the ghost conformal field theory.  The spaces
$\mathcal{M}_1,  \mathcal{M}_2,$ and  $\mathcal{M}_3$ are disjoint subspaces
of the matter CFT whose union gives the total matter CFT
state space.  The  $\mathcal{M}$ subspaces are defined as:
\begin{eqnarray}
\label{sthesplitvematt}
\mathcal{M}_1 : &&  \hbox{primary\,}~~
|0\rangle ~\hbox{and descendents} ,\nonumber\\
\mathcal{M}_2 : &&  \hbox{primaries}~ |k_0 \not=0\rangle ~\hbox{and
descendents}, \\
\mathcal{M}_3 : &&  \hbox{primaries}~ |k_0=0\rangle
~\hbox{different from}~|0\rangle ~\hbox{and descendents} \nonumber \,.
\end{eqnarray}
In the above, primary means Virasoro primary in the matter sector,
and descendent means Virasoro descendent.  The union of the spaces
give the full CFT because for unitary matter CFT's (which we are
assuming our CFT is) the state space can be broken into primaries and
their descendents.  Any matter primary belongs to one of the three
spaces above.  It should also be clear that the primaries in $\mathcal{M}_3$
have positive conformal dimension.

We now claim that the fields in  $\mathcal{H}_2$  and $\mathcal{H}_3$
need not acquire expectation values (they are
$u$ fields); the tachyon condensate is all in $\mathcal{H}_1$.  We
are therefore
defining
the $t$ fields to be  precisely the fields in $\mathcal{H}_1$.
To prove that this is valid, we first note
that  a field in $\mathcal{H}_2$ cannot appear linearly in a term
where all other fields are $t$ fields ({\it i.e.}, fields in
${\mathcal H}_1$).  The reason is simply momentum
conservation.

            A little more work is necessary to show that the
fields in $\mathcal{H}_3$ cannot couple linearly to the fields in
$\mathcal{H}_1$.    Let us begin with the kinetic term.  Since the
BRST operator is composed of terms that include ghost oscillators
and matter Virasoro operators, it maps each $\mathcal{H}_i$ space
into itself.   The primaries in $\mathcal{H}_1$ and $\mathcal{H}_3$
are BPZ orthogonal, so any two states in the descendent towers are
also orthogonal (this is proven by using the BPZ conjugation properties
of Virasoro operators to move them from one state to the other until
some state is annihilated or the whole expression reduces to the
BPZ inner product of the primaries).
For the interaction term a similar argument holds.  First note
that the three string vertex does not couple two matter primaries from
$\mathcal{H}_1$ to a matter primary from $\mathcal{H}_3$.  This is
because in the CFT matter correlator the primaries from $\mathcal{H}_1$ appear
as identity operators, so the whole correlator is proportional to the one-point
function of the primary in $\mathcal{H}_3$, which vanishes because the state
has non-zero dimension. The Virasoro conservation laws on the vertex then
imply that  the coupling of any two states in $\mathcal{H}_1$ to a state in
$\mathcal{H}_3$ must vanish. This completes our proof.

The space $\mathcal{H}_1$ is universal.  It does not depend on the
details of the matter conformal field theory, except for the existence
of a zero-momentum SL(2,R) ground state.   The space can be written
as
\begin{equation}
\mathcal{H}_1 \equiv \hbox{Span} \Bigl\{ L_{-j_1}^m \ldots L_{-j_p}^m\,
b_{-k_1} \ldots b_{-k_q} \, c_{-l_1} \ldots c_{-l_r} \, |0\rangle\Bigr\}
\end{equation}
where
\begin{equation}
j_1 \geq j_2 \geq \ldots \geq j_p \,, ~~j_i \geq 2 \,, ~~ k_i \geq  2
\,, ~~ l_i \geq
  -1\,,
\quad
\hbox{and}\quad  l-q=1 \,.
\end{equation}
The first inequality ensures that the descendents are built unambiguously,
the second inequality is needed because $L_{-1} |0\rangle =0$.
The third and fourth inequalities are  familiar, and the last equality
ensures that the ghost number of the state is one.
\medskip

Let us now explain how twist properties allow us to restrict
$\mathcal{H}_1$ further.  The claim is that we can restrict ourselves
to the twist even subspace of $\mathcal{H}_1$.
Heuristically, this follows from the fact that the two- and
three-string vertices are invariant under reflection, so all terms
linear in twist fields would pick up a change of sign and therefore vanish.
The twist-even space, of course,
contains the zero momentum tachyon $c_{1} |0\rangle$ (recall that
$|0\rangle$ is twist odd, and $\Omega c_{-n} \Omega^{-1} = (-)^n c_{-n}$).
The first two properties in (\ref{l1e4}) ensure that the kinetic
term in the string action does not couple a twist odd field to
a twist even field. We also studied the twist properties of the three string
vertex.  In fact, in an exercise,   we considered a twist even field
$A_+$ and a
twist odd field
$A_-$ ( $\Omega A_\pm =
\pm A$ ) both of which were Grassmann odd (like the string field is).
You then showed that $\langle A_+, A_+, A_-\rangle = 0$  (see
(\ref{fortwistproperty})).  Consider now a general string field
$\Phi \in \mathcal{H}$ and split it into twist even
and twist odd parts  $\Phi = \Phi_+ + \Phi_-$.  When the interaction
vertex is evaluated, the terms  linear in $\Phi_-$ are of the form
$\langle \Phi_+ , \Phi_+, \Phi_-\rangle$ (any other similar looking
term is related to this by cyclicity).  So terms linear on twist
odd fields vanish.    This proves that we can indeed constrain
$\mathcal{H}_1$ further.

The twist eigenvalue of a state is
given as $\Omega = (-1)^N$, where $N$ is the number eigenvalue of the state,
defined with $N=0$ for the zero momentum tachyon.  In terms of level,
states at odd levels are twist odd, and states at even levels are twist even.
So, the twist condition allows us to restrict ourselves to the states of
$\mathcal{H}_1$ at even levels.

\medskip
We now turn to the gauge fixing condition.  This gauge fixing
condition, the Feynman-Siegel gauge condition
$b_0 |\Phi\rangle =0$, restricts further the space $\mathcal{H}_1$.
We will discuss the global
validity of the Siegel gauge later, but here we discuss its
clear validity at the linearized level and within the subspace
$\mathcal{H}_1$ already restricted to states at even level.
First, we show that the gauge condition can be reached
starting from fields that do not satisfy it.
Let $|\Phi\rangle $ be a field such that
$b_0 |\Phi\rangle
\not=0$. Since $|\Phi\rangle$ cannot be of level one, $L_0 |\Phi
\rangle\not= 0$.
Then consider the following gauge equivalent state
\begin{equation}
|\widetilde \Phi\rangle = |\Phi\rangle  - Q  {b_0\over L_0} |\Phi\rangle\,.
\end{equation}
Using $\{b_0 , Q\} = L_0$ one readily checks that $b_0|\widetilde \Phi \rangle
=0$, so the gauge can be reached.  Moreover, we now show that no gauge
transformation remains in this gauge. If there were, there would exist
a non-zero string field in the gauge slice that happens to be pure gauge.
Such field $|\Phi \rangle$ would then satisfy $b_0 |\Phi\rangle =0$,
$L_0|\Phi\rangle \not=0$, and $|\Phi\rangle = Q |\epsilon\rangle$.
Since both $b_0$ and $Q$ annihilate
the state:
\begin{equation}
0= b_0 Q |\Phi\rangle + Q  b_0 |\Phi\rangle = \{ b_0, Q\} |\Phi\rangle
= L_0 |\Phi\rangle\,,
\end{equation}
in contradiction  with the fact that the state does have non-zero
dimension.  The Siegel gauge is clearly a good gauge at the linearized
level and within the twist truncated $\mathcal{H}_1$.

\medskip
Let's now consider briefly the additional truncation that is
allowed by $SU(1,1)$ symmetry (item (4) of our list).  Once we work
in the Siegel gauge, this further truncation is allowed.  This truncation
is only possible because of the particular form of the string vertex.
It would not be allowed for arbitrarily defined star products.
Let us recall
how this symmetry arises in the
cubic open string field theory~\cite{Zwiebach:2000vc}. In the Siegel gauge, the
string field action reads
\begin{equation}
\label{gfaction}
S \sim {1\over 2} \langle \phi|L_0 |\phi\rangle + \,{1\over 3}
             \langle \phi | \langle \phi | \langle \phi \,| \,v_3 \rangle \,.
\end{equation}
The vertex coupling the three string fields is of the form
\begin{equation}
\label{thevertex}
|v_3\rangle \sim  \exp\, (E_{matt}) \,\,\exp \Bigl( - \sum_{r,s = 1}^3
\sum_{n, m
=1}^\infty c_{-n}^r\,X^{rs}_{nm}\, b_{-m}^s \Bigr) \,
|0_1\rangle_{123}\,,
\end{equation}
where we have focused on the ghost sector.
The Neumann coefficients
are known to satisfy~\cite{Zwiebach:2000vc}
\begin{equation}
\label{exchange}
X^{rs}_{nm} = \, {n\over m} \, X^{sr}_{mn} \,, \quad n,m \geq 1\,.
\end{equation}
This  relation is not true for general three string
vertices, but holds for the open string field theory vertex.
Given equation (\ref{exchange}),
the argument of the exponential in $|v_3\rangle_{123}$
can be written as a sum  of terms of the form
($r,s,n,m$, not summed)
\begin{equation}
\label{verpart}
X^{rs}_{nm} \, c_{-n}^r b_{-m}^s + X^{sr}_{mn} \, c_{-m}^s b_{-n}^r
= {1\over m} X^{sr}_{mn} \Bigl( \, n c_{-n}^r b_{-m}^s + m c_{-m}^s
b_{-n}^r
\Bigr)\,.
\end{equation}
The  term in
parenthesis is invariant under the continuous transformations
\begin{eqnarray}
\label{continuous}
b_{-n} (\theta) &=& b_{-n} \cos \theta - n c_{-n} \sin \theta \,,\cr
c_{-n} (\theta) &=& c_{-n} \cos \theta + {1\over n} b_{-n} \sin \theta
\,.
\end{eqnarray}
These transformations, valid for all $n\not= 0$, imply $\{ c_n(\theta),
b_m(\theta) \} = \delta_{n+m}$. One readily finds that they are generated
by an operator $\mathcal{S}_1$:
\begin{equation}
\label{ghtrans}
b_{-n} (\theta) = e^{\theta \mathcal{S}_1} \, b_{-n} e^{-\theta
\mathcal{S}_1}\,,
\quad c_{-n} (\theta) = e^{\theta \mathcal{S}_1} \, c_{-n} e^{-\theta
\mathcal{S}_1}\,,
\end{equation}
where $\mathcal{S}_1$ is given by
\begin{equation}
\label{fgen}
\mathcal{S}_1 = \sum_{n=1}^\infty \Bigl( \, {1\over n} \, b_{-n} b_n
  - n c_{-n}
c_n\Bigr)\,.
\end{equation}
Since the vacuum $|0_1\rangle$ is annihilated by $\mathcal{S}_1$,
the vertex $|v_3\rangle$ is invariant under
this $U(1)$ symmetry: $\exp \Bigl( \theta
(\mathcal{S}_1^{(1)}+\mathcal{S}_1^{(2)}+ \mathcal{S}_1^{(3)}) \, \Bigr) |
v_3\rangle_{123} = | v_3\rangle_{123}$.  Equivalently,
\begin{equation}
\label{finv}
\Bigl(\mathcal{S}_1^{(1)}+\mathcal{S}_1^{(2)}+ \mathcal{S}_1^{(3)} \Bigr)
| v_3\rangle_{123} = 0\,.
\end{equation}
Since the vertex $|v_3\rangle_{123}$ is built from ghost
bilinears
of zero ghost number, we deduce that the ghost number operator $\mathcal{G}$
\begin{equation}
\label{ghgen}
\mathcal{G} = \sum_{n=1}^\infty \Bigl( \, c_{-n}b_n - b_{-n} c_n \, \Bigr).
\end{equation}
            is also conserved:
\begin{equation}
\label{ginv}
\Bigl(\mathcal{G}^{(1)}+\mathcal{G}^{(2)}+ \mathcal{G}^{(3)} \Bigr)
| v_3\rangle_{123} = 0 \,.
\end{equation}
We can then form the  commutator
\begin{equation}
\label{sgen}
[ \mathcal{S}_1\, , \,\mathcal{G} \, ] = 2 \mathcal{S}_2\,, \quad \hbox{with}
\quad
\mathcal{S}_2 = \sum_{n=1}^\infty \Bigl( \, {1\over n} \, b_{-n} b_n
+ n c_{-n}
c_n\Bigr)\,.
\end{equation}
The remaining commutators are readily computed:
\begin{equation}
\label{rcomm}
[ \mathcal{S}_2\, , \,\mathcal{G} \, ] = 2 \mathcal{S}_1\,,\quad
[ \mathcal{S}_1\, , \,\mathcal{S}_2 \, ] = -2 \mathcal{G}\,.
\end{equation}
These relations show that $\{ \mathcal{S}_1, \mathcal{S}_2, \mathcal{G} \}$
generate the algebra of $SU(1,1)$. These generators
are the same as those in the $SU(1,1)$ algebra in Siegel and
Zwiebach~\cite{Siegel:1985tw}.\footnote{Defining $X= (\mathcal{S}_2 -
\mathcal{S}_1)/2$,
$Y=(\mathcal{S}_2 + \mathcal{S}_1)/2$, and $H= \mathcal{G}$ we recover the
conventional definition of the isomorphic (real) Lie algebra $sl(2,R)$, with
brackets $[X,Y]= H,~ [H, X] = 2X, ~ [H,Y] = -2Y$. Note that
$T_+ = - 2X$, where $T_+$ is the operator that multiplies
$b_0$ in the BRST operator.}
Since both $\mathcal{S}_1$ and $\mathcal{G}$ are symmetries of the three string
vertex, we also have
\begin{equation}
\label{sinv}
\Bigl(\mathcal{S}_2^{(1)}+\mathcal{S}_2^{(2)}+ \mathcal{S}_2^{(3)} \Bigr)
| v_3\rangle_{123} = 0 \,.
\end{equation}
In summary, the three string vertex is fully $SU(1,1)$ invariant.

The set of Fock space states built with the action of ghost
and antighost oscillators on the vacuum $|0_1\rangle$
can be decomposed into finite dimensional
irreducible representations of $SU(1,1)$. Note that $(nc_{-n}, b_{-n})$
transforms as a doublet. As usual, from the tensor product of two
doublets one can obtain a nontrivial singlet; this is just
\begin{equation}
\label{singlets}
m b_{-n} c_{-m} +  n b_{-m} c_{-n}\,.
\end{equation}It is now simple to argue that the twist even subspace of
$\mathcal{H}_1$ in the Siegel gauge can be further restricted
to $SU(1,1)$ singlets. Since the kinetic operator
$L_0$ commutes with the $SU(1,1)$ generators,
            the kinetic term cannot couple a non-singlet
to a singlet. Indeed, consider such a term $\langle s| L_0 |a\rangle$,
where $\langle s|$ is a singlet
and $|a\rangle$ is not a singlet.
Given the structure of the representations (completely analogous
to the finite dimensional unitary representations of $SU(2)$), it
follows that there is a state $|b\rangle$ and an $SU(1,1)$ generator
$\mathcal{J}$ such that $|a\rangle = \mathcal{J} |b\rangle$. Therefore
$\langle s| L_0 |a\rangle = \langle s| L_0 \mathcal{J}|b\rangle
= \langle s|\mathcal{J} L_0 |b\rangle =0$, where the last step gives zero
because
$\mathcal{J}$ annihilates the singlet (this requires $bpz (\mathcal{J} )
= \pm \mathcal{J}$, which is true). It remains
to show that the vertex cannot couple a non-singlet to two singlets.
Indeed, with analogous notation we have
\begin{eqnarray}
{}_1\langle s_1|{}_2\langle s_2| {}_3\langle a \, | v_3 \rangle
&&= {}_1\langle s_1|{}_2\langle s_2| {}_3\langle b \, | \mathcal{J}^{(3)}| v_3
\rangle \nonumber\\
&&=- {}_1\langle s_1|{}_2\langle s_2| {}_3\langle b \, | (\mathcal{J}^{(1)}+
\mathcal{J}^{(2)}) |\,v_3 \rangle =0 \,,
\end{eqnarray}
where we used the conservation of $\mathcal{J}$ on the vertex, and on the
last step the $\mathcal{J}$ operators annihilate the singlets.

\medskip
This completes our discussion of the various symmetries and
conditions that can be used to constrain the subspace of the
string state space that acquires vacuum expectation values
in the tachyon vacuum.


\subsection{The nonperturbative vacuum}
\label{sec:vacuum}

Sen's first conjecture states that the string field theory action
should lead to a nontrivial vacuum solution, with energy density
\begin{equation}
  -T_{25} = -\frac{1}{2 \pi^2 g^2}  \,.
\end{equation}
In this subsection we discuss evidence for the validity of this
conjecture in Witten's OSFT.  As mentioned in the introduction, this
result holds exactly in BSFT.

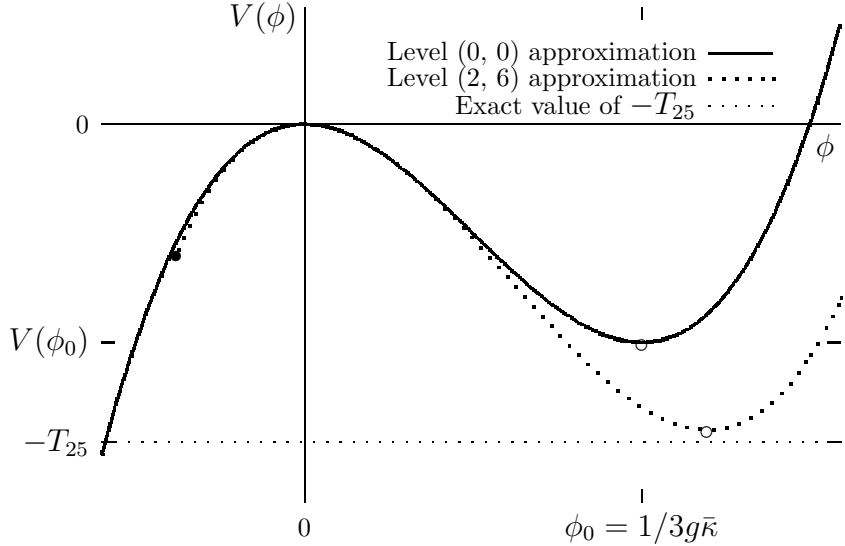
\begin{figure}
\setlength{\unitlength}{0.240900pt}
\ifx\plotpoint\undefined\newsavebox{\plotpoint}\fi
\sbox{\plotpoint}{\rule[-0.200pt]{0.400pt}{0.400pt}}%
\begin{picture}(1380,900)(120,0)
\font\gnuplot=cmr10 at 10pt
\gnuplot
\sbox{\plotpoint}{\rule[-0.200pt]{0.400pt}{0.400pt}}%
\put(280.0,675.0){\rule[-0.200pt]{4.818pt}{0.400pt}}
\put(260,675){\makebox(0,0)[r]{0}}
\put(1419.0,675.0){\rule[-0.200pt]{4.818pt}{0.400pt}}
\put(280.0,332.0){\rule[-0.200pt]{4.818pt}{0.400pt}}
\put(260,332){\makebox(0,0)[r]{$V (\phi_0)$}}
\put(1419.0,332.0){\rule[-0.200pt]{4.818pt}{0.400pt}}
\put(280.0,175.0){\rule[-0.200pt]{4.818pt}{0.400pt}}
\put(260,175){\makebox(0,0)[r]{$-T_{25} $}}
\put(1419.0,175.0){\rule[-0.200pt]{4.818pt}{0.400pt}}
\put(598.0,82.0){\rule[-0.200pt]{0.400pt}{4.818pt}}
\put(598,41){\makebox(0,0){0}}
\put(598.0,840.0){\rule[-0.200pt]{0.400pt}{4.818pt}}
\put(1127.0,82.0){\rule[-0.200pt]{0.400pt}{4.818pt}}
\put(1127,41){\makebox(0,0){$\phi_0 = 1/3g\bar{\kappa}$}}
\put(1127.0,840.0){\rule[-0.200pt]{0.400pt}{4.818pt}}
\put(280.0,675.0){\rule[-0.200pt]{279.203pt}{0.400pt}}
\put(1401,638){\makebox(0,0)[l]{$\phi$}}
\put(480,841){\makebox(0,0)[l]{$V (\phi)$}}
\put(598.0,82.0){\rule[-0.200pt]{0.400pt}{187.420pt}}
\sbox{\plotpoint}{\rule[-0.400pt]{0.800pt}{0.800pt}}%
\put(1213,786){\makebox(0,0)[r]{Level (0, 0) approximation}}
\put(1233.0,786.0){\rule[-0.400pt]{24.090pt}{0.800pt}}
\put(280,156){\usebox{\plotpoint}}
\multiput(281.41,156.00)(0.511,1.892){17}{\rule{0.123pt}{3.067pt}}
\multiput(278.34,156.00)(12.000,36.635){2}{\rule{0.800pt}{1.533pt}}
\multiput(293.40,199.00)(0.512,1.935){15}{\rule{0.123pt}{3.109pt}}
\multiput(290.34,199.00)(11.000,33.547){2}{\rule{0.800pt}{1.555pt}}
\multiput(304.41,239.00)(0.511,1.666){17}{\rule{0.123pt}{2.733pt}}
\multiput(301.34,239.00)(12.000,32.327){2}{\rule{0.800pt}{1.367pt}}
\multiput(316.41,277.00)(0.511,1.621){17}{\rule{0.123pt}{2.667pt}}
\multiput(313.34,277.00)(12.000,31.465){2}{\rule{0.800pt}{1.333pt}}
\multiput(328.41,314.00)(0.511,1.485){17}{\rule{0.123pt}{2.467pt}}
\multiput(325.34,314.00)(12.000,28.880){2}{\rule{0.800pt}{1.233pt}}
\multiput(340.40,348.00)(0.512,1.536){15}{\rule{0.123pt}{2.527pt}}
\multiput(337.34,348.00)(11.000,26.755){2}{\rule{0.800pt}{1.264pt}}
\multiput(351.41,380.00)(0.511,1.304){17}{\rule{0.123pt}{2.200pt}}
\multiput(348.34,380.00)(12.000,25.434){2}{\rule{0.800pt}{1.100pt}}
\multiput(363.41,410.00)(0.511,1.259){17}{\rule{0.123pt}{2.133pt}}
\multiput(360.34,410.00)(12.000,24.572){2}{\rule{0.800pt}{1.067pt}}
\multiput(375.40,439.00)(0.512,1.237){15}{\rule{0.123pt}{2.091pt}}
\multiput(372.34,439.00)(11.000,21.660){2}{\rule{0.800pt}{1.045pt}}
\multiput(386.41,465.00)(0.511,1.078){17}{\rule{0.123pt}{1.867pt}}
\multiput(383.34,465.00)(12.000,21.126){2}{\rule{0.800pt}{0.933pt}}
\multiput(398.41,490.00)(0.511,0.988){17}{\rule{0.123pt}{1.733pt}}
\multiput(395.34,490.00)(12.000,19.402){2}{\rule{0.800pt}{0.867pt}}
\multiput(410.40,513.00)(0.512,0.988){15}{\rule{0.123pt}{1.727pt}}
\multiput(407.34,513.00)(11.000,17.415){2}{\rule{0.800pt}{0.864pt}}
\multiput(421.41,534.00)(0.511,0.807){17}{\rule{0.123pt}{1.467pt}}
\multiput(418.34,534.00)(12.000,15.956){2}{\rule{0.800pt}{0.733pt}}
\multiput(433.41,553.00)(0.511,0.762){17}{\rule{0.123pt}{1.400pt}}
\multiput(430.34,553.00)(12.000,15.094){2}{\rule{0.800pt}{0.700pt}}
\multiput(445.41,571.00)(0.511,0.717){17}{\rule{0.123pt}{1.333pt}}
\multiput(442.34,571.00)(12.000,14.233){2}{\rule{0.800pt}{0.667pt}}
\multiput(457.40,588.00)(0.512,0.639){15}{\rule{0.123pt}{1.218pt}}
\multiput(454.34,588.00)(11.000,11.472){2}{\rule{0.800pt}{0.609pt}}
\multiput(468.41,602.00)(0.511,0.536){17}{\rule{0.123pt}{1.067pt}}
\multiput(465.34,602.00)(12.000,10.786){2}{\rule{0.800pt}{0.533pt}}
\multiput(479.00,616.41)(0.491,0.511){17}{\rule{1.000pt}{0.123pt}}
\multiput(479.00,613.34)(9.924,12.000){2}{\rule{0.500pt}{0.800pt}}
\multiput(491.00,628.40)(0.489,0.512){15}{\rule{1.000pt}{0.123pt}}
\multiput(491.00,625.34)(8.924,11.000){2}{\rule{0.500pt}{0.800pt}}
\multiput(502.00,639.40)(0.674,0.516){11}{\rule{1.267pt}{0.124pt}}
\multiput(502.00,636.34)(9.371,9.000){2}{\rule{0.633pt}{0.800pt}}
\multiput(514.00,648.40)(0.913,0.526){7}{\rule{1.571pt}{0.127pt}}
\multiput(514.00,645.34)(8.738,7.000){2}{\rule{0.786pt}{0.800pt}}
\multiput(526.00,655.40)(0.913,0.526){7}{\rule{1.571pt}{0.127pt}}
\multiput(526.00,652.34)(8.738,7.000){2}{\rule{0.786pt}{0.800pt}}
\multiput(538.00,662.38)(1.432,0.560){3}{\rule{1.960pt}{0.135pt}}
\multiput(538.00,659.34)(6.932,5.000){2}{\rule{0.980pt}{0.800pt}}
\put(549,666.34){\rule{2.600pt}{0.800pt}}
\multiput(549.00,664.34)(6.604,4.000){2}{\rule{1.300pt}{0.800pt}}
\put(561,669.34){\rule{2.891pt}{0.800pt}}
\multiput(561.00,668.34)(6.000,2.000){2}{\rule{1.445pt}{0.800pt}}
\put(573,671.34){\rule{2.650pt}{0.800pt}}
\multiput(573.00,670.34)(5.500,2.000){2}{\rule{1.325pt}{0.800pt}}
\put(584,672.84){\rule{2.891pt}{0.800pt}}
\multiput(584.00,672.34)(6.000,1.000){2}{\rule{1.445pt}{0.800pt}}
\put(596,672.84){\rule{2.891pt}{0.800pt}}
\multiput(596.00,673.34)(6.000,-1.000){2}{\rule{1.445pt}{0.800pt}}
\put(608,671.84){\rule{2.891pt}{0.800pt}}
\multiput(608.00,672.34)(6.000,-1.000){2}{\rule{1.445pt}{0.800pt}}
\put(620,670.34){\rule{2.650pt}{0.800pt}}
\multiput(620.00,671.34)(5.500,-2.000){2}{\rule{1.325pt}{0.800pt}}
\put(631,667.84){\rule{2.891pt}{0.800pt}}
\multiput(631.00,669.34)(6.000,-3.000){2}{\rule{1.445pt}{0.800pt}}
\put(643,664.34){\rule{2.600pt}{0.800pt}}
\multiput(643.00,666.34)(6.604,-4.000){2}{\rule{1.300pt}{0.800pt}}
\multiput(655.00,662.06)(1.432,-0.560){3}{\rule{1.960pt}{0.135pt}}
\multiput(655.00,662.34)(6.932,-5.000){2}{\rule{0.980pt}{0.800pt}}
\multiput(666.00,657.07)(1.132,-0.536){5}{\rule{1.800pt}{0.129pt}}
\multiput(666.00,657.34)(8.264,-6.000){2}{\rule{0.900pt}{0.800pt}}
\multiput(678.00,651.07)(1.132,-0.536){5}{\rule{1.800pt}{0.129pt}}
\multiput(678.00,651.34)(8.264,-6.000){2}{\rule{0.900pt}{0.800pt}}
\multiput(690.00,645.08)(0.825,-0.526){7}{\rule{1.457pt}{0.127pt}}
\multiput(690.00,645.34)(7.976,-7.000){2}{\rule{0.729pt}{0.800pt}}
\multiput(701.00,638.08)(0.913,-0.526){7}{\rule{1.571pt}{0.127pt}}
\multiput(701.00,638.34)(8.738,-7.000){2}{\rule{0.786pt}{0.800pt}}
\multiput(713.00,631.08)(0.774,-0.520){9}{\rule{1.400pt}{0.125pt}}
\multiput(713.00,631.34)(9.094,-8.000){2}{\rule{0.700pt}{0.800pt}}
\multiput(725.00,623.08)(0.674,-0.516){11}{\rule{1.267pt}{0.124pt}}
\multiput(725.00,623.34)(9.371,-9.000){2}{\rule{0.633pt}{0.800pt}}
\multiput(737.00,614.08)(0.611,-0.516){11}{\rule{1.178pt}{0.124pt}}
\multiput(737.00,614.34)(8.555,-9.000){2}{\rule{0.589pt}{0.800pt}}
\multiput(748.00,605.08)(0.674,-0.516){11}{\rule{1.267pt}{0.124pt}}
\multiput(748.00,605.34)(9.371,-9.000){2}{\rule{0.633pt}{0.800pt}}
\multiput(760.00,596.08)(0.599,-0.514){13}{\rule{1.160pt}{0.124pt}}
\multiput(760.00,596.34)(9.592,-10.000){2}{\rule{0.580pt}{0.800pt}}
\multiput(772.00,586.08)(0.543,-0.514){13}{\rule{1.080pt}{0.124pt}}
\multiput(772.00,586.34)(8.758,-10.000){2}{\rule{0.540pt}{0.800pt}}
\multiput(783.00,576.08)(0.539,-0.512){15}{\rule{1.073pt}{0.123pt}}
\multiput(783.00,576.34)(9.774,-11.000){2}{\rule{0.536pt}{0.800pt}}
\multiput(795.00,565.08)(0.539,-0.512){15}{\rule{1.073pt}{0.123pt}}
\multiput(795.00,565.34)(9.774,-11.000){2}{\rule{0.536pt}{0.800pt}}
\multiput(807.00,554.08)(0.539,-0.512){15}{\rule{1.073pt}{0.123pt}}
\multiput(807.00,554.34)(9.774,-11.000){2}{\rule{0.536pt}{0.800pt}}
\multiput(819.00,543.08)(0.489,-0.512){15}{\rule{1.000pt}{0.123pt}}
\multiput(819.00,543.34)(8.924,-11.000){2}{\rule{0.500pt}{0.800pt}}
\multiput(830.00,532.08)(0.539,-0.512){15}{\rule{1.073pt}{0.123pt}}
\multiput(830.00,532.34)(9.774,-11.000){2}{\rule{0.536pt}{0.800pt}}
\multiput(842.00,521.08)(0.539,-0.512){15}{\rule{1.073pt}{0.123pt}}
\multiput(842.00,521.34)(9.774,-11.000){2}{\rule{0.536pt}{0.800pt}}
\multiput(855.40,507.55)(0.512,-0.539){15}{\rule{0.123pt}{1.073pt}}
\multiput(852.34,509.77)(11.000,-9.774){2}{\rule{0.800pt}{0.536pt}}
\multiput(865.00,498.08)(0.539,-0.512){15}{\rule{1.073pt}{0.123pt}}
\multiput(865.00,498.34)(9.774,-11.000){2}{\rule{0.536pt}{0.800pt}}
\multiput(877.00,487.08)(0.491,-0.511){17}{\rule{1.000pt}{0.123pt}}
\multiput(877.00,487.34)(9.924,-12.000){2}{\rule{0.500pt}{0.800pt}}
\multiput(889.00,475.08)(0.489,-0.512){15}{\rule{1.000pt}{0.123pt}}
\multiput(889.00,475.34)(8.924,-11.000){2}{\rule{0.500pt}{0.800pt}}
\multiput(900.00,464.08)(0.539,-0.512){15}{\rule{1.073pt}{0.123pt}}
\multiput(900.00,464.34)(9.774,-11.000){2}{\rule{0.536pt}{0.800pt}}
\multiput(912.00,453.08)(0.539,-0.512){15}{\rule{1.073pt}{0.123pt}}
\multiput(912.00,453.34)(9.774,-11.000){2}{\rule{0.536pt}{0.800pt}}
\multiput(924.00,442.08)(0.599,-0.514){13}{\rule{1.160pt}{0.124pt}}
\multiput(924.00,442.34)(9.592,-10.000){2}{\rule{0.580pt}{0.800pt}}
\multiput(936.00,432.08)(0.489,-0.512){15}{\rule{1.000pt}{0.123pt}}
\multiput(936.00,432.34)(8.924,-11.000){2}{\rule{0.500pt}{0.800pt}}
\multiput(947.00,421.08)(0.599,-0.514){13}{\rule{1.160pt}{0.124pt}}
\multiput(947.00,421.34)(9.592,-10.000){2}{\rule{0.580pt}{0.800pt}}
\multiput(959.00,411.08)(0.674,-0.516){11}{\rule{1.267pt}{0.124pt}}
\multiput(959.00,411.34)(9.371,-9.000){2}{\rule{0.633pt}{0.800pt}}
\multiput(971.00,402.08)(0.543,-0.514){13}{\rule{1.080pt}{0.124pt}}
\multiput(971.00,402.34)(8.758,-10.000){2}{\rule{0.540pt}{0.800pt}}
\multiput(982.00,392.08)(0.774,-0.520){9}{\rule{1.400pt}{0.125pt}}
\multiput(982.00,392.34)(9.094,-8.000){2}{\rule{0.700pt}{0.800pt}}
\multiput(994.00,384.08)(0.674,-0.516){11}{\rule{1.267pt}{0.124pt}}
\multiput(994.00,384.34)(9.371,-9.000){2}{\rule{0.633pt}{0.800pt}}
\multiput(1006.00,375.08)(0.913,-0.526){7}{\rule{1.571pt}{0.127pt}}
\multiput(1006.00,375.34)(8.738,-7.000){2}{\rule{0.786pt}{0.800pt}}
\multiput(1018.00,368.08)(0.700,-0.520){9}{\rule{1.300pt}{0.125pt}}
\multiput(1018.00,368.34)(8.302,-8.000){2}{\rule{0.650pt}{0.800pt}}
\multiput(1029.00,360.07)(1.132,-0.536){5}{\rule{1.800pt}{0.129pt}}
\multiput(1029.00,360.34)(8.264,-6.000){2}{\rule{0.900pt}{0.800pt}}
\multiput(1041.00,354.07)(1.132,-0.536){5}{\rule{1.800pt}{0.129pt}}
\multiput(1041.00,354.34)(8.264,-6.000){2}{\rule{0.900pt}{0.800pt}}
\multiput(1053.00,348.06)(1.432,-0.560){3}{\rule{1.960pt}{0.135pt}}
\multiput(1053.00,348.34)(6.932,-5.000){2}{\rule{0.980pt}{0.800pt}}
\put(1064,341.34){\rule{2.600pt}{0.800pt}}
\multiput(1064.00,343.34)(6.604,-4.000){2}{\rule{1.300pt}{0.800pt}}
\put(1076,337.34){\rule{2.600pt}{0.800pt}}
\multiput(1076.00,339.34)(6.604,-4.000){2}{\rule{1.300pt}{0.800pt}}
\put(1088,333.84){\rule{2.650pt}{0.800pt}}
\multiput(1088.00,335.34)(5.500,-3.000){2}{\rule{1.325pt}{0.800pt}}
\put(1099,331.84){\rule{2.891pt}{0.800pt}}
\multiput(1099.00,332.34)(6.000,-1.000){2}{\rule{1.445pt}{0.800pt}}
\put(1111,330.84){\rule{2.891pt}{0.800pt}}
\multiput(1111.00,331.34)(6.000,-1.000){2}{\rule{1.445pt}{0.800pt}}
\put(1135,330.84){\rule{2.650pt}{0.800pt}}
\multiput(1135.00,330.34)(5.500,1.000){2}{\rule{1.325pt}{0.800pt}}
\put(1146,332.34){\rule{2.891pt}{0.800pt}}
\multiput(1146.00,331.34)(6.000,2.000){2}{\rule{1.445pt}{0.800pt}}
\put(1158,335.34){\rule{2.600pt}{0.800pt}}
\multiput(1158.00,333.34)(6.604,4.000){2}{\rule{1.300pt}{0.800pt}}
\put(1170,339.34){\rule{2.400pt}{0.800pt}}
\multiput(1170.00,337.34)(6.019,4.000){2}{\rule{1.200pt}{0.800pt}}
\multiput(1181.00,344.39)(1.132,0.536){5}{\rule{1.800pt}{0.129pt}}
\multiput(1181.00,341.34)(8.264,6.000){2}{\rule{0.900pt}{0.800pt}}
\multiput(1193.00,350.40)(0.913,0.526){7}{\rule{1.571pt}{0.127pt}}
\multiput(1193.00,347.34)(8.738,7.000){2}{\rule{0.786pt}{0.800pt}}
\multiput(1205.00,357.40)(0.674,0.516){11}{\rule{1.267pt}{0.124pt}}
\multiput(1205.00,354.34)(9.371,9.000){2}{\rule{0.633pt}{0.800pt}}
\multiput(1217.00,366.40)(0.611,0.516){11}{\rule{1.178pt}{0.124pt}}
\multiput(1217.00,363.34)(8.555,9.000){2}{\rule{0.589pt}{0.800pt}}
\multiput(1228.00,375.41)(0.491,0.511){17}{\rule{1.000pt}{0.123pt}}
\multiput(1228.00,372.34)(9.924,12.000){2}{\rule{0.500pt}{0.800pt}}
\multiput(1240.00,387.41)(0.491,0.511){17}{\rule{1.000pt}{0.123pt}}
\multiput(1240.00,384.34)(9.924,12.000){2}{\rule{0.500pt}{0.800pt}}
\multiput(1253.40,398.00)(0.512,0.639){15}{\rule{0.123pt}{1.218pt}}
\multiput(1250.34,398.00)(11.000,11.472){2}{\rule{0.800pt}{0.609pt}}
\multiput(1264.41,412.00)(0.511,0.671){17}{\rule{0.123pt}{1.267pt}}
\multiput(1261.34,412.00)(12.000,13.371){2}{\rule{0.800pt}{0.633pt}}
\multiput(1276.41,428.00)(0.511,0.717){17}{\rule{0.123pt}{1.333pt}}
\multiput(1273.34,428.00)(12.000,14.233){2}{\rule{0.800pt}{0.667pt}}
\multiput(1288.41,445.00)(0.511,0.807){17}{\rule{0.123pt}{1.467pt}}
\multiput(1285.34,445.00)(12.000,15.956){2}{\rule{0.800pt}{0.733pt}}
\multiput(1300.40,464.00)(0.512,0.938){15}{\rule{0.123pt}{1.655pt}}
\multiput(1297.34,464.00)(11.000,16.566){2}{\rule{0.800pt}{0.827pt}}
\multiput(1311.41,484.00)(0.511,0.943){17}{\rule{0.123pt}{1.667pt}}
\multiput(1308.34,484.00)(12.000,18.541){2}{\rule{0.800pt}{0.833pt}}
\multiput(1323.41,506.00)(0.511,1.033){17}{\rule{0.123pt}{1.800pt}}
\multiput(1320.34,506.00)(12.000,20.264){2}{\rule{0.800pt}{0.900pt}}
\multiput(1335.40,530.00)(0.512,1.237){15}{\rule{0.123pt}{2.091pt}}
\multiput(1332.34,530.00)(11.000,21.660){2}{\rule{0.800pt}{1.045pt}}
\multiput(1346.41,556.00)(0.511,1.169){17}{\rule{0.123pt}{2.000pt}}
\multiput(1343.34,556.00)(12.000,22.849){2}{\rule{0.800pt}{1.000pt}}
\multiput(1358.41,583.00)(0.511,1.304){17}{\rule{0.123pt}{2.200pt}}
\multiput(1355.34,583.00)(12.000,25.434){2}{\rule{0.800pt}{1.100pt}}
\multiput(1370.40,613.00)(0.512,1.486){15}{\rule{0.123pt}{2.455pt}}
\multiput(1367.34,613.00)(11.000,25.905){2}{\rule{0.800pt}{1.227pt}}
\multiput(1381.41,644.00)(0.511,1.440){17}{\rule{0.123pt}{2.400pt}}
\multiput(1378.34,644.00)(12.000,28.019){2}{\rule{0.800pt}{1.200pt}}
\multiput(1393.41,677.00)(0.511,1.530){17}{\rule{0.123pt}{2.533pt}}
\multiput(1390.34,677.00)(12.000,29.742){2}{\rule{0.800pt}{1.267pt}}
\multiput(1405.41,712.00)(0.511,1.666){17}{\rule{0.123pt}{2.733pt}}
\multiput(1402.34,712.00)(12.000,32.327){2}{\rule{0.800pt}{1.367pt}}
\multiput(1417.40,750.00)(0.512,1.885){15}{\rule{0.123pt}{3.036pt}}
\multiput(1414.34,750.00)(11.000,32.698){2}{\rule{0.800pt}{1.518pt}}
\multiput(1428.41,789.00)(0.511,1.847){17}{\rule{0.123pt}{3.000pt}}
\multiput(1425.34,789.00)(12.000,35.773){2}{\rule{0.800pt}{1.500pt}}
\put(1123.0,332.0){\rule[-0.400pt]{2.891pt}{0.800pt}}
\sbox{\plotpoint}{\rule[-0.500pt]{1.000pt}{1.000pt}}%
\put(1213,745){\makebox(0,0)[r]{Level (2, 6) approximation}}
\multiput(1233,745)(20.756,0.000){5}{\usebox{\plotpoint}}
\put(1333,745){\usebox{\plotpoint}}
\put(399,477){\usebox{\plotpoint}}
\multiput(399,477)(7.708,19.271){2}{\usebox{\plotpoint}}
\put(415.53,515.06){\usebox{\plotpoint}}
\put(424.91,533.57){\usebox{\plotpoint}}
\put(434.87,551.78){\usebox{\plotpoint}}
\put(445.52,569.59){\usebox{\plotpoint}}
\put(457.19,586.74){\usebox{\plotpoint}}
\put(470.16,602.94){\usebox{\plotpoint}}
\put(483.09,619.09){\usebox{\plotpoint}}
\put(498.01,633.51){\usebox{\plotpoint}}
\put(514.36,646.27){\usebox{\plotpoint}}
\put(531.74,657.59){\usebox{\plotpoint}}
\put(550.57,666.21){\usebox{\plotpoint}}
\put(570.43,672.11){\usebox{\plotpoint}}
\put(590.95,675.00){\usebox{\plotpoint}}
\put(611.66,674.17){\usebox{\plotpoint}}
\put(632.00,670.25){\usebox{\plotpoint}}
\put(651.89,664.42){\usebox{\plotpoint}}
\put(671.22,656.89){\usebox{\plotpoint}}
\put(689.54,647.17){\usebox{\plotpoint}}
\put(707.14,636.16){\usebox{\plotpoint}}
\put(724.30,624.53){\usebox{\plotpoint}}
\put(740.53,611.60){\usebox{\plotpoint}}
\put(756.59,598.48){\usebox{\plotpoint}}
\put(772.21,584.82){\usebox{\plotpoint}}
\put(787.36,570.64){\usebox{\plotpoint}}
\put(802.52,556.48){\usebox{\plotpoint}}
\put(817.19,541.81){\usebox{\plotpoint}}
\put(831.36,526.64){\usebox{\plotpoint}}
\put(845.91,511.85){\usebox{\plotpoint}}
\put(859.70,496.34){\usebox{\plotpoint}}
\put(873.43,480.77){\usebox{\plotpoint}}
\put(887.21,465.26){\usebox{\plotpoint}}
\put(901.00,449.75){\usebox{\plotpoint}}
\put(914.79,434.23){\usebox{\plotpoint}}
\put(928.58,418.72){\usebox{\plotpoint}}
\put(942.37,403.21){\usebox{\plotpoint}}
\put(956.16,387.69){\usebox{\plotpoint}}
\put(969.95,372.18){\usebox{\plotpoint}}
\put(984.22,357.12){\usebox{\plotpoint}}
\put(997.66,341.34){\usebox{\plotpoint}}
\put(1012.19,326.53){\usebox{\plotpoint}}
\put(1026.53,311.54){\usebox{\plotpoint}}
\put(1041.66,297.34){\usebox{\plotpoint}}
\put(1056.74,283.10){\usebox{\plotpoint}}
\put(1072.36,269.44){\usebox{\plotpoint}}
\put(1088.36,256.23){\usebox{\plotpoint}}
\put(1104.96,243.78){\usebox{\plotpoint}}
\put(1121.72,231.55){\usebox{\plotpoint}}
\put(1139.45,220.77){\usebox{\plotpoint}}
\put(1158.02,211.49){\usebox{\plotpoint}}
\put(1176.94,203.02){\usebox{\plotpoint}}
\put(1197.07,197.98){\usebox{\plotpoint}}
\put(1217.57,194.93){\usebox{\plotpoint}}
\put(1238.23,194.65){\usebox{\plotpoint}}
\put(1258.75,197.69){\usebox{\plotpoint}}
\put(1278.64,203.49){\usebox{\plotpoint}}
\put(1297.20,212.75){\usebox{\plotpoint}}
\put(1314.64,223.98){\usebox{\plotpoint}}
\put(1330.58,237.26){\usebox{\plotpoint}}
\put(1345.49,251.68){\usebox{\plotpoint}}
\put(1359.28,267.19){\usebox{\plotpoint}}
\put(1371.94,283.62){\usebox{\plotpoint}}
\put(1383.55,300.82){\usebox{\plotpoint}}
\multiput(1391,312)(10.878,17.677){2}{\usebox{\plotpoint}}
\put(1415.61,354.15){\usebox{\plotpoint}}
\put(1425.26,372.53){\usebox{\plotpoint}}
\put(1434.38,391.17){\usebox{\plotpoint}}
\put(1439,401){\usebox{\plotpoint}}
\sbox{\plotpoint}{\rule[-0.200pt]{0.400pt}{0.400pt}}%
\put(1213,704){\makebox(0,0)[r]{Exact value of $-T_{25}$}}
\multiput(1233,704)(20.756,0.000){5}{\usebox{\plotpoint}}
\put(1333,704){\usebox{\plotpoint}}
\put(280,175){\usebox{\plotpoint}}
\put(280.00,175.00){\usebox{\plotpoint}}
\put(300.76,175.00){\usebox{\plotpoint}}
\put(321.51,175.00){\usebox{\plotpoint}}
\put(342.27,175.00){\usebox{\plotpoint}}
\put(363.02,175.00){\usebox{\plotpoint}}
\put(383.78,175.00){\usebox{\plotpoint}}
\put(404.53,175.00){\usebox{\plotpoint}}
\put(425.29,175.00){\usebox{\plotpoint}}
\put(446.04,175.00){\usebox{\plotpoint}}
\put(466.80,175.00){\usebox{\plotpoint}}
\put(487.55,175.00){\usebox{\plotpoint}}
\put(508.31,175.00){\usebox{\plotpoint}}
\put(529.07,175.00){\usebox{\plotpoint}}
\put(549.82,175.00){\usebox{\plotpoint}}
\put(570.58,175.00){\usebox{\plotpoint}}
\put(591.33,175.00){\usebox{\plotpoint}}
\put(612.09,175.00){\usebox{\plotpoint}}
\put(632.84,175.00){\usebox{\plotpoint}}
\put(653.60,175.00){\usebox{\plotpoint}}
\put(674.35,175.00){\usebox{\plotpoint}}
\put(695.11,175.00){\usebox{\plotpoint}}
\put(715.87,175.00){\usebox{\plotpoint}}
\put(736.62,175.00){\usebox{\plotpoint}}
\put(757.38,175.00){\usebox{\plotpoint}}
\put(778.13,175.00){\usebox{\plotpoint}}
\put(798.89,175.00){\usebox{\plotpoint}}
\put(819.64,175.00){\usebox{\plotpoint}}
\put(840.40,175.00){\usebox{\plotpoint}}
\put(861.15,175.00){\usebox{\plotpoint}}
\put(881.91,175.00){\usebox{\plotpoint}}
\put(902.66,175.00){\usebox{\plotpoint}}
\put(923.42,175.00){\usebox{\plotpoint}}
\put(944.18,175.00){\usebox{\plotpoint}}
\put(964.93,175.00){\usebox{\plotpoint}}
\put(985.69,175.00){\usebox{\plotpoint}}
\put(1006.44,175.00){\usebox{\plotpoint}}
\put(1027.20,175.00){\usebox{\plotpoint}}
\put(1047.95,175.00){\usebox{\plotpoint}}
\put(1068.71,175.00){\usebox{\plotpoint}}
\put(1089.46,175.00){\usebox{\plotpoint}}
\put(1110.22,175.00){\usebox{\plotpoint}}
\put(1130.98,175.00){\usebox{\plotpoint}}
\put(1151.73,175.00){\usebox{\plotpoint}}
\put(1172.49,175.00){\usebox{\plotpoint}}
\put(1193.24,175.00){\usebox{\plotpoint}}
\put(1214.00,175.00){\usebox{\plotpoint}}
\put(1234.75,175.00){\usebox{\plotpoint}}
\put(1255.51,175.00){\usebox{\plotpoint}}
\put(1276.26,175.00){\usebox{\plotpoint}}
\put(1297.02,175.00){\usebox{\plotpoint}}
\put(1317.77,175.00){\usebox{\plotpoint}}
\put(1338.53,175.00){\usebox{\plotpoint}}
\put(1359.29,175.00){\usebox{\plotpoint}}
\put(1380.04,175.00){\usebox{\plotpoint}}
\put(1400.80,175.00){\usebox{\plotpoint}}
\put(1421.55,175.00){\usebox{\plotpoint}}
\put(1439,175){\usebox{\plotpoint}}
\sbox{\plotpoint}{\rule[-0.500pt]{1.000pt}{1.000pt}}%
\put(1127,332){\raisebox{-.8pt}{\circle{15}}}
\put(1229,194){\raisebox{-.8pt}{\circle{15}}}
\put(396,471){\raisebox{-.8pt}{\circle*{15}}}
\end{picture}
\caption[x]{\footnotesize The effective tachyon potential in level (0,
              0) and (2, 6) truncations.  The open circles denote minima in each
              level truncation.  The filled circle denotes a branch point where
              the level (2, 6) truncation approximation reaches the limit of
              Feynman-Siegel gauge validity.}
\label{f:potential}
\end{figure}

The string field theory equation of motion is
\begin{equation}
Q \Psi + g \Psi \star \Psi = 0 \,.
\label{eq:SFT-EOM}
\end{equation}
Despite much work over the last few years, there is still no analytic
solution of this equation of motion\footnote{as of October, 2003}.
There is, however, a systematic approximation scheme, known as level
truncation, which can be used to solve this equation 
numerically~\cite{ks-open}.
The level $(L, I) $ truncation of the full string field theory
involves dropping all
fields at level $N > L$, and disregarding any interaction term between
three fields whose levels add up to a number that
        is greater than $I$.  For
example, the simplest truncation of the theory is the level (0, 0)
truncation.  This
is the truncation which was used in section \ref{firsttestofit}.
Including only
the zero-momentum component of the tachyon field, since
we are looking for a Lorentz-invariant vacuum, the truncated theory is
simply described by a potential for the tachyon zero-mode
\begin{equation}
V (\phi) = -\frac{1}{2}\phi^2 + g \bar{\kappa} \phi^3 \,.
\end{equation}
where $\bar{\kappa} = \kappa/3 = 3^{7/2}/2^6$.  This cubic function,
which was computed in (\ref{eq:CFT-cubic-potential}) using the CFT
approach, and in (\ref{eq:expanded-action}) using the oscillator
approach, is graphed in Figure~\ref{f:potential}.
As discussed in section \ref{firsttestofit}, this potential has a
local minimum~at
\begin{equation}
\phi_0 = \frac{1}{3g \bar{\kappa}}  \,,
\end{equation}
and at this point the potential is
\begin{equation}
V (\phi_0) = -\frac{1}{54}  \frac{1}{g^2 \bar{\kappa}^2}  =
  -\frac{2^{11}}{3^{10}}  \frac{1}{g^2}
\approx (0.68) \left( -\frac{1}{2 \pi^2 g^2}  \right) \,.
\end{equation}
Thus, simply including the tachyon zero-mode gives a
nontrivial vacuum with $68\%$ of the vacuum energy density predicted
by Sen.  This vacuum is denoted by an open circle in Figure~\ref{f:potential}.

At higher levels of truncation, there are a multitude of fields with
various tensor structures.  However, again assuming that we are
looking for a vacuum which preserves Lorentz symmetry, we can restrict
attention to the interactions between scalar fields at zero momentum.  We
will work in Feynman-Siegel gauge to simplify calculations; as shown
in the previous subsection, this gauge is good at least in a local
neighborhood of the point where all fields vanish.  The situation is
further simplified by the existence of the twist symmetry,
which as mentioned in the previous subsection guarantees that no cubic
vertex between (zero-momentum) scalar fields can connect three fields with a
total level which is odd, and thus means that odd fields are not
relevant to diagrams with only external tachyons at tree level.
Therefore, we need only consider even-level scalar fields in looking
for Lorentz-preserving solutions to the SFT equations of motion.  With
these simplifications, in a general level truncation the string field
is simply expressed as a sum of a finite number of terms
\begin{equation}
\Psi_{\rm s} = \sum_{i} \phi_i | s_i \rangle
\label{eq:scalar-expansion}
\end{equation}
where $\phi_i$ are the zero-modes of the scalar fields associated with
even-level states $|s_i \rangle$.  As discussed in the previous
subsection, this set of scalar fields can be further restricted to be
SU(1, 1) singlets in the universal subspace ${\mathcal H}_1$.
For example, including fields up to
level 2, we have
\begin{equation}
\Psi_{{\rm s}} = \phi| 0_1 \rangle+
B \; (\alpha_{-1} \cdot \alpha_{-1})| 0_1 \rangle+
\beta \; b_{-1} c_{-1}| 0_1 \rangle\,.
\end{equation}
In terms of the matter Virasoro generators, the state associated with
the field $B$ is
\begin{equation}
(\alpha_{-1} \cdot \alpha_{-1}) | 0_1 \rangle
= 2L_{-2}| 0_1 \rangle\,,
\end{equation}
which lies in the universal subspace ${\mathcal H}_1$.
The potential for all the scalars appearing in the level-truncated
expansion (\ref{eq:scalar-expansion}) can be simply expressed as a
cubic polynomial in the zero-modes of the scalar fields
\begin{equation}
V = \sum_{i, j}d_{ij} \phi_i \phi_j + g \bar{\kappa}
\sum_{i, j, k}t_{ijk} \phi_i \phi_j \phi_k \,.
\label{eq:scalar-potential}
\end{equation}
Using the expressions for the Neumann coefficients given in Section
5.3, the potential for all the scalar fields up to level $L$ can be
computed in a level  $(L, I)$ truncation.  For example, the potential
in the level $(2, 6)$ truncation is given~by
\begin{eqnarray}
V & = &  -\frac{1}{2}\phi^2 + 26  B^2 -\frac{1}{2}\beta^2
\label{eq:v26}\\
             &  &+ \bar{\kappa} g\left[
\phi^3  -\frac{130}{9} \phi^2 B -\frac{11}{9}  \phi^2 \beta
+ \frac{30212}{243}  \phi B^2
+ \frac{2860}{243}  \phi B \beta
+ \frac{19}{81}  \phi \beta^2
             \right.  \nonumber \\
& &\hspace*{0.4in}\left.
  -\frac{2178904}{6561}  B^3
  -\frac{332332}{6561}  B^2 \beta
  -\frac{2470}{2187}  B \beta^2
  -\frac{1}{81}  \beta^3
             \right] \nonumber
\end{eqnarray}
As an example of how these terms arise, consider the $\phi^2 B$ term.
The coefficient in this term is given by
\begin{equation}g\, \langle V_3 | \, \bigl(| 0_1 \rangle \otimes |
0_1 \rangle \otimes
\alpha_{-1} \cdot \alpha_{-1} | 0_1 \rangle\bigr)  =
  -g \bar{\kappa} \;  (3 \cdot 26) \; V^{11}_{11}
              =  -g \bar{\kappa} \frac{130}{9} \,,
\end{equation}where we have used $V^{11}_{11} = 5/27$.

In the level (2, 6) truncation of the theory, the nontrivial vacuum
is found by simultaneously solving the three quadratic equations found by
setting to zero the derivatives of the potential (\ref{eq:v26}) with respect to
$\phi, B,$ and $\beta$.  There are a number of different solutions to
these equations, but only one is in the vicinity of $\phi = 1/3g
\bar{\kappa}$.  The solution of interest is
\begin{eqnarray}
\phi  \approx  0.39766\, \frac{1}{g \bar{\kappa}}  \,,\quad
B    \approx 0.02045\, \frac{1}{g \bar{\kappa}}\, ,\quad
\beta   \approx  -0.13897\, \frac{1}{g \bar{\kappa}} \,.
\end{eqnarray}
Plugging these values into the potential gives
\begin{equation}
E_{(2, 6)} = -0.95938 \,T_{25} \,,
\end{equation}
or 95.9\% of the result predicted by Sen.
This vacuum is denoted by an open circle in Figure~\ref{f:potential}.

It is a straightforward,  computationally intensive project
to generalize this calculation to higher levels of truncation.  This
calculation was carried out to level (4, 8) by Kostelecky and
Samuel~\cite{ks-open} many years ago.
They noted that the vacuum seemed to be converging, but they lacked any
physical picture to give meaning to this vacuum.
Following Sen's conjectures, the
level (4, 8) calculation was done again using somewhat different methods by Sen
and Zwiebach~\cite{Sen-Zwiebach}, who showed that the energy at this level is
$-0.986 \;T_{25}$.  The calculation was automated by Moeller and
Taylor~\cite{Moeller-Taylor}, who calculated up to level (10, 20),
where there are
252 scalar fields, including all even-level scalar fields up to level 10; this
computation was done using oscillators, without restriction to the universal
subspace.  Up to this level, the vacuum energy converges monotonically, as
shown in Table 1.
\begin{table}[htp]
\begin{center}
\begin{tabular}{|| c  || c |  c ||}
\hline
\hline
level & $ g \bar{\kappa}\langle \phi \rangle$ & $V/T_{25}$\\
\hline
\hline
(0, 0) & 0.3333 & -0.68462\\
\hline
(2, 4) & 0.3957 &  -0.94855\\
(2, 6) & 0.3977 &  -0.95938\\
\hline
(4, 8) & 0.4005 &  -0.98640\\
(4, 12) & 0.4007 & -0.98782\\
\hline
(6, 12) & 0.4004 & -0.99514\\
(6, 18) & 0.4004 & -0.99518\\
\hline
(8, 16) & 0.3999 & -0.99777\\
(8, 20) & 0.3997 & -0.99793\\
\hline
(10, 20) & 0.3992 & -0.99912\\
\hline
\hline
\end{tabular}
\caption[x]{\footnotesize Tachyon VEV and vacuum energy in stable
vacua of level-truncated theory}
\label{t:vacuum}
\end{center}
\end{table}
These numerical calculations indicate that level truncation of string
field theory leads to a good systematic approximation scheme for
computing the nonperturbative tachyon vacuum.  It is also worth noting
that in these computations, level $(L, 2L)$ and $(L, 3L)$
approximations give fairly similar values.

The preceding results were the best values for the vacuum energy at
the time of these original lectures.  More recently, Gaiotto and
Rastelli reported further numerical
results~\cite{Gaiotto-Rastelli-strings,gr-analysis}.  By programming in
C++ instead of mathematica, and by computing using matter Virasoro
operators rather than oscillators, so that only fields in the
universal subspace ${\mathcal H}_1$ were included, they were able to
extend the computation to level (18, 54).  Their results are shown in
Table~\ref{t:vacuum-gr}
\begin{table}[htp]
\begin{center}
\begin{tabular}{|| c |  c ||}
\hline
\hline
level & $V/T_{25}$\\
\hline
\hline
(12, 24) & 0.99979 \\
(12, 36) & 0.99982 \\
\hline
(14, 28) & 1.00016 \\
(14, 42) & 1.00017 \\
\hline
(16, 32) & 1.00037 \\
(16, 48) & 1.00038 \\
\hline
(18, 36) & 1.00049 \\
(18, 54) & 1.00049 \\
\hline
\hline
\end{tabular}
\caption[x]{\footnotesize  Vacuum energy in stable
vacua of level-truncated theory}
\label{t:vacuum-gr}
\end{center}
\end{table}
These results were rather surprising, indicating that
while the energy monotonically approaches $-T_{25}$ up to level 12, at
level (14, 42) the energy drops below $-T_{25}$, and that the
energy continues to decrease, reaching $-1.00049\, T_{25}$ at level
(18, 54).  We will discuss the resolution of this unexpected overshoot
shortly.

First, however, it is interesting to consider the tachyon condensation
problem from the point of view of the effective tachyon potential.  If
instead of trying to solve the quadratic equations for all $N$ of the
fields appearing in (\ref{eq:scalar-potential}), we instead fix the
tachyon field $\phi$ and solve the quadratic equations for the
remaining $N -1$ fields, we can determine a effective potential $V
(\phi)$ for the tachyon field.  This has been done numerically up to
level (16, 48)~\cite{Moeller-Taylor,gr-analysis}.  At each level, the
tachyon effective potential smoothly interpolates between the
perturbative vacuum and the nonperturbative vacuum near $\phi = 0.4/g
\bar{\kappa}$.  For example, the tachyon effective potential at level
(2, 6) is graphed in Figure~\ref{f:potential}.  In all level
truncations other than (0, 0) and (2, 4) (at least up to level (10,
20)), the tachyon effective potential has two branch point
singularities at which the continuous solution for the other fields
breaks down; for the level (2, 6) truncation, these branch points
occur at $\phi \approx -0.127/g \bar{\kappa}$ and $\phi \approx
2.293/g \bar{\kappa}$; the lower branch point is denoted by a solid
circle in Figure~\ref{f:potential}.  As a result of these branch
points, the tachyon effective potential is only valid for a finite
range of $\phi$, ranging between approximately $-0.1/g \bar{\kappa}$
and $ 0.6/g \bar{\kappa}$.  In Section \ref{sec:gauge} we review
results which indicate that these branch points arise because the
trajectory in field space associated with this potential encounters
the boundary of the region of Feynman-Siegel gauge validity.  It seems
almost to be a fortunate accident that the nonperturbative vacuum lies
within the region of validity of this gauge choice.  It is worth
mentioning again here that in the BSFT approach, the tachyon potential
can be computed exactly~\cite{BSFT}.  In this formulation, there is no
branch point in the effective potential, which is unbounded below for
negative values of the tachyon.  On the other hand, the nontrivial
vacuum in the background-independent approach arises only as the
tachyon field goes to infinity, so it is harder to study the physics
of the stable vacuum from this point of view.

Another interesting perspective on the tachyon effective potential is
found by performing a perturbative (but off-shell)
computation of the
coefficients in the tachyon effective potential in the level-truncated theory.
This gives a power series expansion of the effective  potential
\begin{eqnarray}
V (\phi)  & = & \sum_{n = 2}^{ \infty}  c_n (\bar{\kappa} g)^{n-2} \phi^n
\label{eq:v}\\
& = & -\frac{1}{2}\phi^2 + (\bar{\kappa} g) \phi^3 + c_4 (\bar{\kappa} g)^2
\phi^4 + c_5 (\bar{\kappa} g)^3 \phi^5 +\cdots\nonumber
\end{eqnarray}
The coefficients up to $c_{60}$ have been computed in the level
truncations up to (10, 20)~\cite{Moeller-Taylor}.  Because of the
branch point singularity near $\phi = -0.1/g \bar{\kappa}$, this
series has a radius of convergence much smaller than the value of
$\phi$ at the nonperturbative vacuum.  Thus, the energy at the stable
vacuum lies outside the naive range of the potential defined by the
perturbative expansion.

Now, let us return to the problem of the overshoot in energy below
$-T_{25}$ found at level 14 by Gaiotto and Rastelli\footnote{The
material in the remainder of this section was developed only after the
original TASI lectures in 2001, but is included because of its
relevance to the main development in this section.}.  The most
straightforward way of determining whether or not this represents a
real problem for string field theory would be to simply continue the
calculation to higher levels.  Unfortunately, at present this is not
tractable, as the difficulty of computation grows exponentially in the
level.  Thus, we must resort to more indirect methods.  It was found
empirically by Taylor~\cite{WT-perturbative} that the level $L$
approximations of string field theory give on-shell and off-shell
amplitudes with error of order $1/L$.  This work and further
evidence~\cite{Coletti-Sigalov-Taylor,Beccaria-Rampino} indicates that
amplitudes can be very accurately approximated by computing them in
different level $L$ truncations, and matching to a power series in
$1/L$.  Such an approach can be taken to determine highly accurate
values for the coefficients $c_n$ in (\ref{eq:v}).  As noted above,
the resulting power series has a finite radius of convergence, and the
stable vacuum lies beyond this limit.  There is a standard technique,
however, known as the method of Pad\'e approximants, which allows one
to extrapolate a function beyond its naive radius of convergence, if
the function is sufficiently well-behaved in the direction in which it
is extrapolated.  The idea of Pad\'e approximants is to replace a
power series having given coefficients for a fixed number of terms
with a rational function with the same number of coefficients,
choosing the coefficients of the rational function to give a power
series which agrees with the fixed coefficients in the original power
series.  For example, consider the first three terms in the level $L
=2 $
approximation to the tachyon
effective potential (\ref{eq:v}),
\begin{equation}
  -\frac{1}{2}\phi^2 + \kappa g \phi^3-\frac{34}{27}  (\kappa g) \phi^4 \,.
           \label{eq:truncated}
\end{equation}
This truncated expansion
has no local minima, while the Pad\'e approximant
\begin{equation}
P^3_1 (\phi) =\frac{-\frac{1}{2}\phi^2 + \frac{10}{27} \kappa g \phi^3
}{1 + \frac{34}{27}  \kappa g \phi}
\end{equation}
does; this approximant thus represents a better description of the
tachyon potential than the truncated expression (\ref{eq:truncated}).
The advantage of Pad\'e approximants is that they allow one to
incorporate poles into approximations of a function with a desired
local power series behavior.  For a wide class of functions,
successive Pad\'e approximants converge exponentially quickly in the
region where the function is smooth.  Empirically, this seems to be
the case for the tachyon effective potential.  Thus, the energy
minimum at any finite level of truncation can be determined to an
arbitrary degree of accuracy from the leading coefficients in the
potential.  For example, the energy can be computed to 10 digits of
accuracy by including approximately 40 coefficients $c_n$; this
calculation is, however, highly sensitive to the accuracy of the
coefficients~\cite{Ellis-Karliner-WT}.

\begin{figure}[!ht]
\leavevmode
\begin{center}
\epsfxsize = 12 cm \epsfbox{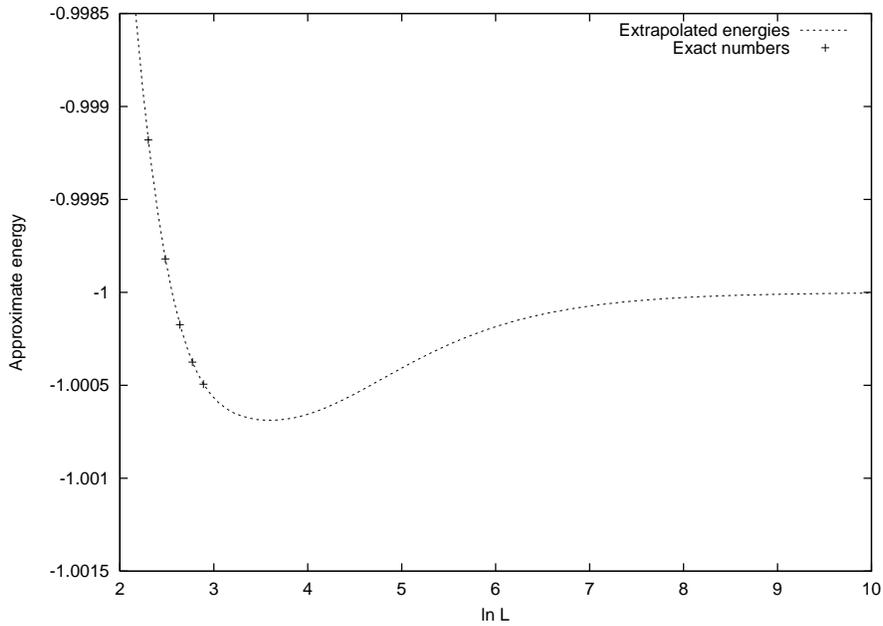}
\end{center}
\caption[x]{\footnotesize
Expected approximations to the vacuum energy at different
            levels of truncation, extrapolated from data at lower levels of
truncation.}
\label{f:e-l}
\end{figure}

Combining Pad\'e approximants with approximations to the coefficients
$c_n$, computed by matching level-truncated results in a $1/L$
expansion, it is possible to predict not only the exact value of the
energy at the stable minimum as $L \rightarrow \infty$, but also to
predict the values of the approximate energy at intermediate values of
$L$.  Such a computation was performed using the level approximated
values of $c_n$ up to level $ (10, 20)$~\cite{WT-Pade}.  By first using
these values to predict the level-approximated values at higher
levels, and then inserting these values into Pad\'e approximants, the
overshoot phenomenon found by Gaiotto and Rastelli was accurately
reproduced.  For example, compared to the value -1.0003678 found by
these authors at level (16, 32), extrapolation from results at levels
$(L, 2L)$ up to (10, 20) gives a predicted value of -1.0003773 at
level (16, 32).  Furthermore, the extrapolated values $E_L$ of the
energy at the stable minimum were found to decrease up to
approximately level 26, and then to increase, approaching an
asymptotic value as $L \rightarrow \infty$ of $E_\infty \approx -1$
with error $\sim 10^{-4}$.  These results suggested that the energy at
the minimum in the level-truncated theory takes the form shown in
Figure~\ref{f:e-l}.

In the calculation just described, there were two sources of error:
1) the coefficients $c_n$ had some numerical inaccuracy, and 2)
there is some error introduced in extrapolating from low
levels of truncation.

This computation was improved by Gaiotto and
Rastelli~\cite{gr-analysis}.  These authors used a different approach:
instead of extrapolating the finite $L$ results for the coefficients
$c_n$, they extrapolated the nonperturbatively computed effective
potential $V (\phi)$ at various values of $\phi$.  Because Pad\'e
approximants are so accurate, for exactly known values of $c_n$ and $V
(\phi)$ this approach is equivalent to the combined
Pad\'e-extrapolation in $c_n$ approach, but generally this approach
trades inaccuracy in $c_n$ for inaccuracy in $V (\phi)$.  In practice,
it is much easier to compute the coefficients $c_n$ exactly than the
nonperturbative effective potential $V (\phi)$, which
requires numerically solving a large system of quadratic equations.
Gaiotto and Rastelli were able, however, to
use their results on $V (\phi)$ at higher levels, which greatly
increased the accuracy of their extrapolations.  They found that
while level $(L, 2L)$ and $(L, 3L)$
approximations tend to be very similar, extrapolations based on level
$(L, 3L)$ truncations seem more robust.  Using data up to
level (16, 48) they found an extrapolated value of $E_\infty\approx
  -1.00003$, differing from $-1$ by an order of magnitude less than the
value of the energy estimated at level  28, where the overshoot is
predicted to be maximal.   This gives compelling support to the
conclusion that the level-truncated approximations to the energy
indeed behave as shown in Figure~\ref{f:e-l}, and approach the value
predicted by Sen as $L \rightarrow \infty$.


\subsection{Gauge fixing}
\label{sec:gauge}

In this subsection we discuss some aspects of the Feynman-Siegel gauge
choice used in most explicit calculations in OSFT to date.  Let us
restrict attention to the zero momentum action for even-level scalar
fields.  This action is invariant under (\ref{eq:SFT-gauge}) with a
general gauge parameter of the form
\begin{equation}\Lambda  =   \sum \mu^a | s_a \rangle
=  \mu_1 b_{-2} | 0_1 \rangle +
\cdots \,.
\end{equation}The ghost number zero states $| s_a \rangle$  are annihilated
by $b_0$, so they do not contain
$c_0$.  The
variation of a general zero-momentum scalar field takes the form
\begin{equation}
\delta \phi_i = D^{ia} \mu_a + g \bar{\kappa}\, T^{ija} \phi_j \mu_a  \,.
\end{equation}
At $\phi_i = 0$, we have the linear
variation $\delta \phi_i = D^{ia}
\mu_a$.  Let $\phi_q$ denote fields
associated with ghost number one states that contain a $c_0$.  For example, at
level two there is a field $\eta$ associated with the state $c_0b_{-2} | 0_1
\rangle$.
At each level, the number of fields $\phi_q$ is clearly
        equal to the number of gauge parameters $\mu_a$; the
corresponding states are simply related by removing
        or replacing the $c_0$.
From
the formula for $Q_B = c_0L_0 + \cdots$, it is easy to verify that
$D^{qa}$ is a linear one-to-one map at each level, so
\begin{equation}
\det D^{qa} \neq 0
\end{equation}
holds at each level.  This is why the Feynman-Siegel gauge, which sets
$\phi_q=0$ at each level
(and which limits us to gauge parameters
associated with states without a $c_0$), is a good gauge choice near $\phi_i
= 0$, as shown in subsection~\ref{sec:constraints-symmetries}.

Let us now consider the gauge transformations at a general point in
field space $\langle\phi_i\rangle$.  We have
\begin{equation}
\delta \phi^i = M^{ia} \mu_a
\end{equation}
where
\begin{equation}
M^{ia} = D^{ia} + g \bar{\kappa} T^{ija} \langle\phi_j\rangle \,.
\end{equation}
Feynman-Siegel gauge breaks down whenever the determinant of this
matrix vanishes
\begin{equation}
\det M^{qa} =0 \,.
\end{equation}
This condition defines a region in field space within which
Feynman-Siegel gauge is valid.  At the boundary of this region, some gauge
transformations give field variations which are tangent to the
Feynman-Siegel gauge-fixed hypersurface.  Some gauge orbits which
cross the Feynman-Siegel gauge surface inside this region will cross
again outside the region, giving a form of Gribov ambiguity.
Furthermore, some gauge orbits never encounter the region of gauge
validity.  Thus, Feynman-Siegel gauge is really only locally valid.

We can study the region of Feynman-Siegel gauge validity in level
truncation, using finite matrices $M^{qa}$.  It is instructive to
consider a simple example of the breakdown of this gauge choice.
Consider dropping all fields other than the tachyon $\phi = \phi_1$
and the field $\eta= \phi_4$.  The gauge transformation rules then
become
\begin{eqnarray}
\delta \phi & = & \mu g \bar{\kappa} \left[ -\frac{16}{9} \phi +
            \frac{128}{81} \eta \right]   \\
\delta \eta & = &  -\mu + \mu g \bar{\kappa} \left[ -\frac{224}{81}
            \phi + \frac{1792}{729}  \eta \right] \,.\nonumber
\end{eqnarray}
In this simple model, $M$ is a one-by-one matrix,
\begin{equation}
M = -\mu (1 + g \bar{\kappa} \frac{224}{ 81} \phi) \,.
\end{equation}
The gauge choice $\eta = 0$ breaks down when $\eta = \delta \eta = 0$
which occurs when
\begin{equation}
\phi = -\frac{1}{g\bar{\kappa}}  \frac{81}{224}   \,.
\end{equation}
It is easy to see that smaller values of $\phi$ are gauge-equivalent
to values of $\phi$ above this boundary value, while some gauge orbits
never intersect the line $\eta = 0$.

The complete action including all  even level (zero momentum)
scalar fields and gauge invariances has been computed up to level (8,
16)~\cite{Ellwood-Taylor-gauge}. One result of this computation is that the
Feynman-Siegel gauge boundary condition $\det M^{qa} = 0$ seems to be very
stable near the origin as the level of truncation is increased.  This
gives some
confidence that there is a well-defined finite region in field space where
Feynman-Siegel gauge is valid, and that the boundary of this region can be
arbitrarily well approximated by level-truncation calculations.  Another
interesting result which can be seen from these calculations is that (to the
precision possible in the level-truncated analysis) the branch points in the
tachyon effective potential arise precisely at those points where the
trajectory in field space associated with the effective potential
crosses the Feynman-Siegel gauge boundary.  Thus, these branch points
are gauge artifacts.  As mentioned previously, the tachyon effective
potential computed from boundary string field theory does not suffer
from such branch point problems.

It would be very desirable, however,
to have an approach which enables one to describe
the full string field space, including configurations which do not have
gauge representatives in the local region of Feynman-Siegel gauge
validity.   Other
gauge choices can be made, but those which have been explored to date
are only minor variations on the Feynman-Siegel gauge choice, and do
not lead to qualitatively different results.  One might have hoped to
isolate the true vacuum without gauge fixing at all, given that level
truncation breaks the gauge symmetry and thus allows a discrete set of
solutions at any level.  This approach, however, is not particularly promising:
the solutions found at each level lie at very different
places on the gauge orbit, and do not approach any natural limit.
Nonetheless, it seems of paramount importance to find some method for
exploring the full field space of the theory.  Currently inaccessible
regions of the field space may contain solutions that have not yet
been found (see subsection~\ref{subsec:thebackgroundsosft}).


\subsection{Lower-dimensional D-branes as solitons}
\label{sec:solitons}

One aspect of the Sen conjectures (item (2) in the list of
section~\ref{conjectures}),
proposes that lower-dimensional D-branes can be viewed as solitons
of the D25-brane string field theory.
The solitons
involve
profiles for the tachyon field which arise because the tachyon
potential is non-trivial.
The tachyon solitons  are lumps, as opposed to kinks, which appear
in superstring field theory solitons.

In this section we will discuss the basic ideas required to
test this conjecture.  We will follow the approach of
M\"oeller, Sen, and Zwiebach~\cite{Moeller-sz} (other attempts~\cite{djmt}
do not use level expansion). In
order to be able to use a level expansion we curl up one spatial coordinate $x$
into a circle of radius $R$ (the corresponding string coordinate is
called $X$).
We will work with $R>1$.
Along this direction, we will wrap a D1-brane.  We will then
consider the possibility that a certain process of tachyon condensation
results in
the D1-brane becoming a D0-brane.  Our use of D1- and D0-branes
is   just a matter of notational ease. Additional D-brane dimensions
could be included.

Recall that the mass of the D1-brane can be written in the form
\begin{equation}
\label{md1wvvwsmdo}
M_{D1} = 2\pi R T_1  = {1\over 2 \pi^2 g^2}\,
\end{equation}
where $g$ is the coupling constant of the open string field theory
that describes the D1-brane:
\begin{equation}
S = -{1\over g^2} \Bigl(  {1\over 2} \langle \Phi, Q \Phi\rangle + {1\over 3}
\langle \Phi,  \Phi * \Phi \rangle \Bigr)  \equiv -{1\over g^2 } \mathcal{V}
(\Phi) \,.
\end{equation}
A few remarks are in order.
In the above string action we have
included into the string coupling factor $(1/g^2)$ the volume
$(2\pi R)$ of the
compact circle where the D1-brane is wrapped.  By doing so, we can still use
a CFT overlap with unit normalization, and the right-hand side
in (\ref{md1wvvwsmdo}) gives the total mass of the brane.    The zero
string field here is supposed to describe the vacuum with a D1-brane stretched
around the circle.  For time-independent string fields (the kind of fields we
consider here),
$\mathcal{V}(\Phi)$  is a potential.  More precisely the potential
energy P.E. associated with a string field is
\begin{equation}
\hbox{P.E.} = - S(\Phi) = {1\over g^2} \mathcal{V} (\Phi)
= (2\pi R T_1) \, 2\pi^2  \, \mathcal{V}(\Phi),
\end{equation}
where we used (\ref{md1wvvwsmdo}).
This potential energy is really the potential energy of field configurations
measured with respect to the D1-brane  background.  Therefore, the total
energy  $E_{tot}$ of the configuration is obtained by adding the energy
of the D1-brane to the above P.E.  We find
\begin{equation}
\label{totenergyfs}
E_{tot}(\Phi) =  (2\pi R T_1) \Bigl( 1 +  2\pi^2  \, \mathcal{V}(\Phi) \Bigr).
\end{equation}
Since we will use the level expansion to
investigate if a D0-brane can be
represented as a lump solution, it is reasonable to
use the level expansion to calculate the mass of the D1-brane, as well.
So, we re-express the energy of the D1-brane in (\ref{totenergyfs}) in terms of
the string field potential at the vacuum.   Let $\Phi
= T_{vac}$ denote the string field of the D1-brane SFT that represents the
tachyon vacuum. Then, we have
$-1 = 2\pi^2 \mathcal{V}(T_{vac})$, and we can rewrite
\begin{equation}
\label{theenergycalcualtesjnvdkj}
E_{tot}(\Phi) =  (2\pi R T_1) \Bigl( 2\pi^2  \, \mathcal{V}(\Phi)-2\pi^2  \,
\mathcal{V}(T_{vac}) \Bigr).
\end{equation}
Indeed, this formula works correctly:  when $\Phi=0$ the total energy
equals the mass of the D1-brane, and when $\Phi = T_{vac}$ the energy
is zero (since the D1-brane has disappeared).

Let $T_{lump}$ denote the lump (string field) solution, which
is expected to represent the D0-brane in the field theory of the
D1-brane.
The energy of the lump solution is obtained from
(\ref{theenergycalcualtesjnvdkj}) for $\Phi = T_{lump}$:
\begin{equation}
\label{theenerglumpcualtesjnvdkj}
E_{lump} = E_{tot}(T_{lump} ) =  (2\pi R T_1) \Bigl( 2\pi^2  \,
\mathcal{V}(T_{lump})-2\pi^2
\, \mathcal{V}(T_{vac}) \Bigr).
\end{equation}
            The tensions
$T_0$ and
$T_1$ of the D0- and the D1-branes are related by $T_0 = 2\pi T_1$
(the D0-brane tension is the D0-brane energy).
We can therefore form the ratio $r(R) $ of the lump energy and the
D0-brane energy
\begin{equation}
\label{predsecconjecture2}
r (R) = {E_{lump}\over T_0} =
             R \Bigl( 2\pi^2  \,
\mathcal{V}(T_{lump})-2\pi^2
\, \mathcal{V}(T_{vac}) \Bigr).
\end{equation}
In the exact solution (or at infinite level), the ratio $r(R)$ should be equal
to one.  This is the content of the second tachyon conjecture.  At
any finite level
$r(R)$ is some slowly varying function of $R$.  Testing
the conjecture for $R \to 1$ is quite difficult, and one must go
to very high level in the computation.  Testing the conjecture for
$R$ very large is also laborious, since many terms enter into
any finite-level expansion.  So, in practice, one chooses some reasonable
value of $R$ (the value $R=\sqrt{3}$ is convenient) and calculates
to a fixed level.

Before reviewing some of the results obtained, let's do
the simplest computation explicitly.
We consider a tachyon field $T(x)$ which is
expanded as
\begin{equation}
T(x) = t_0 + \sum_{n=1}^\infty  t_n \cos (nx/R) \,.
\end{equation}
The corresponding string field is written as
\begin{eqnarray}
|T\rangle &&= t_0 c_1 |0\rangle + \sum_{n=1}^\infty {1\over 2} t_n
\Bigl( e^{inX(0)/R} + e^{-inX(0)/R} \Bigr) \, c_1|0\rangle \,, \nonumber\\
&&= t_0 c_1 |0\rangle + \sum_{n=1}^\infty  {1\over 2}\,  t_n  \Bigl(
c_1|n/R\rangle  + c_1|-n/R\rangle\Bigr) \,.
\end{eqnarray}
We now evaluate the string action, keeping  $t_0$ and the first tachyon
harmonic~$t_1$:
\begin{equation}
|T\rangle = t_0
c_1 |0\rangle +  {1\over 2}\,  t_1  \Bigl( c_1|1/R\rangle  +
c_1|-1/R\rangle\Bigr)
\,.
\end{equation}
Consider first the contribution of $t_1$ to the kinetic term
\begin{eqnarray}
{1\over 2} \langle T, QT\rangle \Bigl|_{t_1} &&=
{1\over 2} {t_1\over 2} \cdot {t_1\over 2}  \Bigl( \langle 1/R|
+ \langle -1/R|\Bigr)  c_{-1} c_0 L_0 c_1 \Bigl( |1/R\rangle +|-1/R\rangle
\Bigr) \nonumber\\
&&= {1\over 4} t_1^2  \Bigl( -1 + {1\over R^2}\Bigr)\,.
\end{eqnarray}
Note that, as mentioned earlier, the overlaps have unit normalization.
Let us now calculate the terms that arise from the interaction. Because
of momentum conservation there are no
$t_1^3 $ or $t_1 t_0^2$ terms.  There is only a $t_1^2 t_0$ coupling, which is
readily calculated as
\begin{equation}
{1\over 3} \cdot 3 \cdot {t_1\over 2} \cdot {t_1\over 2} t_0 \cdot
2\cdot \langle c_1 e^{iX/R} \,, ce^{-iX/R}, c \rangle = {1\over 2} t_0 t_1^2
K^{3 - {2\over R^2}}\,.
\end{equation}
Let us explain the origin of the various factors.  The first $1/3$ is the
one that comes with the interaction term in the action.  The factor
of 3 is because there are three possible places to insert the operator
associated with $t_0$.   The factors of $t_1/2$ and $t_0$ come
from the field expansion, and the factor of two arises because there
are two ways in which the momentum can be conserved.  The correlator
has been evaluated in a way similar to the previous computation that led
to (\ref{tachver}).  Indeed, the only difference is that the conformal
dimension of two of the operators has been shifted from $-1$ to
$-1 + {1\over R^2}$.

Collecting now our results and using the previously
calculated potential for $t_0$ (\ref{eq:CFT-cubic-potential}) we find
\begin{equation}
\mathcal{V} (t_0, t_1) = -{1\over 2} t_0^2 - {1\over 4} \Bigl( 1- {1\over
R^2}\Bigr) t_1^2  + {1\over 3} K^3 t_0^2 + {1\over 2} t_0 t_1^2
K^{3 - {2\over R^2}}\,.
\end{equation}
The original tachyon is still there: it corresponds to the field $t_0$, which
in the present expansion has no momentum.  For $R>1$, the field $t_1$ is
also a tachyon.  This field is present because of the instability to
form a D1-brane.  Indeed, for $R>1$ the energy of the D1-brane is
larger than the energy of the D0-brane, and the decay is possible.
For $R<1$, the D0-brane has more energy than the D1-brane. In this
case, it is
not clear if some high level computation can exhibit the D0-brane
as a solution of the D1-brane field theory.
We return to this problem in the next subsection.

Let's take $R=\sqrt{3}$.  In this case the potential $\mathcal{V}(t_0, t_1)$
has a critical point which represents a lump:  $t_0 \simeq 0.18$ and $t_1 = -
0.34$.  Of course there is also the conventional tachyon vacuum solution
with $t_0 =1/K^3$ and $t_1 =0$.  With these two solutions, one can
readily compute the ratio $r(\sqrt{3})$ in (\ref{predsecconjecture2}).
We find $r(\sqrt{3}) \simeq 0.774$ in this lowest order calculation.
The result is certainly quite good.   This computation is called a
level (1/3; 2/3) computation since the highest level field $t_1$ has
level $1/3 = 1/R^2$, and we kept terms in the potential up to level $2/3$.
A computation at level (2,4) gives $r\simeq 1.02$, and
for level (3,6) one finds $r \simeq 0.994$.  The convergence to the
answer is quite spectacular.  This computation includes the tachyon
harmonics $t_1$, $t_2$, and $t_3$, as well as fields from the second
level and their first harmonics.  No higher
level computations have been done for this problem.
The computations are not completely universal since the Virasoro
structure of the state space depends on the radius of the circle.
For rational values of $R$ one may find null states, so this is why
we took $R$ irrational.  Even for $R$ irrational, not all states
can be written as Virasoro descendents of the vacuum $|0\rangle$.
New primaries (and their descendents) are needed starting at
level 4.

Since we are equipped with the tachyon harmonics, one is able
to construct explicitly the tachyon profile for the lump solution
which represents the D0-brane.  As the level is
increased, the
profile appears to settle into a well-defined limit.  That same
profile appears to arise for various values of the radius $R$ of
the circle used for the computation.   The profile is roughly of the
form
\begin{equation}
T(x) \simeq  a + b \, e^{-x^2/(2\sigma^2)} \,,  \quad a \simeq 0.56, \quad
b\simeq -0.83, \quad \sigma \simeq 1.52 \,.
\end{equation}
The $\sigma$ width of the lump is therefore about $1.5 \sqrt{\alpha'}$.
The significance (or gauge independence) of this  width is not clear.
Nevertheless, it is interesting that D-branes, which are defined
by definite positions in CFT, appear as thick objects in SFT.  Physical
questions regarding D-branes are expected to have identical answers in
the two approaches.

The above computations have been generalized to the case of
lump solutions of codimension two. In this case, we can imagine a
D2-brane wrapped on a torus $T^2$ which decays into a D0-brane.
The results in the level expansion appear to confirm that the
lump solutions do represent D0-branes.  Less accurate results
are obtained; the energy
has only been estimated with about ten percent
accuracy.

\medskip
The above results have simple analogs in field theory~\cite{Harvey-Kraus}.
Consider a simple scalar field theory in $p+1$ spatial dimensions, where we
single out a coordinate $x$ for special treatment:
\begin{equation}
S = \int dt d^py dx\Bigl\{ {1\over 2} \Bigl( {\partial \phi\over
\partial t}\Bigr)^2 - {1\over 2} |\nabla_y \phi|^2  -
{1\over 2} \Bigl( {\partial \phi\over
\partial x}\Bigr)^2 - V(\phi)\Bigr\}\,.
\end{equation}
As you can readily verify, time-independent solitonic solutions $\phi(x)$,
which depend only on the coordinate $x$,  are
obtained by solving the second-order ordinary differential equation
\begin{equation}
\label{eqnprofidlednkn}
{d^2\phi\over dx^2} = V'(\phi(x))\,.
\end{equation}
This equation takes the form of the equation of motion of a unit
mass particle in a one-dimensional potential $-V(x)$.  As an example,
we consider a theory with potential~\cite{Zwiebach-toy}
\begin{equation}
V(\phi) = {1\over 3} (\phi-1)^2 \left(\phi + {1\over 2}\right) \,.
\end{equation}
The potential has a maximum at $\phi=0$ and a local minimum
at $\phi=1$.  At $\phi=0$ the interpretation is that of a D$(p+1)$-brane
with tension
\begin{equation}
T_{p+1} = V(\phi=0) =  {1\over 6}\,.
\end{equation}
As a simple exercise, verify that
\begin{equation}
\phi (x) = 1 - {3\over 2}
\, \hbox{sech}^2 (x/2) \,,
\end{equation}
is a lump solution for this potential.

\noindent
{\em Exercise}:  Show that the lump solution is an object with
tension $T_p = 6/5$.

In string theory the ratio ${1\over 2\pi } {T_p\over T_{p+1}}$ is equal
to one.  In this field theory model with a cubic potential, we find
\begin{equation}
{1\over 2\pi } {T_p\over T_{p+1}} = {1\over 2\pi } {6\over 5} \cdot 6
= {18\over
5\pi} \simeq 1.146\,.
\end{equation}
It is also a familiar result in soliton field theory that the spectrum
of excitations that live on the world-volume of the lump solution
$\bar \phi(x)$
is governed by a Schr\"odinger equation with a potential $V'' (\bar \phi(x))$.
The mass-squared values for the modes that live on the lump coincide
with the Schr\"odinger energies.

There has been some interest in finding potentials that
accurately describe the behavior of the tachyon.  While the kinetic terms
are not standard, the potential
\begin{equation}
\label{pottachecfff}
V(\phi)  = -{1\over 4} \phi^2 \ln \phi^2  \,, \quad  \phi >0\,.
\end{equation}
appears to be an exact effective tachyon potential.  This potential
was obtained~\cite{Minahan:2000ff} in an attempt to construct realistic
tachyon potentials, and was later confirmed to appear in the BSFT approach to
string field theory~\cite{BSFT}.
The tachyon vacuum is at $\phi=0$, and surprisingly
(but correctly!) the tachyon mass goes to infinity at this vacuum.
This is consistent with the conjecture that perturbative open string
degrees of freedom disappear at the tachyon vacuum.

\noindent
{\em Exercise}. Show that $\bar\phi(x) = \exp (-x^2/4)$ is the lump
solution for the potential (\ref{pottachecfff}) and the Schr\"odinger
potential for fluctuations on the lump solution is ${x^2\over 4} -
{3\over 2}$, a
simple harmonic oscillator potential. Finally, confirm that the values of
$m^2$ for the particles that live on the lump are $-1, 0, 1, 2, \ldots$.
This is the expected string spectrum!


\subsection{Open string theory backgrounds}
\label{subsec:thebackgroundsosft}

We mentioned in the last subsection that when the radius $R$ of a
circle on which a D1-brane is compactified becomes small, it is not
known how to represent a D0-brane in the string field theory on the
D1-brane.  When $R < 1$, the energy of the resulting D0-brane is
larger than the energy of the original D1-brane.  Thus, such a
solution would have positive energy with respect to the original
system.  The difficulty of constructing such a D0-brane solution is an
example of a more general, and we believe crucial, question for OSFT:
Does OSFT, either through level truncation or some more sophisticated
analytic approach, admit classical solutions which describe open string
backgrounds with higher energy than the configuration with respect to
which the theory is originally defined?  If OSFT is
to be a truly complete
formulation of string theory, such
solutions must be possible, since  all open string
        backgrounds must be accessible to the theory.

Another problem of this type is to find, either analytically or
numerically, a solution of the OSFT formulated with one D25-brane that
describes {\it two} D25-branes.
It should be just as feasible to go from a vacuum with one
D-brane to a vacuum with two D-branes as it is to go from a
vacuum with one D-brane to the empty vacuum.
Despite some work on this problem~\cite{Ellwood-Taylor-2}, there is as
yet no evidence of a solution.
Several approaches which have
been tried include: {\it i}) following a positive
mass field upward, looking for a stable point; this method seems
to fail because of gauge-fixing problems---the effective
potential often develops a singularity before reaching the energy
$+ T_{25}$, {\it ii}) following the intuition of the RSZ model
(discussed in the following section)
and constructing a gauge transform of the original D-brane
solution which is $\star-$orthogonal to the original D-brane
vacuum.  It can be shown formally that such a state, when added
to the original D-brane vacuum gives a new solution with the
correct energy for a double D-brane; unfortunately, however, we
have been unable to identify such a state numerically in level
truncation.

While so far no progress has been made towards the construction of
solutions with higher energy than the initial vacuum, it is also
interesting to consider the marginal case.  An example of such a
situation is embodied is the problem of translating a single D-brane
of less than maximal dimension in a transverse direction.  It was
shown by Sen and Zwiebach~\cite{Sen-Zwiebach-translate} (in a T-dual
picture) that after moving a D-brane a finite distance of order of the
string length in a transverse direction, the level-truncated string
field theory equations develop a singularity.  Thus, in level
truncation it does not seem possible to move a D-brane a macroscopic
distance in a transverse direction\footnote{Although this can be done
formally~\cite{Kluson}, it is unclear how the formal solution relates
to an explicit expression in the oscillator language.}.  In this case,
a toy model~\cite{Zwiebach-toy} suggests that the problem is that the
infinitesimal marginal parameter for the brane translation ceases to
parameterize the marginal trajectory in field space after a finite
distance, just as the coordinate $x$ ceases to parameterize the circle
$x^2 + y^2 = 1$ near $x = 1$.  Indeed, an explicit
calculation~\cite{Coletti-Sigalov-Taylor} of the
field redefinition needed to take the
OSFT field $A$ associated with the transverse motion to the correct marginal
parameter $a$ shows that this field redefinition has a subleading term
\begin{equation}
A = a + \alpha a^3 + \cdots,
\end{equation}
where $\alpha < 0$.  Thus, as $a$ increases, eventually a point is
reached where $A$ begins to decrease.  This shows that $A$ is not a
good parameter for marginal deformations of arbitrary size.
It would be nice to have a clear understanding of how arbitrary
marginal deformations are encoded in the theory.

            To show that open string field theory is sufficiently general to
address arbitrary questions involving different vacua, it is clearly
necessary to show that the formalism is powerful enough to describe
multiple brane vacua, the D0-brane lump on an arbitrary radius circle,
and translated brane vacua.  It is currently unclear whether the
obstacles to finding these vacua are technical or conceptual.  It may
be that the level-truncation approach is not well-suited to finding
these vacua, and a new approach is needed.


\section{String field theory around the stable vacuum}
\label{sec:VSFT}

The tachyon conjectures state that the classically stable
vacuum is the closed string vacuum.  This
implies that there
should be no open string excitations in this vacuum,
given that the D-brane represented by the original
OSFT has decayed and exists no more.  Without a D-brane
conventional perturbative open string states are not expected
to exist.
If any perturbative states exist in this vacuum, they should be closed
string states, which are only expected to appear in the quantum open
string field theory.

There are two natural questions concerning this conjecture.  First, we
ask: Can it be tested?  For this, we can begin with the original OSFT
on the background of a D25-brane, for example, and use the (numerical)
solution $\Phi_0$ for the tachyon vacuum to  expand the classical
OSFT around the tachyon vacuum and to calculate the
spectrum.  The conjecture requires that no physical states be
encountered. Second, we ask: Is
there a more natural formulation of open string theory around the tachyon
vacuum, in which, for example, the background independence of the
theory might be more manifest?
The theory around the tachyon vacuum,
is, no doubt, rather unusual.  In the tachyon vacuum there
are no apparent physical states, at least none that take any familiar
form.  Physical perturbative states can arise only
from quantum
effects or classically after the theory is shifted to a nontrivial
background that represents some D-brane configuration.

The tachyon vacuum is a rather special
vacuum:  it is the end product of the decay of {\em any} D-brane
configuration. Presumably, the theory at the tachyon vacuum is
independent of the particular version of OSFT used to reach it
upon tachyon condensation,
in the sense that the string field theories associated with different
D-brane configurations should be equivalent under field redefinition
around the stable vacuum of each theory.
If that is the case, there
may exist a theory -- which we can call {\em Vacuum String Field Theory}, or
VSFT -- which formulates the physics of the tachyon vacuum directly,
{\em without} using a D-brane background to reach the tachyon vacuum.

Presently, there is background
dependence in the  formulation of Witten's OSFT; some specific
D-brane background must
be chosen to define the theory, even though this D-brane configuration may be
removed through tachyon condensation.  As a result, even if the theory
is in an abstract sense completely background independent, we are
stuck with some particular choice of ``coordinates'' on the theory
arising from the original choice of background, which may make physics in other
backgrounds rather difficult to disentangle.
The tachyon vacuum is
also a specific background, but it is certainly a choice that is more
canonical than one which picks one out of an infinite number of
D-brane configurations.  There are perhaps two canonical choices: an
infinite number of space-filling D-branes, which has been motivated
from the viewpoint of K-theory~\cite{k-theory}, and a background with
no D-branes whatsoever -- the tachyon vacuum.  In this section we
investigate the second choice.

A strikingly simple formulation
of VSFT was proposed by Rastelli, Sen, and
Zwiebach
(RSZ)~\cite{RSZ}, in which the BRST operator is taken to be purely
contained in the ghost sector.
In this theory, closed-form analytic solutions that represent
D-branes can be found and
take the form of projectors of the star-algebra.  One shortcoming of this VSFT
is that certain computations are singular and require regularization.
It remains to be seen if a regular VSFT exists.

In subsection \ref{sec:vacuum-theory} we describe the form of the OSFT
action when expanded around the classically stable tachyon vacuum.
Subsection \ref{sec:decoupling} describes evidence from Witten's OSFT
that the open
string degrees of freedom truly disappear from the theory in this vacuum.
In \ref{sec:RSZ-model} we introduce and discuss the RSZ model of
VSFT.  Subsection \ref{subsec:slivers} describes an important class of
states in the star algebra: slivers and projectors, which play a key
role in constructing D-branes in the RSZ model, and which may also be
useful in understanding solutions of the Witten theory.  Finally, in
\ref{subsec:closedstringsinosft} we discuss closed strings in OSFT.


\subsection{String field theory in the true vacuum}
\label{sec:vacuum-theory}

We have seen that numerical results from level-truncated string field
theory strongly suggest the existence of a classically stable vacuum solution
$\Phi_0$ to the string field theory equation of motion.  The state
$\Phi_0$, while
still unknown analytically, has been determined numerically to a high degree of
precision.  This state seems like a very well-behaved string field
configuration.  While there is no positive-definite inner product on
the string field Fock space, the state $\Phi_0$ certainly has finite
norm under the natural inner product $\langle V_2| \Phi_0, c_0L_0
\Phi_0\rangle$, and is even better behaved under the product
$\langle V_2| \Phi_0, c_0
\Phi_0\rangle$.  Thus, it is natural to assume that
$\Phi_0$ defines a
classically stable vacuum for the theory, around which we can expand
the action to find a string field theory around the tachyon vacuum.

\medskip
Let $\Phi_0$ be the string field
configuration describing the tachyon vacuum.
This string field satisfies the classical field equation
\begin{equation}\label{feq}
Q \Phi_0 +  \Phi_0 * \Phi_0 = 0 \,.
\end{equation}If $\widetilde\Phi=\Phi -
\Phi_0$ denotes the shifted open string field, then
the cubic string field theory action  (\ref{e1}) expanded around the tachyon
vacuum has the form:
\begin{equation}\label{e2}
S (\Phi_0 + \widetilde\Phi) = S(\Phi_0) \,-\, {1\over g^2}\,\,\bigg[\, {1\over
2} \langle
\,\widetilde\Phi
\,,\, \widetilde Q\,
\widetilde\Phi
\rangle + {1\over 3}\langle \,\widetilde\Phi \,,\, \widetilde\Phi *
\widetilde\Phi \rangle \bigg] \,.
\end{equation}Here $S(\Phi_0)$ is a constant,
which according to the energetics
part of the tachyon conjectures equals the tension
of the D-brane  times its volume
(as before, we assume that the time
interval has unit length so that the action can be identified with the
negative of the potential energy for static configurations).
The kinetic operator $\widetilde Q$ is given in terms
of $Q$ and $\Phi_0$ as:
\begin{equation}\label{e3}
\widetilde Q \widetilde\Phi = Q \widetilde\Phi + \Phi_0 * \widetilde\Phi +
\widetilde\Phi* \Phi_0\, .
\end{equation}More generally, on arbitrary string fields one would define
\begin{equation}\label{e3p}
\widetilde Q A = Q A + \Phi_0 * A - (-1)^{A}  A * \Phi_0\, .
\end{equation}The consistency of the new action (\ref{e2})
is guaranteed from the
consistency of the action in (\ref{e1}). Since neither the inner
product nor the star multiplication have changed,
the identities
in (\ref{l1e2}) still hold. One can also check that the identities
in (\ref{l1e1}) hold when $Q$ is replaced by $\widetilde Q$.
Just as the original action is invariant under the gauge transformations
(\ref{l1e5}), the new action is invariant under
$\delta \widetilde\Phi =\widetilde Q \Lambda + \widetilde\Phi * \Lambda -
\Lambda * \widetilde\Phi$ for any Grassmann-even ghost-number zero state
$\Lambda$.

Since the energy density of the brane represents
a positive cosmological constant,
it is natural to add the constant $-M=-S(\Phi_0)$ to
(\ref{e1}). This will cancel the $S(\Phi_0)$ term in
(\ref{e2}), and will make manifest
the
expected zero energy density in the final vacuum without D-brane. For the
analysis around this final vacuum it suffices therefore to
study the action
\begin{equation}\label{e2p}
S_0 (\widetilde\Phi) \equiv \,-\, {1\over g^2}\,\,\bigg[\, {1\over 2} \langle
\,\widetilde\Phi
\,,\, \widetilde Q\,
\widetilde\Phi
\rangle + {1\over 3}\langle \,\widetilde\Phi \,,\, \widetilde\Phi *
\widetilde\Phi \rangle \bigg] \,.
\end{equation}This string field theory around the stable vacuum has
precisely the
same form as Witten's original cubic string field theory, only with a
different BRST operator $\widetilde Q$, which so far is only determined
numerically.  While this is insufficient for a complete formulation,
it suffices to test the conjecture that open string excitations disappear
in the tachyon vacuum, as we will discuss in Section~\ref{sec:decoupling}

\medskip
The numerical solution for $\Phi_0$ provides a numerical definition
of the string field theory around the tachyon vacuum.  How could we
do better?  If we had a closed form solution $\Phi_0$ available, the problem
of formulating SFT around the tachyon vacuum would be significantly
simplified.  It is not clear, however, that the resulting formulation
would be the best possible one.
Previous experience with background deformations (small and
large) in SFT indicates that even if we knew $\Phi_0$ explicitly and
constructed $S_0(\widetilde\Phi)$ using eq.(\ref{e2p}), this may not
be the most
convenient form of the action.
Typically a nontrivial field
redefinition is
necessary to bring the shifted SFT action to the canonical form
representing the new background~\cite{Sen:1993mh}.
In fact, in some cases, such as
in the formulation of open SFT for D-branes with various values of
magnetic fields, it is possible to formulate the various SFT's
directly~\cite{Kawano:1999fw,Srednicki},
but the nontrivial classical solution relating
theories
with different magnetic fields are not known. This suggests that if a
simple form exists for the SFT action around the tachyon vacuum it might
be easier to guess it than to derive it.

In fact, this is exactly the approach to the formulation  of vacuum
string field
theory (VSFT) taken by Rastelli, Sen, and Zwiebach (RSZ)~\cite{RSZ}.  These
authors postulate that at the tachyon vacuum the
action takes the form
\begin{equation}\label{e2findpxx}
\mathcal{S} (\Phi) \equiv \,-\, K_0\,\,\bigg[\, {1\over 2} \langle
\,\Phi \,,\, \mathcal{Q}\, \Phi
\rangle + {1\over 3}\langle \,\Phi \,,\, \Phi *
\Phi \rangle \bigg] \,,
\end{equation}where the new kinetic operator
$\mathcal{Q}$ is an operator build solely out of ghosts fields.
If this gives a consistent theory at the tachyon vacuum, they argue,
their choice of $\mathcal{Q}$ must be field redefinition equivalent
to the $\widetilde Q$ that arises directly by shifting the original OSFT action
with the tachyon solution $\Phi_0$.
We discuss the RSZ model in  section~\ref{sec:RSZ-model}.


\subsection{Decoupling of open strings}
\label{sec:decoupling}

It may seem surprising to imagine that {\it all} the perturbative open
string degrees of freedom will vanish at a particular point in field
space, since these are all the
apparent
degrees of freedom available in the
theory.  This is not a familiar phenomenon from quantum field theory.  To
understand how the open strings can decouple, it may be helpful to
begin by considering the simple example of the (0, 0) level-truncated
theory.  In this theory, the quadratic terms in the action become
\begin{equation}
  -\int d^{26} p \;
\phi (-p) \left[ \frac{p^2 -1}{2}  + g \bar{\kappa} \left(
            \frac{16}{27}  \right)^{p^2} \cdot 3 \langle \phi \rangle \right]
\phi (p) \,.
\end{equation}
Taking
$\langle \phi \rangle = 1/3 \bar{\kappa} g$, we find that the
quadratic term is a transcendental expression which does not vanish
for any real value of $p^2$.  Thus, this theory has no poles, and the
tachyon has decoupled from the theory.  Of course, this is not the
full story, as there are still finite complex poles.  It does, however
suggest a mechanism by which the nonlocal parts of the action (encoded
in the exponential of $p^2$) can remove physical poles.

To get the full story, it is necessary to continue the analysis to
higher level.  At level 2, there are 7 scalar fields,
the tachyon and the 6 fields associated with the Fock space states
\begin{eqnarray}
(\alpha_{-1} \cdot \alpha_{-1})| 0_1, p \rangle &  &
b_{-1} \cdot c_{-1}| 0_1, p \rangle \nonumber\\
c_{0} \cdot  b_{-1}| 0_1, p \rangle &  &
(p \cdot \alpha_{-2})| 0_1, p \rangle \\
(p \cdot \alpha_{-1})^2| 0_1, p \rangle &  &
(p \cdot  \alpha_{-1}) c_0b_1| 0_1, p \rangle  \nonumber
\end{eqnarray}
Note that in this analysis we cannot fix Feynman-Siegel gauge, as we
only believe that this gauge is valid for the zero-modes of the scalar
fields in the vacuum $\Psi_0$.  An attempt at analyzing the spectrum
of the theory in Feynman-Siegel gauge using level truncation has been
made~\cite{ks-open}, but gave no sensible results\footnote{Note added:
Recently, Giusto and Imbimbo have carried out a more detailed analysis
of the spectrum around the stable vacuum in Feynman-Siegel gauge
\cite{Giusto-Imbimbo}.  Their more careful analysis shows that
spurious poles found in \cite{ks-open} are cancelled when the
truncation level is sufficiently high.  This approach gives a nice
confirmation of the results of \cite{Ellwood-Taylor-spectrum}, while
working in a fixed gauge, and extends these results by having
sensitivity to cohomology associated with states which are closed for
all momentum, but not exact at discrete values of momentum (type A
states in the notation of \cite{Giusto-Imbimbo}).}.  Diagonalizing the
quadratic term in the action on the full set of 7 fields of level
$\leq 2$, we find~\cite{Ellwood-Taylor-spectrum} that poles develop at
$M^2 = 0.9$ and $M^2 = 2.0$ (in string units, where the tachyon has
$M^2 = -1$).  These poles correspond to states satisfying $\tilde{Q}
\tilde{\Psi} = 0$.  The question now is, are these states physical?
If they are exact states, of the form $\tilde{\Psi} = \tilde{Q}
\tilde{\Lambda}$, then they are simply gauge degrees of freedom.  If
not, however, then they are states in the cohomology of $\tilde{Q}$
and should be associated with physical degrees of freedom.
Unfortunately, we cannot precisely determine whether the poles we find
in level truncation are due to exact states, as the level-truncation
procedure breaks the condition $\tilde{Q}^2 = 0$.  Thus, we can only
measure {\it approximately} whether a state is exact.  A detailed
analysis of this question was carried out by Ellwood and
Taylor~\cite{Ellwood-Taylor-spectrum}.  In their paper, all terms in
the SFT action of the form $\phi_i \; \psi_j (p) \; \psi_k (-p)$ were
determined, where $\phi_i$ is a scalar zero-mode, and $\psi_{j, k}$
are nonzero-momentum scalars.  In addition, all gauge transformations
involving at least one zero-momentum field were computed up to level
(6, 12).  At each level up to $L = 6$, the ghost number 1 states in
the kernel ${\rm Ker} \;\tilde{Q}^{(1)}_{(L, 2L)}$ were computed.  The
extent to which each of these states lies in the exact subspace was
measured using the formula
\begin{equation}
\% \;{\rm exactness} = \sum_{i}\frac{(s \cdot e_i)^2}{ (s \cdot s)}
\end{equation}
where $\{e_i\}$ are an orthonormal basis for ${\rm Im} \;
\tilde{Q}^{(0)}_{ (L, 2L)}$, the image of $\tilde{Q}$ acting on the
space of ghost number 0 states in the appropriate level truncation.
(Note that this measure involves a choice of inner product on the Fock
space; several natural inner products were tried, giving roughly
equivalent results).  The result of this analysis was that up to the
mass scale of the level truncation, $M^2 \leq L -1$, all the states in
the kernel of $\tilde{Q}^{(1)}$ were $\geq 99.9\%$ within the exact
subspace, for $L \geq 4$.  This result seems to give very strong
evidence for Sen's third conjecture that there are no perturbative
open string excitations around the stable classical vacuum $\Psi_0$.
This analysis was only carried out for even level scalar fields; it
would be nice to check that a similar result holds for odd-level
fields and for tensor fields of arbitrary rank.

Another more abstract argument that there are no open string states in
the stable vacuum was given by Ellwood, Feng, He and Moeller~\cite{efhm}.
These authors argued that in the stable vacuum, the identity state $|
I \rangle$
in the SFT star algebra, which satisfies
$ I \star A = A$ for a very general class of string fields
$A$, seems to be an exact state,
\begin{eqnarray}
| I \rangle
           & = &  \tilde{Q} | \Lambda
\rangle\,. \label{eq:i-exact}
\end{eqnarray}
If indeed the identity is exact, then it follows
immediately that the cohomology of $\tilde{Q}$ is empty, since
$\tilde{Q}A = 0$ then implies that
\begin{eqnarray}
A  =   (\tilde{Q} \Lambda)  \star A
            =  \tilde{Q} (\Lambda \star A) -\Lambda \star \tilde{Q} A
=  \tilde{Q} (\Lambda \star A) \,.
\end{eqnarray}
So to prove that the cohomology of $\tilde{Q}$ is trivial, it suffices
to show that $\tilde{Q} | \Lambda \rangle = | I \rangle$.
While there are some subtleties  involved with the identity
string field, Ellwood {\it et al.} found a very elegant expression for
this field,
\begin{eqnarray}
| I \rangle & = &
\left( \cdots e^{\frac{1}{8} L_{-16}} e^{\frac{1}{4} L_{-8}}
e^{\frac{1}{2} L_{-4}}\right)
e^{L_{-2}} | 0 \rangle\,.
\end{eqnarray}
(Recall that $| 0 \rangle= b_{-1}| 0_1 \rangle$.)
They then looked numerically for a state $| \Lambda \rangle$
satisfying (\ref{eq:i-exact}).  For example, truncating at level $L = 3$,
\begin{eqnarray}
| I \rangle & = & | 0 \rangle + L_{-2}| 0 \rangle
+ \cdots\label{eq:identity-2}\\
           & = & | 0 \rangle-b_{-3} c_{1}| 0 \rangle -2b_{-2}c_0| 0 \rangle
           +\frac{1}{2} (\alpha_{-1} \cdot \alpha_{-1})| 0 \rangle
+ \cdots\nonumber
\end{eqnarray}
while the only candidate for $|\Lambda\rangle$ is
\begin{equation}
| \Lambda \rangle = \alpha \; b_{-2}| 0 \rangle,
\end{equation}
for some constant $\alpha$.  The authors of ~\cite{efhm} showed that
the state (\ref{eq:identity-2}) is best approximated as exact when
$\alpha \sim 1.12$; for this value, their measure of exactness becomes
\begin{equation}
\frac{\left| \tilde{ Q} |\Lambda \rangle -| I \rangle \right
            |}{| I |}  \rightarrow 0.17,
\end{equation}
which the authors interpreted as a $17\%$ deviation from exactness.
Generalizing this analysis to higher levels, they found at levels 5,
7, and 9, a deviation from exactness of $11\%, 4.5\%$ and $3.5\%$
respectively.  At level 9, for example, the identity field has 118
components, and there are only 43 gauge parameters, so this is a
highly nontrivial check on the exactness of the identity.  Like the
results of Ellwood and Taylor~\cite{Ellwood-Taylor-spectrum}, these
results strongly support the conclusion that the cohomology of the
theory is trivial in the stable vacuum.  In this case, the result
applies to fields of all spins and all ghost numbers.

Given that the Witten string field theory seems to have a classical
solution with no perturbative open string excitations, in accordance
with Sen's conjectures, it is quite interesting to ask what the
physics of the  string field theory in the
stable vacuum
should describe.  One natural assumption might be that this theory
should include closed string states in its quantum spectrum.
Unfortunately, addressing this question requires performing
calculations in the quantum theory around the stable vacuum.  Such
calculations are quite difficult (although progress in this direction
has been made by Minahan in the $p$-adic version of the theory~\cite{Minahan}).
Even in the perturbative vacuum, it is difficult to systematically
study closed strings in the quantum string from theory.  We discuss
this question again in the final subsection of this section.


\subsection{Pure ghost Vacuum String Field Theory}
\label{sec:RSZ-model}

Our discussion in Section~\ref{sec:vacuum-theory} suggests that
a VSFT may be formulated as a cubic string field theory, with
some new choice $\mathcal{Q}$ for the kinetic operator.
The choice of $\mathcal{Q}$ will be required to satisfy the following
properties:
\begin{itemize}

\item
The operator $\mathcal{Q}$ must be of ghost number one
and must satisfy the conditions (\ref{l1e1})
that guarantee gauge invariance of the string action.

\item  The operator $\mathcal{Q}$ must have vanishing cohomology.

\item The operator $\mathcal{Q}$ must be universal, namely, it must
be possible to
write without reference to the brane boundary conformal field theory.

\end{itemize}

The first condition is unavoidable; the theory must be gauge
invariant if it is to be consistent.  The second condition is
reasonable, but perhaps stronger than needed: all we probably
know is that there should be no cohomology at ghost number one,
which is the ghost number at which physical states appear.
The third constraint is the most stringent one.  It
           implies that VSFT is an intrinsic theory that can be formulated
without using an auxiliary D-brane.

\smallskip
The simplest possible choice is $\mathcal{Q}=0$, which
gives the
purely cubic version of open string field theory~\cite{CUBIC}. Indeed, it
has
long been tempting to identify the tachyon vacuum with a theory where
the kinetic operator vanishes because, lacking the kinetic term,
the string field gauge symmetries are not spontaneously broken.
Nevertheless, there are well-known complications with this identification.
The string field shift $\bar
\Phi$ that relates the cubic to the purely cubic OSFT
           appears to satisfy $Q \bar\Phi=0$ as well as $\bar\Phi*\bar
\Phi=0$. The
tachyon condensate definitely does not satisfy these two identities.
We therefore search
for nonzero $\mathcal{Q}$.

We can satisfy the three requirements by letting
${\mathcal{Q}}$ be constructed purely from ghost operators. In
particular we claim
that the ghost number one operators
\begin{equation}\label{cn}
{\mathcal{C}}_n \equiv  c_n +  (-)^n \, c_{-n}  \,, \quad n=0,1,2,\cdots
\end{equation}satisfy the properties
\begin{eqnarray} \label{ecp}
&& {\mathcal{C}}_n {\mathcal{C}}_n  = 0, \nonumber \\
&&{\mathcal{C}}_n (A * B) = ({\mathcal{C}}_n A) * B + (-1)^{A} A *
({\mathcal{C}}_n B)\,, \\
&& \langle \, {\mathcal{C}}_n A , B \,\rangle = - (-)^A \langle A ,
{\mathcal{C}}_n B \rangle
\,.\nonumber
\end{eqnarray}
The first
property is manifest.
The last property
follows
because under BPZ conjugation $c_n \to (-)^{n+1} c_{-n}$.
The second property
follows from the
conservation laws~\cite{Rastelli-Zwiebach}
\begin{equation}\langle V_3| ({\mathcal{C}}_n^{(1)} + {\mathcal{C}}_n^{(2)} +
{\mathcal{C}}_n^{(3)} ) = 0\,.
\end{equation}These conservation laws arise by consideration of
integrals of the form
$\int dz  c(z) \varphi(z)$ where $\varphi(z) (dz)^2$ is a globally
defined quadratic differential.

Each of the operators ${\mathcal{C}}_n$ has
vanishing cohomology.  To see this note that  for each
$n$ the operator $B_n ={1\over 2} (b_n +  (-)^n \, b_{-n}) $ satisfies
$\{ {\mathcal{C}}_n , B_n \} = 1$.  It then
follows that whenever ${\mathcal{C}}_n \psi=0$, we
have $\psi = \{  {\mathcal{C}}_n , B_n \} \psi = {\mathcal{C}}_n (
B_n \psi)$, showing that
$\psi$ is  ${\mathcal{C}}_n$ trivial.
Since they are built from ghost oscillators, all ${\mathcal{C}}_n$'s
are manifestly universal.

It is clear from the structure of the consistency conditions that we can take
$\mathcal{Q} = \sum_{n=0}^\infty a_n \, {\mathcal{C}}_n$, where the
$a_n$'s are constant
coefficients.   As we will see below, many properties of the RSZ theory
follow simply from the fact that $\mathcal{Q}$ is pure ghost.  But,
there are some
computations that may require a choice of $\mathcal{Q}$ (more on this
later).  The
work of Hata and Kawano~\cite{Hata-Kawano}  gave the clue for the choice of
$\mathcal{Q}$ taken by RSZ:
\begin{eqnarray} \label{ep1}
           \mathcal{Q} &=&{1\over 2i} (c(i) - \bar c(i)) = {1\over 2i} (c(i)
  - c(-i)) =
\sum_{n=0}^\infty
(-1)^n {\mathcal{C}}_{2n}\, , \nonumber\\
&=& c_0 - (c_2 + c_{-2}) +  (c_4 + c_{-4}) - \cdots \,.
\end{eqnarray}
Since the canonical zero-time open string in the complex $z$-plane is the
half-circle $|z|=1$ that lies on the upper half plane, the operator
$c(i)$ represents
a ghost insertion precisely at the open string midpoint. This is the
most delicate
point on the open string  given that the three string interaction is
a world-sheet
with a curvature singularity at the point where the three string
midpoints meet.
The other
operator
$c(-i)$ is needed in order that
$\mathcal{Q}$ is twist invariant (see the first equation in  (\ref{l1e4})).
With this choice of
${\mathcal Q}$,
       the string field action is written  as
\begin{equation}\label{eg1}
S = -K_0 \Big[ {1\over 2} \langle \Phi, \mathcal{Q} \Phi\rangle +
{1\over 3} \langle \Phi, \Phi * \Phi \rangle \Big]\, ,
\end{equation}where the overall
normalization $K_0$  turns out to be infinite.
Although the constant $K_0$ can be absorbed into a rescaling of $\Psi$,
this changes the normalization of
${\mathcal Q} $. We shall instead choose a
convenient normalization of $\mathcal{Q}$ and keep the constant $K_0$ in the
action as in eq.(\ref{eg1}).

In this VSFT the ansatz was made that any D$p$-brane
           solution takes the factorized form~\cite{RSZ-2}
\begin{equation}\label{eo3}
\Phi = \Phi_m \otimes \Phi_g\, ,
\end{equation}where $\Phi_g$
denotes a state obtained by acting with the ghost
oscillators on the SL(2,R) invariant vacuum of the ghost CFT, and
$\Phi_m$  is a
state obtained by acting with matter operators on the SL(2,R)
invariant vacuum of the matter CFT.
Let us denote by
$*^g$ and $*^m$ the star product in the ghost and matter sector
respectively.
Eq.(\ref{e2}) then factorizes as
\begin{equation}\label{eo4}
\mathcal{Q} \Phi_g = - \Phi_g *^g \Phi_g \,,
\end{equation}and
\begin{equation}\label{eo5}
\Phi_m = \Phi_m *^m \Phi_m\, .
\end{equation}This last equation is particularly simple:  it states
that $\Phi_m$
is a projector
(a projector $P$ in an algebra with product $*$ is an element that
satisfies $P* P = P$).
The equation for $\Phi_g$ appears to be more complicated.

For
any string field configuration
$\Phi$ that satisfies the equation of motion, the action is given by
\begin{equation}
S = -{K_0\over 6}  \langle \Phi, \mathcal{Q} \Phi \rangle \,,
\end{equation}
and with the ansatz  (\ref{eo3}) this becomes
\begin{equation}
S = -{K_0\over 6}  \langle \Phi_g |  \mathcal{Q} \Phi_g \rangle
\langle \Phi_m |
\Phi_m\rangle\,,
\end{equation}
Here the inner products are the BPZ ones for the separate matter and
ghost conformal field theories.  For any static
solution, the action is equal to minus the potential energy.  If we
are  describing a
D$p$-brane, the action is equal to minus the volume of the brane
times the tension of the
brane.

To proceed further it is assumed that the ghost part $\Phi_g$ is
universal for all
D$p$-brane solutions. Under this assumption the ratio of energies
associated with two different D-brane solutions with matter parts
$\Phi_m'$ and $\Phi_m$ respectively, is given by:
\begin{equation}\label{eo7}
{E'\over E} = {\langle \Phi_m' | \Phi_m'\rangle_m \over \langle \Phi_m |
\Phi_m\rangle_m} \, .
\end{equation}Thus the ghost part drops out of this calculation.
The inner products in the above right-hand side
include brane volume factors, which once removed, give
us brane tensions. Equation
(\ref{eo7}) has allowed
some important tests of VSFT.
If solutions $\Phi_m'$ and $\Phi_m$ are available, one
can calculate the ratio of tensions of D-branes. Since the ratios
are known, one has a test of VSFT.   The solutions, as mentioned before, are
projectors of the star algebra.  The D25-brane solution, for example,
can be represented by the sliver state $|\Xi\rangle$, which is the first
example of a star-algebra projector that was discovered.  The sliver state
can be constructed for any conformal field theory (a brief discussion is
given in the following subsection).
Similarly, D$p$-brane solutions can be obtained as modified slivers,
and numerical verification that the correct ratio of tensions emerges
was obtained~\cite{RSZ-2}.  Subsequently,  and equipped with a better
understanding
of the star-algebra, Okuyama~\cite{Okuyama:2002tw} was able to demonstrate
analytically that the correct ratio of tensions emerges.

In a series of stimulating papers~\cite{Hata-Kawano,VSFT-tension}, 
Hata, Kawano,
and Moriyama, showed that the relationship $2\pi^2  g^2 T_{25} =1$ between the
D25-brane tension  and the string coupling can be tested in VSFT
without knowledge of
the  explicit form of the purely ghost
$\mathcal{Q}$. In other words, the normalization of the action, the infinite
constant $K_0$,  does not feature in the computation.  This is  easy to see.
The D-brane tension,  which is proportional to the value of the
action evaluated
on the sliver solution, is linearly proportional to $K_0$.
In order to calculate the string coupling, Hata and Kawano proposed to look
for the tachyon state on the D-brane; this state should appear as a fluctuation
around the sliver solution. With this tachyon state, the string
coupling $g$ can
be obtained as the coupling for three on-shell tachyons.   The effective
action for the tachyon fluctuation $t$ would take the form
\begin{equation}
           K_0 \Bigl(  \alpha\,{1\over 2} \, t\, (\partial^2 + 1) t  + {1\over
3}\, \beta\, t^3 \Bigr)\,,
\end{equation}
where $\alpha$ and $\beta$ are calculable finite constants.  The
field rescaling
$t= T/ \sqrt{K_0 \alpha}$ brings this action to canonical form
\begin{equation}
           \,{1\over 2} \, T\, (\partial^2 + 1) T  + {1\over 3}\, {\beta\over
\sqrt{K_0 \alpha} }\, T^3\,,
\end{equation}
and the string coupling can be read $g = \beta/\sqrt{K_0 \alpha}$.
Since $T_{25} \sim
K_0$, the relation $2\pi^2  g^2 T_{25} =1$ does not involve $K_0$.
The original
computations, however, did not work out, because the tachyon state had been
incorrectly identified~\cite{0111153}.  In a remarkable 
work~\cite{Okawa}, Okawa
gave a correct identification of the tachyon state and demonstrated that
the relation between the string coupling and the brane tension works
out correctly.  Still both the string coupling and the brane tension
are singular.

\medskip
It is interesting to wonder what features
    of VSFT that depend on the particular
choice of pure ghost operator $\mathcal{Q}$.  It appears that
a completely regular definition
of the spectrum of strings around D-brane solutions may involve
$\mathcal{Q}$.  Indeed,  Okawa has recently demonstrated that
the knowledge of $\mathcal{Q}$  is necessary to produce VSFT
solutions that give rise to a string coupling and brane tension both
of which are finite~\cite{okawatbp1}.
The specific form of $\mathcal{Q}$  may
also be needed for  the calculation of closed string
amplitudes using VSFT.   It is clear, however, that the choice in (\ref{ep1})
is rather special.  We remarked earlier that the equation (\ref{eo4}) for the
ghost part of the solution is not just a projector equation.  It
turns out, however,
that there is a twist of the ghost CFT of $(b,c)$ in which the
antighost becomes
a field of dimension one and ghost becomes a field of dimension zero. The
new CFT has central charge $c=-2$.  If $\mathcal{Q}$ is given by
(\ref{ep1}), the solution
of (\ref{eo4}) is simply the sliver state of the twisted conformal field
theory.~\cite{RSZ-closed}

We conclude this subsection with
some comments on
regularization and the singular aspects of VSFT.  Arguments by
Gross and Taylor~\cite{Gross-Taylor-II}, and by Schnabl (unpublished)
indicated that the brane tension associated with VSFT solutions is
zero for any finite $K_0$.  Numerical experiments confirm these
arguments.  A regulation scheme
was developed by Gaiotto {\em et.al} ~\cite{RSZ-closed} in which
$K_0$ is replaced by $K_0(a)$, and the gauge-fixed kinetic operator
of VSFT  is made $a$-dependent in such a way that for for infinite $a$
the pure ghost operator is recovered.  The $K_0(a)$ divergence
as $a\to \infty$ is determined from the requirement that the D-brane
tension is correctly reproduced.   The regulated theory appears to be
well defined, but universality is lost in the regulation.  On the other
hand, the analysis of the regulated equations led to the discovery
of another special projector of the star algebra: the butterfly
state~\cite{RSZ-closed,Schnabl,RSZ-projectors}.

We noted in section~\ref{sec:vacuum-theory} that
after a shift to the tachyon vacuum the open string field theory on
a D25-brane becomes a cubic string
field theory with kinetic operator $\widetilde Q$.  This operator is
not made solely
of ghosts.  We would expect, however,  that the RSZ theory, if fully
correct, is
field redefinition equivalent to the theory with $\widetilde Q$.
If we consider the action (\ref{e2p}),
a  homogeneous field redefinition of the type
\begin{equation}\label{jkl}
\widetilde\Phi = \, e^K \, \Phi\,,
\end{equation}has special properties if $K$ is a ghost number zero Grassmann
even operator that satisfies the following relations
\begin{eqnarray} \label{efderp}
&& K (A * B) = (K A) * B +  A * (K B)\,, \nonumber  \\
&& \langle \, K A , B \,\rangle = - \langle A , K B \rangle \,.
\end{eqnarray}
These properties guarantee that the form of the
cubic term is unchanged, and
that, after the field redefinition, the action takes the form
\begin{equation}\label{e2findp}
S (\Phi) = \,-\, {1\over g^2}\,\,\bigg[\, {1\over 2} \langle
\,\Phi \,,\, {\widehat Q}\, \Phi
\rangle + {1\over 3}\langle \,\Phi \,,\, \Phi *
\Phi \rangle \bigg] \,,
\end{equation}where
\begin{equation}
\label{kjh}
{\widehat Q} =  \, e^{-K} \, \widetilde Q \, e^K\,.
\end{equation}It is a good exercise to verify that equations (\ref{efderp})
guarantee that $\widetilde Q$ satisfies the properties listed
in  (\ref{l1e1}).  Therefore the new action is consistent.

The operator $\widetilde Q$ is, by construction,  regular, while
$\widehat Q$, which we
want to be equal to the VSFT operator $\mathcal{Q}$, should be an infinite
constant times a ghost insertion at the string midpoint
(the infinite constant is necessary because $g$ is finite).  A large
class of string
reparameterizations that leave the open string midpoint invariant
can be constructed with operators $K$ that satisfy the relations
(\ref{efderp}).
A reparameterization in which a finite  part of the string is squeezed
into an infinitesimal neighborhood of the string midpoint will turn a regular
$\widetilde Q$ that contains a term linear in the ghost field, into an operator
$\widehat Q$ whose leading term is precisely
           a divergent ghost insertion at the
string midpoint.~\cite{RSZ-closed}    This happens because the term linear
in the ghost field is the term with an operator of lowest possible dimension,
and a squeezing transformation, will transform  this
negative-dimension operator
with an infinite factor.  It is thus plausible that a singular
squeezing transformation
relates the string field theory around the tachyon vacuum
to the RSZ theory.


\subsection{Slivers and projection operators}
\label{subsec:slivers}


{}From the point of view of the RSZ approach to VSFT just discussed,
projection operators  of the star algebra play a
crucial role in the construction of solutions of the theory.
Such projection operators may also be useful in understanding
solutions in the original Witten theory.
Quite a bit of work has been
done on constructing and analyzing  projectors in the star
algebra since the RSZ model was originally proposed.  Without
going into the technical details,
we now briefly review some of the important features of projectors.

\begin{figure}[!ht]
\leavevmode
\begin{center}
\epsfxsize = 11 cm \epsfbox{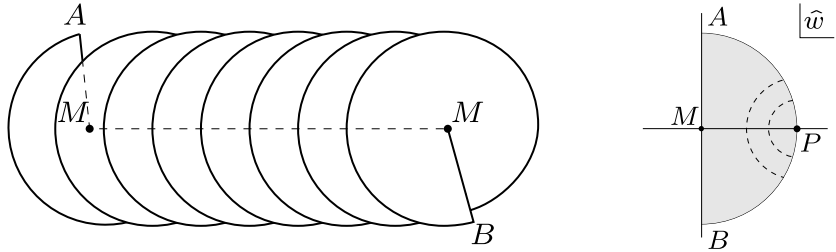}
\end{center}
\caption[]{\footnotesize  The
sliver appears as a cone with infinite excess angle-- namely,
an infinite helix. The segments $AM$ and $BM$ represent the
left-half and the right-half of the string.  The local
coordinate patch,
represented by the shaded half disk shown to the right, must be
glued in to form the complete surface.
} \label{f9}
\end{figure}

The first matter projector which was explicitly constructed is the
``sliver'' state.  This state was identified as a  conformal field theory
surface state by Rastelli and Zwiebach~\cite{Rastelli-Zwiebach}. As such,
there is a surface associated with the state: a disk with one puncture
on the boundary and a specified local coordinate at the puncture.  This
conformal field theory picture gives a complete state; it includes both
the matter and the ghost part of the state.  Moreover, the state can be
constructed for any conformal field theory:
\begin{equation}\label{slivasvirdesc}
|\Xi\rangle = \exp
\Bigl( -{1\over 3} L_{-2} + {1\over 30} L_{-4} - {11\over 1890} L_{-6} +
{34\over 467775} L_{-8}
           + \cdots \Bigr) | 0\rangle \,.
\end{equation}The geometrical picture of the sliver state is shown in
figure~\ref{f9}.
The full punctured disk is the glued surface obtained
by attaching  the infinite helix and the coordinate patch,
which carries  the puncture~$P$.
There are many
alternative pictures of the sliver.

\begin{figure}[!ht]
\leavevmode
\begin{center}
\epsfxsize = 10 cm \epsfbox{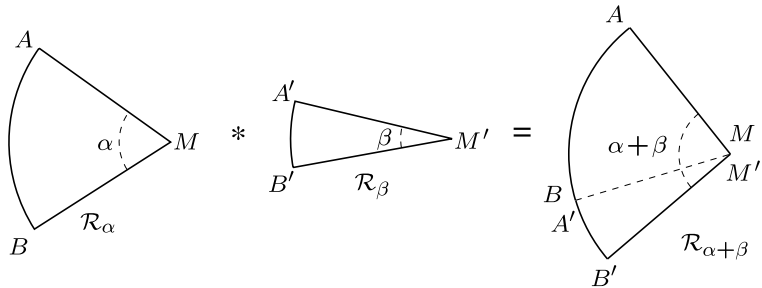}
\end{center}
\caption[]{\footnotesize The star multiplication of a sector
state with angle $\alpha$ to a sector state with angle
$\beta$ gives a sector state with angle $\alpha+ \beta$.
Sector states are just another presentation of wedge
states.
} \label{f8}
\end{figure}

To understand why the sliver state squares to itself one must
have a picture of star multiplication for surface states.
A full discussion~\cite{Rastelli:2001vb} would take too long, but the rough
idea is easily explained.
The sliver state is essentially the limit $\lim_{n \rightarrow \infty}
(| 0 \rangle)^n$, where multiplication is performed via the star product.
A surface state in a BCFT can be viewed (by excising the
coordinate patch) as a disk whose boundary has two parts: a part in which
the boundary condition that defines the BCFT is imposed, and a
part which represents an open string.  To star-multiply two surface
states, one glues the right-half of the string in the first surface to the
left-half of the string in the second surface; the resulting surface is the
surface that represents the star product.   A particularly simple class
of surface states are sector states or wedge states. One such state
$\mathcal{R}_\alpha$
is shown to the left of figure~\ref{f8}.  The BCFT boundary condition
applies to the curved boundary of
the sector. The radial segment AM is
the left-half of the open string and the radial segment $MB$ is the right-half
of the open string.  The sector state is defined by the angle
$\alpha$ at the string
midpoint $M$.  In the figure we show the multiplication of $\mathcal{R}_\alpha$
and $\mathcal{R}_\beta$.  The result is a sector state
$\mathcal{R}_{\alpha + \beta}$ with total angle
$\alpha + \beta$.  The sliver state $\Xi$ is the wedge state
$\mathcal{R}_\infty$
with infinite angle.   It is then clear that the star product of two
slivers is still
a wedge state of infinite angle, and thus also a sliver.   The state
obtained in the
limit when the angle is equal to zero is in fact the identity state of the star
algebra. It is manifestly clear that the product of any surface state with the
identity gives the surface state.  The identity state can also be written
as an exponential of Virasoro operators acting on the vacuum.
In fact,
as mentioned in section \ref{sec:decoupling},
a very curious result was
found~\cite{efhm}:
\begin{eqnarray}
|\mathcal{I} \rangle & = & \left( \prod\limits_{n=2}^{\infty}
               \exp\left\{- \frac{2}{2^n} L_{-2^n}\right\} \right)
         e^{L_{-2}} |0 \rangle \nonumber \\
&=&
\ldots
\exp(-\frac{2}{2^3} L_{-2^3}) \exp(-\frac{2}{2^2} L_{-2^2})
         \exp(L_{-2}) |0 \rangle  \,,
\end{eqnarray}
with the Virasoro operators of higher level stacking to the left.
We thus confirm that the identity is also a Virasoro descendent of the
vacuum.

In an independent construction, Kostelecky and 
Potting~\cite{Kostelecky-Potting}
constructed a state  $\Psi_m$ of the matter sector of the D25-brane BCFT that
squared to itself (up to a proportionality constant). The construction used
the oscillator
language.    This matter state
takes the form of a squeezed state
\begin{equation}
|\Psi_m\rangle = \mathcal{N} \,\exp \left[\frac{1}{2}\,
           a^{\dagger} \cdot S \cdot a^{\dagger} \right]| 0 \rangle \,.
\label{eq:sliver-squeezed}
\end{equation}
By requiring that such a state satisfy the projection equation $\Psi
\star \Psi = \Psi$, and by making some further assumptions about the
nature of the state, an explicit formula for the matrix $S$ was
found in terms of the matrix $X$ from
(\ref{eq:matX})~\cite{Kostelecky-Potting}. Evidence quickly emerged
that the state constructed by these authors is the matter sector of
the sliver state, and a proof was given by Okuda~\cite{Okuda:2002fj}.

\begin{figure}[!ht]
\leavevmode
\begin{center}
\epsfysize=5cm
\epsfbox{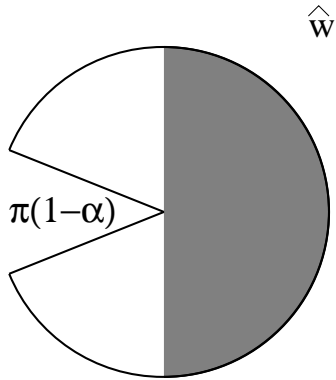}
\end{center}
\caption{\footnotesize  The butterfly state arises in the limit where
$\alpha\to 1$ and the angle indicated in the figure vanishes.} \label{f2butter}
\end{figure}

There are many other projectors that also have a simple
picture as surface states~\cite{Schnabl,RSZ-projectors,Fuchs:2002zz}.
In these projectors, the open string midpoint approaches
(or even coincides with) the boundary of the surface
where the boundary condition is applied.
One particularly useful projector, which arises in the numerical
solution of  VSFT, is  the so-called {\em butterfly} state $\mathcal{B}$.
This is a very interesting state, whose picture is shown in
Figure~\ref{f2butter}.
When one glues two butterfly surfaces in the manner  required by
star-multiplication, the resulting surface does not appear to be, at
first sight,
another butterfly. Nevertheless, the resulting surface is conformally
equivalent
to a butterfly, and this is, in fact, all that is needed in order to have a
projector.   It has been demonstrated that the butterfly is the state
that can be represented as the tensor product $|0\rangle \otimes |0\rangle$,
where $|0\rangle $ is the vacuum of the half-string state
space~\cite{RSZ-projectors}.
Generally, any state of the form $| a \rangle \otimes | a \rangle$
where $| a \rangle$ is the same state in the left and right
half-string Fock spaces is a projector~\cite{Gross-Taylor-I}.
The butterfly has a remarkably simple expression
as a Virasoro descendent of the vacuum
\begin{equation}
\label{butterflyasvirdes}
|\mathcal{B}\rangle = \exp \Bigl( -{1\over 2} L_{-2} \Bigr) |0\rangle \,.
\end{equation}

Projectors  have many properties which are reminiscent
of D-branes.  This relationship between projection operators and
D-branes is familiar from noncommutative field theory, where
projectors also play the role of D-brane solitons~\cite{gms}.
This connection becomes quite concrete in the presence of a background
B field~\cite{Moore-Taylor,b-VSFT}.
In the RSZ theory,
states that describe  an arbitrary but fixed
configuration of D-branes are constructed
by tensoring the matter
projector for the appropriate BCFT  with a fixed ghost state
that satisfies the ghost
equation of motion (\ref{eo4}).   Particular projectors
like the sliver can be constructed which are localized in any number
of space-time dimensions, corresponding to the codimension of a
D-brane.  Under gauge transformations, a rank one projector can be
rotated into an orthogonal rank one projector, so that configurations
containing multiple branes can be constructed as higher rank
projectors formed from the sum of orthogonal rank
one projectors~\cite{RSZ-3,Gross-Taylor-I}.
This gives a very suggestive picture of
how arbitrary D-brane configurations can be constructed in string
field theory.

While this picture is quite compelling, however, there
are some technical obstacles which make this still a somewhat
incomplete story.  In the RSZ model,
singularities appear due to the separation of the matter and ghost sectors.
Moreover, projectors are, in general, somewhat singular states.
For example, the matrix $S$ associated to the matter part of
the sliver state has eigenvalues of
$\pm 1$ for any D$p$-brane~\cite{Moore-Taylor,RSZ-projectors}.  Such
eigenvalues cause
these states to be non-normalizable elements of the matter Fock
space
In the Dirichlet directions, this lack of normalizability
occurs because the
state is essentially localized to a point and is analogous to a delta
function.  In the Neumann directions, the singularity manifests as a
``breaking'' of the strings composing the D-brane, so that the
functional describing the projector state is a product of a function
of the string configurations on the left and right halves of the
string, with no connection mediated through the midpoint.  These
geometric singularities seem to be generic features of the  matter part of
any projector, not just the sliver state~\cite{Schnabl,RSZ-projectors}.  The
singular geometric features of projectors, which can be traced
to the fact that the open string midpoint approaches the boundary,
makes certain calculations in the RSZ theory  somewhat complicated, as all
singularities must be regulated.  Singularities do not
seem to appear in the Witten theory, where the BRST operator and
the numerically calculated solutions seem to behave smoothly at the string
midpoint.  On the other hand, it may be that further study of the
projectors will lead to analytic progress on
the Witten theory, as discussed in a recent paper by Okawa~\cite{okawatbp2}.


\subsection{Closed strings in open string theory}
\label{subsec:closedstringsinosft}

We have discussed in earlier sections the fact that
open string field theory, formulated on the background
of a certain BCFT appears to capture many other open
string backgrounds as solutions of the theory.
Apart from its  singular features, the RSZ theory
admits any BCFT as a solution of the theory.  One
important question remains:  Can closed string backgrounds
be incorporated in open string field theory?  The question
can be answered both in the context of OSFT and in the
context of the RSZ model.  As we will discuss,
there is very little
concrete evidence as yet  that this can be done
in any of the two approaches.
We therefore ask a simpler question:  Can closed string states
be seen in open string field theory?
The answer here is yes, both in OSFT, and in VSFT
(modulo the usual singularities),
although so far this has been understood only in certain limited contexts.

As has been known since the earliest days of the subject, closed
strings appear as poles in perturbative open string scattering
amplitudes.  This was demonstrated explicitly for Witten's theory by
exhibiting the closed string poles arise in the one-loop 2-point
function~\cite{fgst}
(although in this calculation, spurious poles also
appear which complicate the interpretation).  More recently, in a
similar calculation the closed string tadpole generated by the D-brane
was identified in the one-loop open string 1-point
function~\cite{Ellwood-Shelton-Taylor}.  While in principle this type
of argument
can be used to construct all on-shell closed string amplitudes through
factorization, it is much less clear how to think of asymptotic or
off-shell closed
string states in this context.  If Witten's theory is well-defined as a quantum
theory, it would follow from unitarity that the closed string states
should also
arise in some natural sense as asymptotic states of the quantum open
string field
theory.  It is currently rather unclear, however, whether, and if so
how, this might be realized.  There are subtleties in the quantum
formulation of the theory which have never completely been
resolved~\cite{Thorn,Ellwood-Shelton-Taylor}, although most of the
problems of the quantum theory seem to be generated by the closed
string tachyon, and may be absent in a supersymmetric theory.  Both
older SFT literature~\cite{Strominger-closed,Shapiro-Thorn} and recent
work~\cite{Gerasimov-Shatashvili,Moore-Taylor,RSZ-closed,Hashimoto-Itzhaki}
have suggested ways in which closed strings might be incorporated into
the open string field theory, but a definitive resolution of this
question is still not available.

In the RSZ model,  one description of {\em on-shell} closed string
states is reasonably
natural~\cite{RSZ-closed,Hashimoto-Itzhaki,Ambjorn:2002im,Drukker:2002ct}
and scattering amplitudes have been
computed~\cite{Alishahiha:2002as,Takahashi:2003kq}.
For each on-shell closed string vertex operator $V$  one can construct a
gauge-invariant open string
state $\mathcal{O}_V(\Phi)$, where $\Phi$ is the open string field,
and the gauge invariance is the open string gauge invariance.
The world-sheet picture of the state is that of an amputated semi-infinite
strip whose edge represents the open strings, the two halves of which
are glued and the closed string operator is inserted at the conical
singularity.  Given a set of gauge invariant operators associated
with a set of on-shell closed string vertex operators, the RSZ
correlator of the gauge invariant operators appears to give,
up to proportionality factors that need regulation,  the
on-shell closed string amplitude on a surface {\em without}
boundaries.
This result uses a nontrivial and unusual decomposition of
the moduli space of Riemann surfaces without boundaries~\cite{RSZ-closed}.
The decomposition, is related to, but distinct from the one used
in Witten's theory  to cover the moduli space of Riemann surfaces
that have at least one boundary. Other decompositions have been
discussed by Drukker~\cite{Drukker:2002ct}.

If it were possible to encode {\em off-shell} physics naturally
into open string field theory it would be reasonable to hope
that closed string backgrounds could be changed by suitable
expectation values of open string fields
although this would presumably be a subtle effect in the quantum
theory, and difficult to compute explicitly.  Attaining a description
of the full closed string landscape~\cite{landscape} using quantum
OSFT is clearly an optimistic scenario,
but it need not be farfetched;
it may represent an extension
of the AdS/CFT correspondence, in which the CFT side is changed
from SYM into the full open string field theory.
If, as it may be,
it turns out to be that  the closed string sector of the theory
is  encoded  in a singular fashion in OSFT,
        one may be  better off directly working with
closed string field theory~\cite{Zwiebach:1992ie},  or with
open/closed string field
theory~\cite{Zwiebach:1997fe}.
Because of the nonpolynomiality of these theories, it is
not known at  present if
level expansion can be used to extract nonperturbative information.
At any rate, it would be useful to have a clear picture of how far one
can incorporate closed string physics from the open string point of
view. Even if this cannot be realistically achieved in our current models of
SFT, understanding the difficulties involved may help us in our
search for a better formulation of the theory.


\section{Conclusions}
\label{sec:conclusions}

The work described in these lectures has brought the understanding of
string field theory to a new level.  We now have fairly conclusive
evidence that open string field theory can successfully describe
distinct vacua with very different geometrical properties, which are
not related to one another through a marginal deformation.  The
resulting picture, in which a complicated set of degrees of freedom
defined primarily through an algebraic structure, can produce
different geometrical backgrounds as different solutions of the
equations of motion, represents an important step beyond perturbative
string theory.  Such a framework is clearly necessary to discuss questions of a
cosmological nature in string theory. For such questions,
however,  one must generalize from the work described here in which the theory
describes distinct {\it open} string backgrounds, to a formalism
where different
{\it closed} string backgrounds also appear as solutions of the equations.
Ideally, we would like to have a formulation of
string/M-theory in which all the currently understood vacua can arise
in terms of a single well-defined set of degrees of freedom.

It is not yet clear, however, how far it is possible go towards this
goal using the current formulations of string field theory.  It may be
that the correct lesson to take from the work described here is simply
that there {\it are} nonperturbative formulations in which distinct
vacua can be brought together as solutions of a single classical
theory, and that one should search for some deeper fundamental
algebraic formulation where geometry, and even the dimension of
space-time emerge from the fundamental degrees of freedom in the same
way that D-brane geometry emerges from the degrees of freedom of
Witten's open string field theory.  A more conservative scenario,
however, might be that we could perhaps use the current framework of
string field theory, or some limited refinement thereof, to achieve
this goal of providing a universal nonperturbative definition of
string theory and M-theory.  Following this latter scenario, we propose
here a series of questions aimed at continuing the recent developments
in open string field theory as far as possible towards this ultimate
goal.   It is not certain that this research program can be carried to
its conclusion, but it will be very interesting to see how far
open string field theory can go in reproducing important
nonperturbative aspects of string theory.

\medskip
There are, in our mind, two very important concrete problems
related to Witten's string field theory that so far have resisted solution:

\begin{enumerate}

\item[1)] Finding
            an analytic description of the tachyonic vacuum.  Despite several
            years of work on this problem, great success with numerical
            approximations, and some insight from the RSZ vacuum string field
            theory model, we still have no closed form expression for the
            string field $\Phi_0$ which represents the tachyon vacuum
            in  Witten's open string field theory.  It seems almost
            unbelievable that there is not some elegant analytic solution to
            this problem.  An analytic
            solution would almost certainly greatly enhance our understanding of
            this theory and would lead to other significant advances.

\item[2)] Finding certain open string backgrounds as solutions of open
string field theory.  As discussed in section~\ref{subsec:thebackgroundsosft},
we do not know how to obtain a background with multiple
D-branes starting with a background with one D-brane. Nor we
know how to obtain the background which represents a D0-brane
using the background of a D1-brane with lower energy.
It is currently unclear whether the obstacles to finding these
vacua are conceptual or technical.

\end{enumerate}

There are other questions that are probably important to the
future development of string field theory.  These represent,
in our opinion, subjects that merit investigation:

\begin{enumerate}

\item[1)]  Is there a regular formulation of VSFT ?
Such a version of the theory may have further similarities with BSFT
and could turn out to be a complete and flexible formulation
of open string field theory.

\item[2)] How do closed string backgrounds appear
            in open string field theory?  While OSFT and VSFT
            appear to give somewhat singular/intractable descriptions
            of closed string physics,  some better understood, or new,
version of open string theory might provide
            a tractable description of closed string physics.
             Another possibility
            is that  closed string fields are needed in addition to
open string fields; this is the case
            in  light-cone open string field theory and in
covariant open/closed string field theory.

\item[3)] What are the new features of superstring field theory?
The status of the tachyon conjectures for the superstring
has been reviewed by
Ohmori~\cite{Ohmori:2003vq}.
The large set of symmetries of superstring theory makes them, in many
cases, more tractable than bosonic string theories.  Nevertheless,
as of yet, there is no clear sense in which superstring field theory
is simpler than bosonic string field theory~\cite{smet}.
There are also significant conceptual problems that have not
allowed a formulation of vacuum superstring field theory~\cite{Marino:2001ny}.

\item[4)]  How do we describe time-dependent tachyon dynamics?
String field theory gives clear and concrete evidence for the
Sen conjectures.  Although
we have not studied this subject in the present review, there is
much interest in the process by which the tachyon rolls from the
unstable critical point down to the tachyon vacuum.~\cite{rolling-tachyon}
In fact, the early attempts to describe the rolling of the tachyon
in Witten's string field theory~\cite{Moeller:2002vx,Fujita:2003ex} appear to
be in contradiction with the results that follow from conformal
field theory.

\end{enumerate}

            It is  challenging to imagine a
            single set of degrees of freedom which could encode, in different
            phases, all the possible string backgrounds we are familiar with,
including those associated with M-theory.
In principle, a
            nonperturbative background-independent formulation of type II string
            theory should allow one to take the string coupling to infinity in
            such a way that the fundamental degrees of freedom of the theory
remain at some finite point in the configuration space.
This would lead to the vacuum associated with M-theory
            in flat space-time.  It would be quite remarkable if this can be
            achieved in the framework of string field theory.  Given the
            nontrivial relationship between string fields and low-energy
            effective degrees of freedom,
such a result need not be farfetched.
If this picture could be successfully implemented, it
would  give a very satisfying
representation of the
complicated network   of dualities of string and M-theory in terms of
            a single underlying set of degrees of freedom.

\section*{Acknowledgments}
We would like to thank all our collaborators: Nathan Berkovits, Erasmo
Coletti, Michael Douglas, Ian Ellwood, David Gaiotto, David Gross,
Hong Liu, Nicolas Moeller, Greg Moore, Leonardo Rastelli, Ashoke Sen,
Jessie Shelton, and Ilya Sigalov, who helped us understand many of the
issues described in these lectures.  Thanks to the organizers and
students of TASI 2001, who provided a stimulating environment for
these lectures and much relevant feedback.  WT would like to thank the
Kavli Institute for Theoretical Physics for hospitality and support
during the final stages of writing up these lectures. BZ would like
to thank the Harvard University Physics Department for hospitality.
This work was
supported by the DOE through contract \#DE-FC02-94ER40818.

\end{document}